\documentclass[a4paper,journal]{IEEEtran}
 
\usepackage{sty/25OJCOMS_REMS} 
 
\begin{document}


\title{Efficient and Physically-Consistent Modeling \\ of  Reconfigurable Electromagnetic Structures}
\author{
\IEEEauthorblockN{Alexander Stutz-Tirri, Georg Schwan, and Christoph Studer}
\thanks{The authors are with the Department of Information Technology and Electrical Engineering, ETH Zurich, Switzerland. (email: alstutz@ethz.ch, gschwan@ethz.ch,  and studer@ethz.ch)}	
}
\maketitle

\begin{abstract}
Reconfigurable electromagnetic structures (REMSs), such as reconfigurable reflectarrays~(RRAs) or reconfigurable intelligent surfaces~(RISs), hold significant potential to improve \new{the spectral efficiency of} wireless communication \new{systems} and \new{the accuracy of wireless} sensing systems. 
Even though several REMS modeling approaches have been proposed in recent years, the literature lacks models that are both computationally \emph{efficient} and \emph{physically consistent}.
\new{
As a result, algorithms that control the reconfigurable elements of REMSs (e.g., the phase shifts of a RIS) are often built on simplistic and thus inaccurate models.}
To enable physically accurate REMS-parameter tuning, we present a new framework for efficient and physically consistent modeling of \emph{general} REMSs.
Our modeling method combines a circuit-theoretic approach with a new formalism that describes a REMS's interaction with the electromagnetic~(EM) waves in its far-field region.
Our modeling method enables efficient computation of the entire far-field radiation pattern for \emph{arbitrary} configurations of the REMS reconfigurable elements once a \emph{single} full-wave EM simulation of the non-reconfigurable parts of the REMS has been performed. 
The predictions made by \new{our} framework align with the physical laws of classical electrodynamics and model effects caused by inter-antenna coupling, non-reciprocal materials, polarization, ohmic losses, matching losses, influence of metallic housings, noise from low-noise amplifiers, and noise arising in or received by antennas. 
In order to validate the efficiency and accuracy of our modeling approach, we~(i) compare our modeling method to EM simulations and~(ii) conduct a case study involving an RRA that enables simultaneous multiuser beam- and null-forming using a new, computationally efficient, and physically accurate parameter tuning~algorithm.
\\
\end{abstract}


\maketitle


\section{Introduction}
\label{sec:Introduction}
\IEEEPARstart{R}{econfigurable} electromagnetic structures~(REMSs) have sparked significant interest in the wireless communication and sensing literature over the past years. 
Prominent instances of REMSs are reconfigurable reflectarrays (RRAs)~\cite{dardari_dynamic_scatteringarrays_for_simultaneous_electromagnetic_processing_and_radiation_in_holographic_mimo_systems,hao_deng_cao_yin_sarabandi_a_high_aperture_efficiency_1_bit_reconfigurable_reflectarray_antenna_with_extremely_low_power_consumption,zhang_chen_feng_a_dual_polarized_reconfigurable_reflectarray_antenna_based_on_dual_channel_programmable_metasurface,jamali_tulino_mueller_intelligent_surface_aided_transmitter_for_millimeter_wave_ultra_massive_mimo,yang_yang_gao_a_1_bit_10_10_reconfigurable_reflectarray_antenna,hum_perruisseau_carrier_reconfigurable_reflectarrays_and_array_lenses_for_dynamic_antenna_beam_control},  
reconfigurable array lenses~(RALs)~\cite{cao_deng_yin_hao_sarabandi_1_bit_reconfigurable_transmit_and_reflect_array_antenna_using_patch_ground_structure,venkatesh_lu_saeidi_sengupta_a_high_speed_programmable_and_scalable_terahertz_holographic_metasurface_based_on_tilded_cmos_chips,lau_hum_reconfigurable_transmitarray_design_approaches_for_beamforming_applications}, and reconfigurable intelligent surfaces~(RISs)~\cite{di_renzo_zappone_debbah_alouini_smart_radio_environments_empowered_by_reconfigurable_intelligent_surfaces,wu_zhang_zheng_you_zhang_intelligent_reflecting_surface_aided_wireless_communications,wu_zheng_you_zhu_shen_shao_mei_di_zhang_basar_song_di_renzo_luo_zhang_intelligent_surface_empowered_wireless_network_recent_advances_and_the_road_to_6g,an_yuen_guan_di_renzo_dabbah_poor_hanzo_two_dimensional_direction_of_arrival_estimation_using_stacked_intelligent_metasurfaces,an_di_renzo_debbah_yuen_stacked_intelligent_metasrufaces_for_multiuser_beamforming_in_the_wave_domain,zhang_zhang_di_di_renzo_han_poor_song_holographic_integrated_sensing_and_communicatio,zhang_a_wireless_communication_scheme_based_on_space_and_frequency_division_multiplexin_using_digital_metasurfaces,huang_zhang_yang_sun_zhao_luo_reconfigurable_metasurface_for_multifunctional_control_of_electromagnetic_waves}; other, less popular forms of REMSs, also exist; see, e.g.,~\cite{babakhani_etal_transmitter_architectures_based_on_near_field_direct_antenna_modulation}.
Moreover, also non-reconfigurable reflectarrays and non-reconfigurable array lenses have experienced a revival in recent years; see, e.g.,~\cite{zu_optically_and_radiofrequency_transparent_metadevices,shao_shen_xia_a_hybrid_antenna_metasurface_architecture_for_mmwave_and_thz_massive_mimo,naseri_hum_a_dual_band_dual_circularly_polarized_reflectarray_for_k_band_space_applications,martinez_de_rioja_encinar_preliminary_simulations_of_a_18_m_reflectarray_antenna_in_a_geostationary_satellite_to_generate_multi_spot_coverage}.
RRAs comprise one or more actively fed antennas and a nearby array of passive but reconfigurable \new{reflective} elements. 
The reconfigurability of these reflective elements enables dynamic control over the antenna's behavior, mainly the RRA's beampattern, in order to accomplish several goals, such as maximizing gain, reducing side lobes, or minimizing cross-polarization.
RALs consist of arrays of reconfigurable passive elements that function similarly to optical lenses, manipulating electromagnetic~(EM) waves passing through the structure to achieve similar goals as RRAs.
RISs,~which~have gained overwhelming popularity in recent years, typically correspond to planar EM structures with reconfigurable elements and are capable of dynamically manipulating their \new{reflective} behavior. 
Typically, RISs are positioned between transmitter(s) and receiver(s) to dynamically improve the wireless channel for communication or the accuracy of sensing tasks.

The popularity of REMS-related research has led to a significant rise in interest in REMS modeling in an \emph{efficient} and \emph{physically consistent} manner. In what follows, with \emph{efficient} we refer to models~(i) whose parameters can be obtained with acceptable computational effort at design time and~(ii) that enable real-time prediction of the REMS behavior at manageable complexity; with \emph{physically consistent} we refer to the alignment of the model's predictions with the physical laws of classical electrodynamics. This includes effects caused by inter-antenna coupling, non-reciprocal materials, polarization, ohmic losses, matching losses, influence of metallic housings, noise from low-noise amplifiers, and noise arising in or received by antennas. 
While \emph{full-wave EM simulations}\footnote{Full-wave EM simulations numerically solve all of Maxwell's equations without any simplifying assumptions. An example of a full-wave EM simulator is Ansys HFSS~(short for High-Frequency Structure Simulator)~\cite{ansys_hffs}.} can be seen as the gold standard when conducting EM simulations, performing a separate simulation for each configuration of a REMS's reconfigurable elements is computationally prohibitive, as a single simulation for a specific set of configuration parameters can take hours or even days. 
For example, high complexity is an issue when modeling RISs, which must be quite large to have a noticeable impact on the propagation conditions.
Simulating such large REMSs with conventional full-wave EM simulations demands extensive computational resources~(memory as well as CPU time) and is prone to numerical stability issues. 
Consequently, modeling methods that rely on numerous full-wave EM simulations~(e.g., for each possible parameter configuration of an RIS) cannot be considered efficient in the above-mentioned sense.

\new{
Even though a wide variety of efficient REMS modeling approaches has been proposed in recent years, the literature lacks REMS models that are both efficient \emph{and} physically consistent (see~Section~\ref{sec:state_of_the_art}). 
One example implication of the lack of efficient and physically consistent modeling methods is that existing beamforming algorithms designed to optimize the reconfigurable elements of REMSs are often based on inaccurate assumptions, preventing them from achieving optimal performance. 
Another example implication is that system analysis is often conducted using models that neglect significant effects, such as inter-antenna coupling or polarization effects, which can lead to inaccurate results.
In this paper, we resolve the lack of efficient and physically consistent REMS modeling approaches by proposing an efficient and physically consistent model for general REMS.\footnote{Conventional single- and multi-antenna transceivers and other \emph{non-reconfigurable} EM systems are special cases of REMSs and can therefore also be modeled with our framework.}
}

\subsection{Contributions}
Our main contributions are as follows. 
\begin{itemize}
    \item We propose a new, mathematically rigorous formalism to model an arbitrary REMS's interaction with the~(incoming and outgoing) EM waves in its far-field region. 
    \item We propose an \emph{efficient} and \emph{physically consistent} REMS model by incorporating recent findings on circuit-theoretic models with our REMS-far-field-interaction formalism. 
    \item We show how the model parameters required by our REMS-far-field-interaction formalism can be extracted from a \emph{single} full-wave EM simulation.
    \item We validate the proposed REMS-far-field-interaction formalism and the method for obtaining the required model parameters in two scenarios by comparing them with Ansys HFSS simulations.
    \item We demonstrate the efficiency and accuracy of our REMS modeling approach with a case study of a planar RRA with two actively fed antennas that enable simultaneous multiuser beam- and null-forming using a computationally efficient and physics-aware algorithm.
    \item Finally, we discuss the limitations of our framework and outline potential future avenues.
\end{itemize}
\subsection{State of the Art}\label{sec:state_of_the_art}
Various REMS models have been proposed in the literature. 
These models, which are often heuristically motivated, are typically highly efficient but lack either physical consistency or generality; that is, they are only physically consistent for certain specific scenarios.
For example, the models discussed \hbox{in~\cite{tang_chen_dai_han_wireless_communications_with_reconfigurable_intelligent_surface, demir_bjornson_user_centric_cell_free_massive_mimo_with_ris_integrated_antenna_arrays, tiwari_caire_a_new_old_idea_beam_steering_reflectarrays_for_efficient_sub_thz_multiuser_mimo, cui_yin_channel_estimation_for_ris_aided_mmwave_communications_via_3d_positioning,ozdogain_bjornson_larsson_intelligent_reflecting_surfaces} }ignore the REMS's internal coupling and polarization effects; the model used in~\cite{li_cui_information_metamaterials_from_effective_media_to_real_time_information_processing_systems,moccia_liu_wu_casteldi_coding_metasurfaces_for_diffuse_scattering} is based on the assumption that the REMS is composed of identical and periodically arranged unit cells.
Similarly, the inhomogeneous sheets of surface impedance model discussed in~\cite{shabir_di_renzo_zappone_debbah_electromagnetically_consistent_optimization_algorithms_for_the_global_design_of_ris,di_renzo_danufane_tretyakov_communication_models_for_reconfigurable_intelligent_surfaces} is only physically consistent if the REMS consists of unit cells with dimensions and inter-cell spacing significantly smaller than the wavelength of interest.
In contrast, our REMS model accounts for coupling and polarization effects, and is not limited to REMSs that are composed of identical, periodically arranged, or small unit~cells.

\newpage
In~\cite{johnson_bowen_kundtz_bily_discrete_dipole_approximation_model_for_control_and_optimization_of_a_holographic_metamaterial_antenna}, Johnson \emph{et al.} proposed a discrete-dipole approximation~(DDA)-based model for far-field pattern prediction. 
Since the DDA decreases the model complexity, this approach is more efficient in predicting a REMS' behavior than a full-wave EM simulation. 
But there is no free lunch for this approach: the increased efficiency comes at the expense of reduced generality of the model and/or a loss of physical consistency.
For example, Diebold \emph{et al.} in~\cite{diebold_pande_gregg_smith_reflectarray_design_using_a_discrete_dipole_framework} use the DDA to model the behavior of an RRA. 
Specifically, they approximate patch antenna elements with two dipole antennas, achieving high modeling efficiency and physical consistency; however, their model is limited to representing the behavior of the specific patch-antenna-element types used in the paper.
In contrast, our REMS model enables efficient modeling of \emph{general} structures while preserving physical consistency.
In~\cite{han_yin_marzetta_couling_matrix_based_beamforming_for_superdirective_antenna_arrays}, Han \emph{et al.} proposed a coupling-matrix-based REMS model for far-field pattern prediction, which was subsequently utilized by Ji \emph{et al.} in~\cite{ji_huang_chen_sha_dai_he_zhang_yuen_dabbah_electromagnetic_hybrid_beamforming_for_holographic_mimo_communications}.
This coupling-matrix-based modeling approach leverages the linearity of classical electrodynamics to model internal coupling.
However, this model is limited to transmitters or receivers, i.e., cannot capture the scattering behavior of a RIS, for example, and cannot predict the power that is absorbed by the coupled antennas.
In contrast, our REMS model is able to describe the behavior of RISs and can be used to predict the power absorbed by the~(possibly coupled) antennas.

%
In~\cite{wallace_jensen_mutual_coupling_in_mimo_wireless_systems}, Wallace and Jensen postulated that wireless communication systems can be represented using circuit theory. 
In~\cite{ivrlac_nossek_toward_a_circuit_theory_of_communication}, Ivrlač and Nossek demonstrated that circuit-theoretic system models enable both an efficient and physically consistent modeling of the end-to-end channel between transmitters' power amplifiers~(PAs) and receivers' low noise amplifiers (LNAs). 
In~\cite{babakhani_etal_transmitter_architectures_based_on_near_field_direct_antenna_modulation}, Babakhani \emph{et al.} proposed a circuit-theory-based model for wireless communication systems with reconfigurable transceivers. 
In~\cite{gradoni_di_renzo_end_to_end_mutual_coupling_aware_communication_model_for_reconfigurable_intelligent_surfaces}, Gradoni and Di~Renzo proposed a general circuit-theory-based model for wireless communication systems with a reconfigurable channel.
Further circuit-theory-based models for wireless systems with REMSs can be found in~\cite{nerini_gradoni_clerckx_physics_compliant_modeling_and_scaling_laws_of_multi_ri_aided_mimo_systems,li_clerckx_non_reciprocal_beyond_diagonal_ris_multiport_network_models_and_performance_benerfits_in_full_duplex_systems,nerini_shen_li_di_renzo_clerckx, nossek_semmler_joham_utschick_physically_consistent_modeling_of_wireless_links_with_reconfigurable_intelligent_surfaces_using_multiport_network_analysis, nerini_clerckx_physically_consistent_modeling_of_stacked_intelligent_metasurfaces_implemented_with_beyond_diagonal_ris,balasuriya_mezghani_hossain_physically_consistent_multi_band_massive_mimo_systems,del_hougne_physics_compliant_diagonal_representation_of_beyond_diagonal_ris,shen_clerckx_murch_modeling_and_architecture_design_of_reconfigurable_intelligent_surfaces_using_scattering_parameter_network_analysis}. 
These existing models describe end-to-end wireless channel between the transmitter PAs and the receiver LNAs.
Each new physical arrangement of REMSs~(e.g., transceivers) requires recalculating the model parameters, which is generally inefficient.
In contrast, our model allows each REMS in the system to be characterized separately without first specifying its position and orientation.
As we will show, these characterizations can be used to \emph{efficiently} model the system-level behavior of a wireless communication or sensing system with \emph{arbitrarily arranged} REMSs.

In~\cite{kerns_plane_wave_scattering_matrix_theory_of_antennas_and_antenna_antenna_interactions}, Kerns proposed a mathematical formalism to describe the interaction of a single~(non-reconfigurable) antenna with incoming and outgoing EM waves at a reference plane.
Kerns modeled the EM waves as plane waves, i.e., he utilized a plane-wave basis\footnote{In this paper, we use \emph{basis} to refer to the basis of a \emph{linear space}.} to describe the incoming and outgoing EM waves. 
Similarly, Lewis in~\cite{lewis_spherical_wave_source_scattering_matrix_analysis_of_antennas_and_antenna_antenna_interaction} and Hansen \hbox{in~\cite[Ch.~2]{hansen_spherical_near_field_antenna_measurements}} demonstrate that the antenna-to-far-field interaction can be formalized using spherical harmonics as basis for the far-field EM waves. 
For a recent reference utilizing a spherical-harmonics-based approach to model antennas, we refer to \mbox{Shi \emph{et al.}~\cite{shi_pan_gu_liang_zuo_generalized_scattering_matrix_of_antenna}.}
In contrast to Kerns, Lewis, and Hansen, our proposed REMS-far-field-interaction formalism relies on \emph{spherical waves} as the basis for the incoming and outgoing EM waves. 
While the choice of basis is application-dependent, our spherical-wave-based formalism offers significant advantages: 
In contrast to a plane-wave-based approach, our formalism does not require defining a reference plane and therefore naturally facilitates~seamless descriptions of the incoming and outgoing EM waves in \emph{all} directions. 
Additionally, in contrast to a spherical-harmonics-based approach, our formalism avoids computationally complex operations during synthesis~(i.e., for calculating the EM wave radiated in a specific direction), such as computing associated Legendre polynomials.\footnote{\new{The additional computational resources required to compute associated Legendre polynomials, compared to performing the operations necessary when using spherical waves, depend on the implementation and the approximations used but are expected to be at least an order of magnitude greater.}}$^{,}$\footnote{\new{The spherical harmonics basis is countably infinite, whereas the spherical waves basis is uncountably infinite. The implication of this subtle difference may require further study.}}

\begin{rem}[Spherical Waves]
    Although various authors, including Lewis and Hansen, use the term ``spherical waves'' to refer to spherical harmonics, we use the term ``spherical waves'' to refer to waves of the form shown in~\fref{fig:superposition_of_incomming_and_outgoing_wave_in_far_field}. 
    Our terminology allows to distinguish between the concepts of spherical harmonics and spherical waves and is used by other authors, such as Nieto-Vesperinas in~\cite[Sec.~2.1]{nieto_vesperinas_scattering_and_diffraction_in_physical_optics}. 
\end{rem}
%


%
Possibly closest to our work is the modeling approach put forward in~\cite{mezghani_bellili_hossain_reconfigurable_intelligent_surfaces_for_quasi_passive_mmwave_and_thz_networks,wijekoon_mezghani_alexandropoulos_hossain_physically_consistent_modeling_and_optimization_of_non_local_ris_assisted_multi_user_mimo_communication_systems,mezghani_akrout_castellanos_saab_hochwald_health_nossek_reincorporating_circuit_theory_into_information_theory}. 
The authors of these papers use both circuit theory and linear scattering theory to model the behavior of antenna arrays and RISs.
However, it is not immediately evident what basis is used to describe the incident EM waves.\footnote{While Kerns and Dayhoff~\cite{kerns_theory_of_diffraction_in_microwave_interferometry} are cited, their plane-wave formalism would typically necessitate specifying a reference plane for the space-side waves.}
Consequently, it remains unclear how models of multiple antenna arrays or RISs can be combined to predict the wireless channel properties.
Polarization effects cannot be described by the modeling approach proposed in~\cite{mezghani_bellili_hossain_reconfigurable_intelligent_surfaces_for_quasi_passive_mmwave_and_thz_networks,wijekoon_mezghani_alexandropoulos_hossain_physically_consistent_modeling_and_optimization_of_non_local_ris_assisted_multi_user_mimo_communication_systems,mezghani_akrout_castellanos_saab_hochwald_health_nossek_reincorporating_circuit_theory_into_information_theory};
the specific models proposed in~\cite{mezghani_bellili_hossain_reconfigurable_intelligent_surfaces_for_quasi_passive_mmwave_and_thz_networks,mezghani_akrout_castellanos_saab_hochwald_health_nossek_reincorporating_circuit_theory_into_information_theory} are restricted to reciprocal systems;
and for the specific model proposed in~\cite{wijekoon_mezghani_alexandropoulos_hossain_physically_consistent_modeling_and_optimization_of_non_local_ris_assisted_multi_user_mimo_communication_systems}, the rationale behind treating both the incoming and outgoing EM waves as elements in a~$M$-dimensional vector space, where~$M$ is the number of RIS elements, is not fully clarified. 
Furthermore, the authors neither theoretically nor empirically demonstrate that their model is physically consistent, nor do they explain how the model parameters can be obtained in practice.
In contrast,~(i) we introduce a mathematically rigorous formalism for the REMS far-field interaction derived from first principles,~(ii) we demonstrate how multiple REMS models can be combined to predict properties of the wireless channel,~(iii) our REMS model can predict polarization effects, and~(iv) our model is able to describe the behavior of non-reciprocal systems. 
Furthermore, we show how the model parameters for our REMS model can be obtained from a \emph{single} full-wave EM simulation, and we numerically validate our model and provide a concrete application example of our framework. 
\subsection{Paper Outline}
The rest of the paper is organized as follows. 
In~\fref{sec:modeling}, we introduce our REMS model, including a rigorous mathematical formalism to model the REMS's interaction with EM waves in its far-field region. 
Moreover, we discuss aspects of the proposed REMS model such as linear input-output relationships, the calculation of the wireless channel between two REMS, various power metrics, a new gain metric, reciprocity relations, and, finally, the extraction of model parameters from a single full-wave EM simulation.
In~\fref{sec:experimental_validation}, we perform simulations to validate~(i) the proposed REMS-far-field-interaction formalism and~(ii) the method for extracting the necessary model parameters proposed in~\fref{sec:modeling}.
In~\fref{sec:case_study}, we showcase a practical application of our REMS model through a case study featuring a planar RRA and a new and computationally efficient beam- and null-forming algorithm.
In~\fref{sec:limitations}, we discuss the limitations of the proposed REMS modeling approach.
We conclude in~\fref{sec:conclusions} by summarizing the main contributions and discussing potential future avenues. 
All technical proofs are relegated to~\fref{app:proofs}.

\subsection{Notation}\label{sec:notation}
We use boldface and uppercase boldface for general vectors~(e.g.,~$\vect{a}$) and general matrices~(e.g.,~$\mat{A}$), respectively; we use pink sans-serif~(e.g.,~$\phs{s}$) and pink sans-serif boldface~(e.g.,~$\phv{s}$) to represent phasors~(see~\fref{defi:phasor}) and vectors containing phasors, respectively. 
The superscripts~$^\T$ and~$^\He$ represent the transpose~(e.g.,~$\mat{A}^\T$) and conjugate transpose~(e.g.,~$\mat{A}^\He$), respectively.
Given a Hilbert space~$\mathcal{H}$, we use~$\langle \vect{a},\vect{b} \rangle_\mathcal{H}$ to denote the inner product of~$\vect{a},\vect{b}\in\mathcal{H}$, where we use the convention of linearity in the first argument~$\vect{a}$; to simplify notation, we define the shortcut~$[\vect{a}]_\vect{b}\triangleq\langle\vect{a},\vect{b}\rangle_\mathcal{H}$.
The Euclidean norm is denoted with~$\|\cdot\|_2$ or~$\|\cdot\|_\mathcal{H}$. 
The cross product of two three-dimensional vectors~$\vect{a}$ and~$\vect{b}$ is denoted by~$\vect{a}\times \vect{b}$, and the Nabla operator is written as~$\nabla$.
We refer to the~$\ell$th element of the vector~$\vect{a}$ as~$\vect{a}_{\ell}$ and the element in the~$\ell$th row and~$k$th column of the matrix~$\mat{A}$ as~$\mat{A}_{\ell,k}$.
Given a vector~$\vect{a}$, we use~$\diag(\vect{a})$ to denote the diagonal matrix with the elements of~$\vect{a}$ on its main diagonal.
The zero matrix in~$\mathbb{C}^{N\times M}$ is denoted with~$\mat{0}_{N,M}$. 
To simplify notation, we occasionally omit explicitly specifying the dimension of zero matrices~(or vectors).
We use blackboard bold font for operators~(e.g.,~$\mathbb{S}$). 
Given~$N\in\mathbb{N}$, we define the set~$[N]\triangleq\{1,\ldots,N\}$.
For a set~$\mathcal{S}$, we denote its cardinality by~$|\mathcal{S}|$.
We denote the Dirac delta by~$\delta(\cdot)$ and the Kronecker delta by~$\delta_{\ell,k}$, which equals~$1$ if~$\ell = k$ and~$0$ otherwise.
We define the imaginary unit by~$j^2 = -1$.
For a complex number~$z\in\mathbb{C}$, the complex conjugate is~$\overline{z}$, the real part is~$\Re\{z\}$, and the imaginary part is~$\Im\{z\}$. 

To simplify notation, we refrain from explicitly writing physical units, except when specifying numerical values. 
All physical equations are formulated based on SI units.

\newpage
\section{REMS Modeling Method}\label{sec:modeling}

We now introduce our circuit-theoretical REMS model, whose basic structure is depicted in~\fref{fig:model_basic}. 
We then discuss various aspects of the proposed model such as the linear input-output relationships, the calculation of the wireless channel between two REMS, various power metrics, a new gain metric, and reciprocity relationships.
Furthermore, we present our approach for extracting model parameters through a \emph{single} full-wave EM simulation.
\subsection{REMS Model Overview}
The proposed circuit-theoretic REMS model describes how a REMS interacts with the communication or sensing system in which it is embedded.
The REMS can function as a passive structure that manipulates EM waves, act as a transmitter or a receiver, or operate in any combination of these modes.
Our model relies on the following key assumptions. 

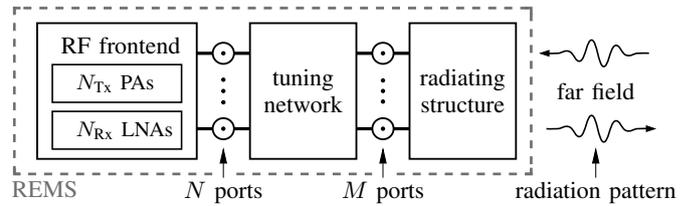
\begin{figure}[tp]
    \centering
    {
    \small
    \begin{tikzpicture}

        \draw  [line width=1pt, color=grey] (-1.5,-1.1) rectangle (5.3,1.1);
        \draw  [line width=1pt,fill=white,color=white] (-1.5-.1,-1.1+.1) rectangle (5.3+.1,1.1-.1);
        \draw  [line width=1pt,fill=white,color=white] (-1.5+.1,-1.1-.1) rectangle (5.3-.1,1.1+.1);
        draw  [line width=1pt,fill=white,color=white] (-1.5-.1,-1.1+.1) rectangle (5.3-.1,1.1+.1);
        \draw  [line width=1pt, dashed, color=grey] (-1.5,-1.1) rectangle (5.3,1.1);

        \draw  [line width=0pt, color=white, fill=white] (1.5,-1,1) rectangle (1.4,-.9);
        \draw  [line width=0pt, color=white, fill=white] (3.6,-1,1) rectangle (3.5,-.9);
        \node[shift={(-1.1,-1.3)}, align=center, color=grey] at (0,0) {\small REMS};

        \draw[line width=1.2pt] (0, 0.5) -- (5, 0.5);
        \draw[line width=1.2pt] (0, -0.5) -- (5, -0.5);
        
        \draw [shift={(5.5,.5)}, line width=.6pt] plot[smooth, tension=.8] coordinates {(0,0) (.15,0) (.3,.025) (.45,-.1) (.6,.175) (.75,-.175) (.9,.1) (1.05,-.025) (1.2,0) (1.35,.0) };
        \draw [shift={(5.5,0.5)}, color=black, arrows = {-Stealth[inset=0, length=4pt, angle'=35]}] (0,0) -- (-0.1,0);
        \draw [shift={(5.5,-.5)}, line width=.6pt] plot[smooth, tension=.8] coordinates {(0,0) (.15,0) (.3,.025) (.45,-.1) (.6,.175) (.75,-.175) (.9,.1) (1.05,-.025) (1.2,0) (1.35,.0) };
        \draw [shift={(5.5,-0.5)}, color=black, arrows = {-Stealth[inset=0, length=4pt, angle'=35]}] (1.35,0) -- (1.45,0);
        \node[shift={(6.15,0)}, align=center] at (0,0) {\small far field};

        \draw  [line width=.7pt, anchor=center, fill=white] (-1.2,-.9) rectangle (.9,.9);
        \draw  [shift={(0,.1)}, line width=.6pt, anchor=center, fill=white] (-1,-.25) rectangle (.7,.25);
        \node[shift={(-.9,.1)}, left, anchor=west] at (0,0) {\footnotesize~$N_\up{Tx}$ PAs};
        \draw  [shift={(0,-0.52)}, line width=.6pt, anchor=center, fill=white] (-1,-.25) rectangle (.7,.25);
        \node[shift={(-.9,-0.52)}, left, anchor=west] at (0,0) {\footnotesize~$N_\up{Rx}$ LNAs};
        \node[shift={(-.1,0.6)}] at (0,0) {\small RF frontend};

        \draw  [shift={(2.3,0)}, line width=.7pt, anchor=center, fill=white] (-.7,-.9) rectangle (.7,.9);
        \node[shift={(2.3,0)}, align=center] at (0,0) {\small tuning\\\small network};

        \draw  [shift={(4.4,0)}, line width=.7pt, anchor=center, fill=white] (-.7,-.9) rectangle (.7,.9);
        \node[shift={(4.4,0)}, align=center] at (0,0) {\small radiating\\\small structure};

        \draw  [shift={(1.25,0.5)}, line width=.7pt, fill=white] (0,0) ellipse (0.14 and 0.14);
        \draw  [shift={(1.25,0.5)}, line width=.5pt, fill=black] (0,0) ellipse (0.02 and 0.02);
        \draw  [shift={(1.25,-0.5)}, line width=.7pt, fill=white] (0,0) ellipse (0.14 and 0.14);
        \draw  [shift={(1.25,-0.5)}, line width=.5pt, fill=black] (0,0) ellipse (0.02 and 0.02);

        \draw  [shift={(1.25,0)}, line width=.5pt, fill=black] (0,0) ellipse (0.02 and 0.02);
        \draw  [shift={(1.25,.15)}, line width=.5pt, fill=black] (0,0) ellipse (0.02 and 0.02);
        \draw  [shift={(1.25,-.15)}, line width=.5pt, fill=black] (0,0) ellipse (0.02 and 0.02);
        
        \draw  [shift={(3.35,0.5)}, line width=.7pt, fill=white] (0,0) ellipse (0.14 and 0.14);
        \draw  [shift={(3.35,0.5)}, line width=.5pt, fill=black] (0,0) ellipse (0.02 and 0.02);
        \draw  [shift={(3.35,-0.5)}, line width=.7pt, fill=white] (0,0) ellipse (0.14 and 0.14);
        \draw  [shift={(3.35,-0.5)}, line width=.5pt, fill=black] (0,0) ellipse (0.02 and 0.02);

        \draw  [shift={(3.35,0)}, line width=.5pt, fill=black] (0,0) ellipse (0.02 and 0.02);
        \draw  [shift={(3.35,.14)}, line width=.5pt, fill=black] (0,0) ellipse (0.02 and 0.02);
        \draw  [shift={(3.35,-.14)}, line width=.5pt, fill=black] (0,0) ellipse (0.02 and 0.02);

        \node[shift={(1.25,-1.35)}, color=black] at (0,0) {\small$N$ ports};
        \draw[shift={(1.25,-1.35)}, line width=0.3pt, color=black, arrows = {-Stealth[inset=0, length=6pt, angle'=25]}] (0,.2) -- (0,.6);

        \node[shift={(3.35,-1.35)}, color=black] at (0,0) {\small$M$ ports};
        \draw[shift={(3.35,-1.35)}, line width=0.3pt, color=black, arrows = {-Stealth[inset=0, length=6pt, angle'=25]}] (0,.2) -- (0,.6);

        \node[shift={(6.15,-1.35)}, color=black] at (0,0) {\small radiation pattern};
        \draw[shift={(6.15,-1.35)}, line width=0.3pt, color=black, arrows = {-Stealth[inset=0, length=6pt, angle'=25]}] (0,.2) -- (0,.6);
        
    \end{tikzpicture}}
\caption{Basic structure of the proposed REMS model. The model is divided into the RF frontend, the tuning network, and the radiating structure. The RF frontend consists of~$N_\up{Tx}$ power amplifiers (PAs) and~$N_\up{Rx}$ low-noise amplifiers (LNAs). The RF frontend is connected to the tuning network via~$N=N_\up{Tx}+N_\up{Rx}$ (circuit-theoretic) ports; and the tuning network is connected to the radiating structure via~$M$ ports. The radiating structure models all interactions between the signals at these~$M$ ports and the (incoming and outgoing) EM waves in the REMS's far-field region.}
\label{fig:model_basic}
\end{figure}

\begin{asm}[PA and LNA Characteristics]\label{asm:basic_assumption_PA_LNA}
Each power amplifier (PA; for the transmitters) can be represented with an active, noiseless, linear, time-invariant two-port. Each low-noise amplifier (LNA; for the receivers) can be represented with a passive, noisy, linear, time-invariant two-port. 
(See~\cite[Sec.~4.1.1]{engberg_noise_theory_of_linear_and_nonlinear_circuits} for a definition of noiseless and noisy linear time-invariant two-ports.)
Additionally, the output impedances of all PAs and the input impedances of all LNAs have a strictly positive real part.
\end{asm}
We are interested in modeling the output stages \mbox{of~$N_\textnormal{Tx}\!\in\!\mathbb{Z}_{\geq 0}$} PAs and the input stages \mbox{of~$N_\textnormal{Rx}\in\mathbb{Z}_{\geq 0}$} LNAs.
~\fref{asm:basic_assumption_PA_LNA} implies that these PA output stages and LNA input stages can be modeled with~$N=N_\textnormal{Tx}+N_\textnormal{Rx}$ separate submodels, each with one port\footnote{We use \emph{port} to refer to a circuit-theoretic port unless otherwise specified. Analogously, \emph{multiport} refers to a circuit-theoretic multiport.}.
We will henceforth refer to the collection of all those PA and LNA submodels as the \emph{radio-frequency (RF) front-end} of the REMS (see the left-hand side in~\fref{fig:model_basic}).

\begin{asm}[Radiating Elements Characteristics]\label{asm:basic_assumption_radiating_part}
The EM radiating\footnote{We use the term \emph{radiating} to refer to structures that interact with their surrounding EM fields (i.e., electromagnetically non-transparent objects), which also includes purely passive objects.} elements (e.g., antennas or metallic housing) are fixed at design time and can be represented as a passive, noisy, linear, time-invariant system. 
Additionally, the connection between the rest of the REMS and the electromagnetically radiating elements can be represented by a finite number of ports.
(See, e.g.,~\cite[Sec.~3]{engberg_noise_theory_of_linear_and_nonlinear_circuits} for an introduction to noisy linear multiports.)
\end{asm}
\fref{asm:basic_assumption_radiating_part} implies that we can model the EM radiating elements by a separate submodel, which we will henceforth refer to as the \emph{radiating structure} (see the right-hand side in~\fref{fig:model_basic}).
Furthermore, it follows that this submodel represents all interactions of the REMS with its environment, including the (incoming and outgoing) EM waves in the REMS's far-field region.
The number of ports connecting the radiating structure with the rest of the REMS model is denoted by~$M\in\mathbb{Z}_{\geq 0}$.
In general,~$M$ can be less than, equal to, or greater than~$N$. 

After removing the transmitters, receivers, and radiating structure, we are left with the reconfigurable elements (e.g., phase shifters) and fixed non-radiating structures (e.g., a matching network). 
We refer to these remaining parts of the REMS as the \emph{tuning network}.  For this part, we make the following assumption. 
\begin{asm}[Tuning Network Characteristics]\label{asm:basic_assumption_tuning_network}
For a fixed configuration of the reconfigurable elements, the tuning network can be modeled as a passive LTI system.
\end{asm}
\fref{asm:basic_assumption_tuning_network} implies that, for a fixed configuration of the reconfigurable elements, the tuning network can be modeled as an~$N+M$-port (see the center in \fref{fig:model_basic}). 

\input{tikzfigs/fig-model_detail}

\subsection{REMS Model Details}
\label{sec:model_detail}
In the following paragraphs, we provide a detailed description of the proposed REMS model illustrated in~\fref{fig:model}. 
\subsubsection{Parameterization of Signals and Multiports}
The proposed REMS model is intended for performing narrow-band analyses, i.e., each frequency of interest is analyzed separately.
Consequently, for a given analysis frequency, all signals are treated as sinusoidal time functions, which are fully characterized by a phasor. 
\begin{defi}[Phasor] \label{defi:phasor}
For a given sinusoidal time function~\mbox{$s(t)=A \cos(2\pi f t+\phi_0)$}, where~$t\in\mathbb{R}$ is the time (in seconds),~$f\in\mathbb{R}$ the frequency (in Hertz),~$A\in\mathbb{R}_{\geq0}$ is the amplitude, and~$\phi_0$ the phase, we define this signal's \emph{phasor} as the complex-valued number given by \mbox{$\phs{s}=\frac{1}{\sqrt{2}}Ae^{j\phi_0}$}, which represents the root mean square (RMS) value and phase of this signal. We represent phasors with pink sans-serif (e.g.,~$\phs{s}$) and vectors containing phasors with pink sans-serif boldface (e.g.,~$\phv{s}$).
\end{defi}

There exists a range of parametrization methods that can be utilized to describe multiports.
We use a scattering parameter (S-parameter)-based description for the following reasons:  
\mbox{(i)~S-parameters} are suitable for representing \emph{all} possible linear multiports (including direct connections between ports, short circuits, open circuits, etc.), which is in contrast to other commonly-used descriptions such as those based on impedance (Z-parameters) or admittance (Y-parameters);
(ii)~A multiport characterized by S-parameters can be directly converted into a signal-flow graph, as demonstrated by Pozar in~\cite[pp.~194--202]{pozar_microwave_engineering}, which we will utilize in~Section~\ref{sec:input_output_relationships} to efficiently derive input-output relationships.
For a more detailed discussion on the parametrization of multiports in circuit-theory-based wireless communication models, we refer to Nerini \emph{et al.} in~\cite{nerini_shen_li_di_renzo_clerckx}.
Following Kurokawa in \cite{kurokawa_power_waves_and_scattering}, we define the incoming and outgoing \emph{circuit-theoretic power waves} as follows.
\begin{defi}[Circuit-Theoretic Power Waves]\label{defi:circuit_theoretic_power_waves}
    Given a multiport with~$L$ ports. 
    For~$\ell\in [L]$, we define the circuit-theoretic power wave traveling into the multiport on the~$\ell$th port as
    \begin{align}
        \phs{a}_\ell&\triangleq \frac{1}{2\sqrt{R_0}}(\phs{v}_\ell+R_0 \phs{i}_\ell)
        \label{eq:definition_power_waves_a}
    \intertext{
    and the circuit-theoretic power wave traveling out of the multiport on the~$\ell$th port as
    }
        \phs{b}_\ell&\triangleq \frac{1}{2\sqrt{R_0}}(\phs{v}_\ell-R_0\phs{i}_\ell),
        \label{eq:definition_power_waves_b}
    \end{align}
    where~$\phs{v}_\ell$ and~$\phs{i}_\ell$ are the phasors of the voltage and current at port~$\ell$, respectively. Here,~$R_0\in\mathbb{R}_{>0}$ is an arbitrary reference impedance ($\SI{50}{\ohm}$ is a commonly-used choice).
\end{defi}

\subsubsection{RF Frontend} 
If the REMS acts as an active transmitter, receiver, or both (i.e., a transceiver), then the RF frontend models the output stage of the PAs and the input stage of the LNAs. 
For REMSs that do not operate in any of these modes (e.g., a passive RIS), one can omit the RF frontend from the model.

For~$\ell\in[N_\up{Tx}]$, we represent the output stage of the~$\ell$th PA by its Th\'evenin equivalent (see the top-left side of~\fref{fig:model}), comprising an ideal voltage source and a source impedance~$Z_{\up{Tx},\ell}$, where the voltage source is represented by the phasor~$\phs{v}_{\up{Tx},\ell}$.
In our model, we neglect any noise generated in the PAs, as the noise power is typically much lower than that of the amplified signal. Nonetheless, our model could be extended to include noise originating in the PAs.

For~$\ell\in[N_\up{Rx}]$, we represent the input stage of the~$\ell$th LNA by its input impedance~$Z_{\up{Rx},\ell}$ (see the bottom-left side of~\fref{fig:model}).
Since we assume each LNA behaves as a noisy two-port (of which we only include one port in our model), it is necessary to include two noise sources to model the noise originating from each LNA; see~\cite[pp.~41--45]{engberg_noise_theory_of_linear_and_nonlinear_circuits} for the details. 
Consequently, for the~$\ell$th LNA, we model the noise with a current and a voltage source, where these noise sources are represented by the complex-valued random variables~$\phs{i}_{\Gamma,\ell}$ and~$\phs{v}_{\Gamma,\ell}$, respectively.  

Any potential RF switch or circulator that could be used within a transceiver is not modeled as part of the RF frontend, but rather modeled in the tuning network.
We denote the vector of outgoing and incoming power waves (from the RF frontend's perspective) present at the ports that connect the RF frontend with the tuning network as ~$\phv{a}_\up{T}\in\mathbb{C}^N$ and~$\phv{b}_\up{T}\in\mathbb{C}^N$, respectively. 
Moreover, we label the first~$N_\up{Tx}$ elements of~$\phv{a}_\up{T}$ and~$\phv{b}_\up{T}$ as~$\phv{a}^{\up{Tx}}_{\up{T}}$ and~$\phv{b}^{\up{Tx}}_{\up{T}}$, respectively. Similarly, the remaining~$N_\up{Rx}$ elements are denoted as~$\phv{a}_{\up{T}}^{\up{Rx}}$ and~$\phv{b}_{\up{T}}^{\up{Rx}}$.
To simplify notation, we define the following vectors and matrices: 
\begin{align}
        \phv{v}_\up{Tx}
        &\hspace{-1pt}\triangleq
        [\phs{v}_{\up{Tx},1}
        \cdots\,
        \phs{v}_{\up{Tx},N_\up{Tx}}
        ]^\T,
    \\
        \phv{v}_\up{Rx}
        &\hspace{-1pt}\triangleq
        [\phs{v}_{\up{Rx},1}
        \cdots\,
        \phs{v}_{\up{Rx},N_\up{Rx}}
        ]^\T,
    \\
        \mat{Z}_\up{Tx}
        &\hspace{-1pt}\triangleq
        \diag\big([Z_{\up{Tx},1}
        \cdots\,
        Z_{\up{Tx},N_\up{Tx}}]^\T\big),
    \\
        \mat{Z}_\up{Rx}
        &\hspace{-1pt}\triangleq
        \diag\big([Z_{\up{Rx},1}
        \cdots\,
        Z_{\up{Rx},N_\up{Rx}}]^\T\big),
    \\
        \phv{v}_\Gamma
        &\hspace{-1pt}\triangleq
        [\phs{v}_{\Gamma,1}
        \cdots\,
        \phs{v}_{\Gamma,N_\up{Rx}}
        ]^\T,
    \\
        \phv{i}_\Gamma
        &\hspace{-1pt}\triangleq
        [\phs{i}_{\Gamma,1}
        \cdots\,
        \phs{i}_{\Gamma,N_\up{Rx}}
        ]^\T.
\end{align}

\subsubsection{Tuning Network}
The tuning network models the REMS's (i) reconfigurable elements (e.g., phase shifters or switches) and (ii) fixed non-radiating parts (e.g., a matching network).
We represent the tuning network by a multiport with a total \mbox{of~$N+M$} ports, which is fully characterized by the scattering matrix~\mbox{$\mat{S}_\up{T}\in\mathbb{C}^{(N+M)\times (N+M)}$}. 
In particular, it holds that
\begin{align}
    \begin{bmatrix} \phv{b}_\up{T} \\ \phv{a}_\up{R} \end{bmatrix} = 
    \underbrace{
    \begin{bmatrix}
    \mat{S}_{\up{T}_{\up{T}\up{T}}} & \mat{S}_{\up{T}_{\up{T}\up{R}}} \\
    \mat{S}_{\up{T}_{\up{R}\up{T}}} & \mat{S}_{\up{T}_{\up{R}\up{R}}}
    \end{bmatrix}
    }_{\triangleq \,\mat{S}_\up{T}}
    \begin{bmatrix} \phv{a}_\up{T} \\ \phv{b}_\up{R} \end{bmatrix}\!.
    \label{eq:matrix_ST}
\end{align}
Here,~$\phv{a}_\up{R}\in\mathbb{C}^{M}$ and~$\phv{b}_\up{R}\in\mathbb{C}^{M}$ represent the outgoing and incoming waves (from the tuning network's perspective) present at the ports connecting the tuning network to the radiating structure, respectively; 
$\mat{S}_{\up{T}_\up{TT}}$ is the scattering matrix characterizing the linear dependence of the outgoing waves~$\phv{b}_\up{T}$ on the incoming waves~$\phv{a}_\up{T}$, etc.
As a consequence of assuming the tuning network to be passive, the largest singular value of~$\mat{S}_\up{T}$ cannot exceed~$1$.
If the tuning network is reciprocal, then it also holds that~$\mat{S}_\up{T}^\T=\mat{S}_\up{T}$.

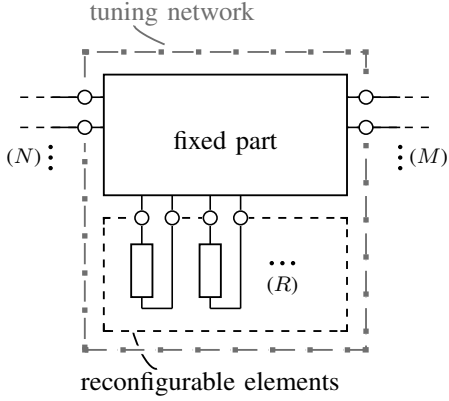
\begin{figure}[tp]
    \centering
    {
    \normalsize
    \begin{tikzpicture}
    \clip (-4.4,-3.4) rectangle (4.4,1.9);

    \draw  [shift={(0,0)}, line width=2pt, color=grey] (-1.85,-1.7-1.05) rectangle (1.85,1.2);
    \draw  [shift={(0,0)}, line width=2pt, color=white, fill=white] (-1.85-.08,-1.7-1.05+.072) rectangle (1.85+.08,1.2-.072);    
    \draw  [shift={(0,0)}, line width=2pt, color=white, fill=white] (-1.85+.08,-1.7-1.05-.072) rectangle (1.85-.08,1.2+.072);

    \draw[line width=.6pt] (-2.3, -.2+.4) -- (2.3, -.2+.4);
    \draw[line width=.6pt] (-2.3, .2+.4) -- (2.3, .2+.4);
    \draw[line width=.58pt, dashed] (-2.7, -.2+.4) -- (2.7, -.2+.4);
    \draw[line width=.58pt, dashed] (-2.7, .2+.4) -- (2.7, .2+.4);

    \draw [shift={(-.4,0)}, line width=.7pt] plot[smooth, tension=1.3, color=black] coordinates {(-.85,-2.35) (-.7,-2.8) (-.5,-3)};
    \draw  [line width=.7pt, color=black, fill=white] (-1.6,-.7) rectangle (1.6,.9);
    \draw  [shift={(-.6,0)}, line width=.0pt, color=white, fill=white] (-1,-2.8) rectangle (0,-2.7);
    
    \draw  [shift={(0,0)}, line width=.7pt, color=grey, dashed, dash pattern=on 0pt off 4pt on 8pt off 2pt] (-1.85,-1.7-1.05) rectangle (1.85,1.2);
    \draw  [shift={(0,0)}, line width=2pt, color=grey, dashed, dash pattern=on 2pt off 12pt] (-1.85,-1.7-1.05) rectangle (1.85,1.2);

    \draw[color=grey] [shift={(-.4,4.2)}, line width=.7pt] plot[smooth, tension=1.2, color=red] coordinates {(-.8,-2.65) (-.68,-2.76) (-.4,-2.9)};
    \node[anchor=center] at (0.0,0) {\normalsize fixed part};

    \draw  [shift={(0,-1.6-.2)}, line width=.7pt, color=black] (-1.6,-.7) rectangle (1.6,.8);
    \draw  [shift={(0,-1.6-.2)}, line width=.7pt, color=white, fill=white] (-1.7,-.6) rectangle (1.7,.7);
    \draw  [shift={(0,-1.6-.2)}, line width=.7pt, color=white, fill=white] (-1.5,-.71) rectangle (1.5,.81);
    \draw  [shift={(0,-1.6-.2)}, line width=.7pt, color=black, dashed, dash pattern=on 3pt off 3pt] (-1.6,-.7) rectangle (1.6,.8);

    \node[anchor=center] at (-.7,1.7) {\color{grey} \normalsize tuning network};

    \node[anchor=center] at (0-.2,-3.2) {\normalsize reconfigurable elements};

    \draw  [shift={(0,.4)}, line width=.0pt, color=white, fill=white] (-1.9,-.4) rectangle (-1.8,.4);
    \draw  [shift={(0,.4)}, line width=.6pt, fill=white] (-1.85,.2) ellipse (0.09 and 0.09);
    \draw  [shift={(0,.4)}, line width=.6pt, fill=white] (-1.85,-.2) ellipse (0.09 and 0.09);

    \draw  [shift={(0,.4)}, line width=.0pt, color=white, fill=white] (1.9,-.4) rectangle (1.8,.45);
    \draw  [shift={(0,.4)}, line width=.6pt, fill=white] (1.85,.2) ellipse (0.09 and 0.09);
    \draw  [shift={(0,.4)}, line width=.6pt, fill=white] (1.85,-.2) ellipse (0.09 and 0.09);

    \draw   [shift={(2.3,-.2)}, line width=.5pt, fill=black] (0,.14)  ellipse (0.02 and 0.02);
    \draw   [shift={(2.3,-.2)}, line width=.5pt, fill=black] (0,0) ellipse (0.02 and 0.02);
    \draw   [shift={(2.3,-.2)}, line width=.5pt, fill=black] (0,-0.14) ellipse (0.02 and 0.02);
    \node   [shift={(2.3,-.2)}, anchor=west] at (-.0,0) {\footnotesize ($M$)};

    \draw   [shift={(-2.3,-.2)}, line width=.5pt, fill=black] (0,.14)  ellipse (0.02 and 0.02);
    \draw   [shift={(-2.3,-.2)}, line width=.5pt, fill=black] (0,0) ellipse (0.02 and 0.02);
    \draw   [shift={(-2.3,-.2)}, line width=.5pt, fill=black] (0,-0.14) ellipse (0.02 and 0.02);
    \node   [shift={(-2.3,-.2)}, anchor=east] at (-.0,0) {\footnotesize ($N$)};

    \draw[shift={(-.9,0)}, shift={(-.2,-1.7)}, line width=.6pt] (0,1) -- (0,-.5);
    \draw[shift={(-.9,0)}, shift={(.2,-1.7)}, line width=.6pt] (0,1) -- (0,-.5);
    \draw[shift={(-.9,0)}, shift={(.2,-1.7)}, line width=.6pt] (-.4,-.5) -- (0,-.5);
    \draw  [shift={(-.9,0)}, line width=.0pt, color=white, fill=white] (-.3,-.9) rectangle (.3,-1.1);
    \draw  [shift={(-.9,0)}, line width=.6pt, fill=white] (-.2,-1) ellipse (0.09 and 0.09);
    \draw  [shift={(-.9,0)}, line width=.6pt, fill=white] (.2,-1) ellipse (0.09 and 0.09);
    \draw  [shift={(-.9,0)}, shift={(-.2,-1.7)}, line width=.7pt, anchor=center, fill=white] (-.12/3*3.5,-.35) rectangle (.12/3*3.5,.35);

    \draw[shift={(0.0,0)}, shift={(-.2,-1.7)}, line width=.6pt] (0,1) -- (0,-.5);
    \draw[shift={(0.0,0)}, shift={(.2,-1.7)}, line width=.6pt] (0,1) -- (0,-.5);
    \draw[shift={(0.0,0)}, shift={(.2,-1.7)}, line width=.6pt] (-.4,-.5) -- (0,-.5);
    \draw  [shift={(0.0,0)}, line width=.0pt, color=white, fill=white] (-.3,-.9) rectangle (.3,-1.1);
    \draw  [shift={(0.0,0)}, line width=.6pt, fill=white] (-.2,-1) ellipse (0.09 and 0.09);
    \draw  [shift={(0.0,0)}, line width=.6pt, fill=white] (.2,-1) ellipse (0.09 and 0.09);
    \draw  [shift={(0.0,0)}, shift={(-.2,-1.7)}, line width=.7pt, anchor=center, fill=white] (-.12/3*3.5,-.35) rectangle (.12/3*3.5,.35);

    \draw   [shift={(0.75,-1.6)}, line width=.5pt, fill=black] (-.14,0)  ellipse (0.02 and 0.02);
    \draw   [shift={(0.75,-1.6)}, line width=.5pt, fill=black] (0,0) ellipse (0.02 and 0.02);
    \draw   [shift={(0.75,-1.6)}, line width=.5pt, fill=black] (.14,0) ellipse (0.02 and 0.02);
    \node   [shift={(0.75,-1.65)}, anchor=north] at (-.0,0) {\footnotesize ($R$)};
        
    \end{tikzpicture}}
\caption{For many scenarios, it makes sense to further separate the tuning network into a fixed part and a reconfigurable part, as the reconfigurable part can often be represented with a diagonal structure~\cite{del_hougne_physics_compliant_diagonal_representation_of_beyond_diagonal_ris}.}
\label{fig:model_possible_extension_tuning_network}
\end{figure}

\begin{rem}
    Depending on the REMS and its application, it may be beneficial to further divide the tuning network into subparts.
    For instance, when modeling an RRA, it might be useful to separate the part of the tuning network that represents the actively fed antennas' matching network from the part that represents the reconfigurable passive EM elements. 
\end{rem}
\begin{rem}\label{rem:del_hougne}
    Del Hougne recently pointed out in~\cite{del_hougne_physics_compliant_diagonal_representation_of_beyond_diagonal_ris} that the actual reconfigurable elements themselves can usually be treated as one-ports (i.e., as general impedances).
    Therefore, even when the reconfigurable part takes the form of an arbitrary fully connected network (sometimes referred to as a ``beyond-diagonal'' structure~\cite{nerini_clerckx_physically_consistent_modeling_of_stacked_intelligent_metasurfaces_implemented_with_beyond_diagonal_ris}), the actual reconfigurable elements can be represented by~$R \in \mathbb{N}$ uncoupled reconfigurable impedances, as shown in~\fref{fig:model_possible_extension_tuning_network}.
    Dividing the tuning network into a fixed and a reconfigurable part as shown in~\fref{fig:model_possible_extension_tuning_network} might be favorable, as existing algorithms that optimize the REMS's configuration parameters are often based on the assumption that the reconfigurable elements have such a diagonal structure.
\end{rem}

\subsubsection{Radiating Structure and Far Field}\label{sec:radiating_structure_and_far_field}
The radiating structure encompasses the radiating elements of the REMS (such as antennas) as well as electromagnetically radiating objects (like metal housings) in its immediate vicinity.
Additionally, noise arising in or received by the radiating elements is modeled in this submodel.
According to Ivrlač and Nossek~\cite{ivrlac_nossek_toward_a_circuit_theory_of_communication}, extrinsic noise can be included by adding the~$M$ voltage sources depicted in \fref{fig:model}, which represent the open-circuit noise voltages.
It follows from basic circuit theory and~\fref{defi:circuit_theoretic_power_waves} that
\begin{align}
    \phv{a}_{\tilde{\up{R}}}
    =
    \phv{a}_\up{R}+\frac{1}{2\sqrt{R_0}}\phv{v}_\Upsilon \quad \text{and} \quad    
    \phv{b}_\up{R}
    =
    \phv{b}_{\tilde{\up{R}}} - \frac{1}{2\sqrt{R_0}}\phv{v}_\Upsilon,
\label{eq:extrinsic_noise_equation}
\end{align}
where we introduce the noise vector \mbox{$\phv{v}_\Upsilon
    \triangleq
    [\phs{v}_{\Upsilon,1}
    \dots,
    \phs{v}_{\Upsilon,M}
    ]^\T,
$}
formed by the phasors of the extrinsic noise sources.

\tikzset{
  ring shading/.code args={from #1 at #2 to #3 at #4}{
    \def\colin{#1}
    \def\radin{#2}
    \def\colout{#3}
    \def\radout{#4}
    \pgfmathsetmacro{\proportion}{\radin/\radout}
    \pgfmathsetmacro{\outer}{.8818cm}
    \pgfmathsetmacro{\inner}{.8818cm*\proportion}
    \pgfmathsetmacro{\middle}{\inner/2+\outer/2}
    \pgfdeclareradialshading{ring}{\pgfpoint{0cm}{0cm}}%
    {
      color(0pt)=(white);
      color(\inner)=(#1);
      color(\outer)=(#3)
    }
    \pgfkeysalso{/tikz/shading=ring}
  },
}

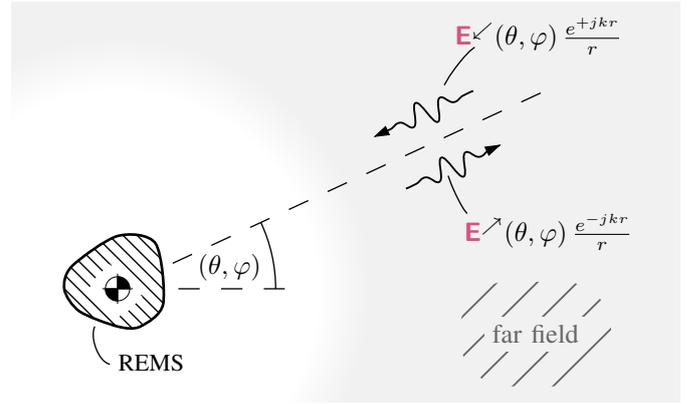
\begin{figure}[tp]
    \centering
    {
    \small
    \begin{tikzpicture}
        \clip (-4.4,-3.5) rectangle (4.4,1.8);

        \draw[line width=1pt,fill=grey,color=grey_10] (-4.4,-3.5) rectangle (4.4,3.5);
        \draw[grey_10, fill=white] (-3,-2) circle (3.3);
        \shade[even odd rule,ring shading={from white at 2.6 to grey_10 at 3.3}]
        (-3,-2) circle (2.5) circle (3.3);

        \begin{scope}[shift={(2.5,-2.6)}]
            \draw[rounded corners=18, pattern color=black_60, grey_10, pattern=diagonal behind text] (-1,-.7) rectangle (1,.7);
            \node[anchor=center, color=black, fill=grey_10] at (0,0) {\color{black_60} \normalsize far field};
        \end{scope}

        \draw [shift={(-3,-2)}, line width=1pt,pattern=south east lines my] plot[smooth cycle] coordinates {(-.5,-.3) (-.7,-.01) (-.6,.3) (0,.7) (.5,.6) (.6,-.2) (.3,-.5) (0,-.5)};
        \node [shift={(-3.1,-2)},anchor=south west, align=left, fill=white] at (-90:1.2) {\small REMS};
        \draw [shift={(-3,-2)}, line width=.5pt] plot[smooth, tension=1.4] coordinates {(-.3,-.5) (-.28,-.8) (-.1,-1)};
        
        \centerOfC at (-3,-2)

        \begin{scope}[shift={(-3,-2)}]
            \draw[line width = .5pt, black, dash pattern=on 8pt off 8pt] (25:.8) -- (25:6.2);
            \draw[line width = .5pt, black, dash pattern=on 8pt off 8pt] (0:.8) -- (0:2.2);
            \draw[line width=0.5pt] (0:2.08) arc (0:25:2.08);
            \node [color=black] at (11:1.5) {\normalsize$(\theta,\varphi)$};
        \end{scope}

        \begin{scope}[shift={(25:4)}]
            \begin{scope}[shift={(-3,-2)}]
                \begin{scope}[rotate=25]
                    \begin{scope}[shift={(0,.4)}]
                        \draw [line width=.7pt] plot[smooth, tension=.8] coordinates {(0,0) (.15,0) (.3,.025) (.45,-.1) (.6,.175) (.75,-.175) (.9,.1) (1.05,-.025) (1.2,0) (1.35,.0) };
                        \draw [arrows = {-Stealth[inset=0, length=6pt, angle'=35]}] (0,0) -- (-0.1,0);
                    \end{scope}
                    \begin{scope}[shift={(0,-.4)}]
                        \draw [line width=.7pt] plot[smooth, tension=.8] coordinates {(0,0) (.15,0) (.3,.025) (.45,-.1) (.6,.175) (.75,-.175) (.9,.1) (1.05,-.025) (1.2,0) (1.35,.0) };
                        \draw [arrows = {-Stealth[inset=0, length=6pt, angle'=35]}] (1.37,0) -- (1.38,0);
                    \end{scope}
                \end{scope}
            \end{scope}          
        \end{scope}

        \node [shift={(1.2,2.2)},anchor=south west, align=left] at (-90:1.2) {\normalsize~$\phv{E}^\swarrow (\theta,\varphi)\,\frac{e^{+jkr}}{r}$};
        \draw [shift={(1.5,2.2)}, line width=.5pt] plot[smooth, tension=1.4] coordinates {(.2,-1)  (0,-1.2) (-.2,-1.5)};
        \node [shift={(1.325,-.4)},anchor=south west, align=left] at (-90:1.2) {\normalsize~$\phv{E}^\nearrow (\theta,\varphi)\,\frac{e^{-jkr}}{r}$};
        \draw [shift={(1.6,0)}, line width=.5pt] plot[smooth, tension=1.4] coordinates {(-.25,-.5) (-.125,-.8) (0,-1)};
    
    \end{tikzpicture}}
\caption{EM waves in the REMS's far-field region, represented as the superposition of incoming (convergent) and outgoing (divergent) spherical waves.
The REMS is positioned at the origin of the coordinate system.}
\label{fig:superposition_of_incomming_and_outgoing_wave_in_far_field}
\end{figure}

In order to mathematically describe the EM waves present in the far field, we use the physicist's convention for spherical coordinate systems~\cite{ISO_quantities_and_units_2_mathematics}, with~$r$ as the \emph{radial distance},~$\theta$ as the \emph{polar angle}, and~$\varphi$ as the \emph{azimuthal angle}.
For each coordinate ($r,\,\theta,\,\varphi)$, we denote the local orthogonal unit vectors in the directions of increasing~$r$,~$\theta$, and~$\varphi$ as~$\hat{\vect{r}}$,~$\hat{\vect{\theta}}$, and~$\hat{\vect{\varphi}}$, respectively.
Moreover, we denote the canonical basis vectors as~$\hat{\vect{x}}$,~$\hat{\vect{y}}$, and~$\hat{\vect{z}}$.
Using this coordinate system, we define the electrical field phasor vector as
\begin{align}
    \phv{E}(r,\theta,\varphi)
    &\triangleq
    \begin{bmatrix}
        \phs{E}_{\hat{\vect{x}}}(r,\theta,\varphi) 
        \\
        \phs{E}_{\hat{\vect{y}}}(r,\theta,\varphi) 
        \\
        \phs{E}_{\hat{\vect{z}}}(r,\theta,\varphi) 
    \end{bmatrix}\!,
\end{align}
where~$\phs{E}_{\hat{\vect{x}}}(r,\theta,\varphi)$,~$\phs{E}_{\hat{\vect{y}}}(r,\theta,\varphi)$, and~$\phs{E}_{\hat{\vect{z}}}(r,\theta,\varphi)$ represent the phasors of the respective components of the electric fields.\footnote{\new{We adhere to the usual convention by denoting the electric field phasor vector with the capital letter~$\phv{E}$, despite deviating from the notation used elsewhere in this paper.}}
As demonstrated by Nieto-Vesperinas in~\cite[Eq.~5.12]{nieto_vesperinas_scattering_and_diffraction_in_physical_optics}, far away from the REMS, in particular as~$r\rightarrow\infty$, we have
\begin{align}
    \lim_{r\rightarrow\infty}
    r
    \phv{E}(r,\theta,\varphi)
    &=
    \phv{E}^\swarrow(\theta,\varphi)e^{+jkr}
    \,+\,
    \phv{E}^\nearrow(\theta,\varphi)e^{-jkr}.
    \label{eq:far_field_superposition}
\end{align}
Here,~$k\triangleq 2 \pi f\sqrt{\mu_0\epsilon_0}$ is the free-space wavenumber,~$\mu_0$ is the free-space permeability, and~$\epsilon_0$ is the free-space permittivity. 
Furthermore,~$\phv{E}^\swarrow(\theta,\varphi)$ and~$\phv{E}^\nearrow(\theta,\varphi)$ are complex-valued three-dimensional vectors that do not depend on the radial distance~$r$. 
Equation \eqref{eq:far_field_superposition} can be interpreted as the superposition of an incoming \emph{convergent spherical wave} and an outgoing \emph{divergent spherical wave}, as illustrated in~\fref{fig:superposition_of_incomming_and_outgoing_wave_in_far_field}.
Here,~$\phv{E}^\swarrow(\theta,\varphi)$ and~$\phv{E}^\nearrow(\theta,\varphi)$ represent the parts of the respective electric field phasor vectors that do not depend on~$r$. 
Because we analyze the EM field in free space, far from any source, it follows directly from the free-space Maxwell's equations that (i) the complete EM field is fully characterized by the electric field and (ii) the electric field components in the propagation direction are zero:
\begin{align}
\label{eq:radial_component_is_zero}
[\phv{E}^\swarrow(\theta,\varphi)]_{\hat{\vect{r}}}=
[\phv{E}^\nearrow(\theta,\varphi)]_{\hat{\vect{r}}}&=\SI[per-mode = symbol]{0}{\volt}.
\end{align}
Here, we use the shorthand notation~$[\phv{a}]_\vect{u}\triangleq\langle\phv{a},\vect{u}\rangle_{\mathbb{C}^3}$. 
As a consequence, the EM fields in the REMS far-field region are fully characterized by the \emph{far-field power wave patterns} introduced in~\fref{defi:far_field_power_wave_pattern}, which are elements of a Hilbert space, as shown in~\fref{prop:hilbert_space}.
A proof of~\fref{prop:hilbert_space} can be found in~\fref{app:proof:hilbert_space}.
\begin{defi}[Far-field Power Wave Patterns]\label{defi:far_field_power_wave_pattern}
    {\color{white}{abcdefghi}}
    Let~$\Omega\triangleq[0,\pi]\times[0,2\pi)$. 
    We define the outgoing far-field power wave pattern 
    \begin{align}
        \phv{a}_\up{F}:\Omega\rightarrow\mathbb{C}^2,
        \,
        (\theta,\varphi)\mapsto
        \frac{1}{\sqrt{Z_0}}
        \begin{bmatrix}
            [\phv{E}^\nearrow(\theta,\varphi)]_{\hat{\vect{\theta}}}
            \\
            [\phv{E}^\nearrow(\theta,\varphi)]_{\hat{\vect{\varphi}}}
        \end{bmatrix}
        \label{eq:definition_far_field_power_waves_a}
        \intertext{
        and the incoming far-field power wave pattern
        }
        \phv{b}_\up{F}:\Omega\rightarrow\mathbb{C}^2,
        \,
        (\theta,\varphi)\mapsto
        \frac{1}{\sqrt{Z_0}}
        \begin{bmatrix}
            [\phv{E}^\swarrow(\theta,\varphi)]_{\hat{\vect{\theta}}}
            \\
            [\phv{E}^\swarrow(\theta,\varphi)]_{\hat{\vect{\varphi}}}
        \end{bmatrix}\!,
        \label{eq:definition_far_field_power_waves_b}
    \end{align}
    where~$Z_0\triangleq \sqrt{\mu_0/\epsilon_0}$ is the free-space impedance.
\end{defi}

\begin{rem}[Polarization]
    The far-field power wave patterns introduced in~\fref{defi:far_field_power_wave_pattern} fully describe polarization, as one can interpret~$\phv{a}_\up{F}(\theta,\varphi)$ and~$\phv{b}_\up{F}(\theta,\varphi)$ as the (with~$\frac{r}{\sqrt{Z_0}}$ scaled) Jones vectors~\cite[pp.~334--341]{jones_a_new_calculus_for_the_treatment_of_optical_systems, pedrotti_introduction_to_optics} of the (approximately plane) waves present at~$(r,\theta,\varphi)$. 
\end{rem}

\begin{prop}\label{prop:hilbert_space}
    Let~$P^\nearrow$ \new{be the} power radiated into the far-field region and~$P^\swarrow$ \new{be the} power incoming from the far-field region.
    If both~$P^\swarrow$ and~$P^\nearrow$ are finite, then the outgoing far-field power wave pattern~$\phv{a}_\up{F}$ and the incoming far-field power wave pattern~$\phv{b}_\up{F}$ are elements of the Hilbert space\footnote{To be specific, the space we use here, is the~$L^2$-space induced by the measure space~$\big( \Omega,\,\mathcal{A},\,\mu \big)$ and the Hilbert space~$\mathbb{C}^2$, where~$\mathcal{A}$ is the respective~$\sigma$-algebra of Lebesgue measurable sets, and~$\mu$ is the measure defined by~$\mu:\mathcal{A}\rightarrow[0,\infty],\,X\mapsto\iint_{(\theta,\varphi)\in X} \sin(\theta)\, \up{d}\theta\, \up{d}\varphi$.}~$L^2$, induced by the functions~$\phv{t}:\Omega\rightarrow\mathbb{C}^2$ that satisfy
    \begin{align}
        \|\phv{t}\|_{L^2}^2\triangleq
         \oiint\displaylimits_{\Omega} \|\phv{t}(\theta,\varphi)\|^2\,\sin(\theta)\,\up{d}(\theta,\varphi)
        &\,< \infty.
        \label{eq:definition_L2_norm}
    \end{align}
    Specifically, it follows that~$\phv{a}_\up{F},\phv{b}_\up{F}\in L^2$.
\end{prop}

\begin{rem}\label{rem:Intensity}
    For~$(\theta,\varphi)\in\Omega$, it is easy to show that~$\|\phv{a}_\up{F}(\theta,\varphi)\|_2^2$ and~$\|\phv{b}_\up{F}(\theta,\varphi)\|_2^2$ correspond to the transmitted and received radiation intensity~$I(\theta,\varphi)$, respectively, i.e., the transmitted/received power per unit solid angle.
\end{rem}
We have derived that the interface variables of the radiating structure are the circuit-theoretic power waves~$\phv{a}_{\tilde{\up{R}}},\phv{b}_{\tilde{\up{R}}}\in\mathbb{C}^M$ at the interface \new{toward} the tuning network, and the far-field power wave patterns~$\phv{a}_\up{F},\phv{b}_\up{F}\in L^2$ at the interface to the far field and the rest of the system in which the REMS is embedded.
It follows from the linearity of Maxwell's equations that the outgoing (from the radiating structure's perspective) power waves~$\phv{b}_{\tilde{\up{R}}}$ and~$\phv{a}_\up{F}$ are related to the incoming power waves~$\phv{a}_{\tilde{\up{R}}}$ and~$\phv{b}_\up{F}$ as follows: 
\begin{align}
    \begin{bmatrix} \phv{b}_{\tilde{\up{R}}} \\ \phv{a}_\up{F} \end{bmatrix} = 
    \underbrace{
    \begin{bmatrix}
    \oper{S}_{\up{R}_{\up{R}\up{R}}} & \oper{S}_{\up{R}_{\up{R}\up{F}}} \\
    \oper{S}_{\up{R}_{\up{F}\up{R}}} & \oper{S}_{\up{R}_{\up{F}\up{F}}}
    \end{bmatrix}
    }_{\triangleq\,\oper{S}_\up{R}}
    \begin{bmatrix} \phv{a}_{\tilde{\up{R}}} \\ \phv{b}_\up{F} \end{bmatrix}.
    \label{eq:matrix_SF}
\end{align}
Here, the four ``submatrices'' of~$\oper{S}_\up{R}$ are bounded\footnote{Boundedness follows from realizing that the Euclidean norms \mbox{of~$\phv{b}_\up{R},\phv{a}_\up{R}\in\mathbb{C}^M$} and the~$L^2$-norms \mbox{of~$\phv{b}_\up{F},\phv{a}_\up{F}\!\in \!L^2$} represent the power transported by these waves as shown in the proof of~\fref{prop:hilbert_space}.} linear operators, which we specify next. 
The \emph{inter-element coupling operator}~$\oper{S}_{\up{R}_{\up{R}\up{R}}}\hspace{-1mm}:\mathbb{C}^M\rightarrow\mathbb{C}^M$ can simply be represented by the matrix~$\mat{S}_{\up{R}_{\up{R}\up{R}}}\in\mathbb{C}^{M\times M}$.
The \emph{transmitting operator}~$\oper{S}_{\up{R}_{\up{F}\up{R}}}:\mathbb{C}^M\rightarrow L^2$ is a finite-rank operator with rank not larger than~$M$.
Hence, its action on~$\phv{s}\in\mathbb{C}^M$ can be represented as 
\begin{align}
    \big(
    \oper{S}_{\up{R}_{\up{F}\up{R}}}\phv{s}
    \big)
    (\theta,\varphi)
    &=
    \sum_{m\in [M]} 
    \vect{s}_{\up{R}_{\up{F}\up{R}}}(m;\theta,\varphi)
    \,
    [\phv{s}]_m,
    \label{eq:basic_form_SFFT}
\end{align}
where for~$\Omega\triangleq [0,\pi]\times[0,2\pi)$ the kernel~$\vect{s}_{\up{R}_{\up{F}\up{R}}}\hspace{-1mm}:[M]\times \Omega \rightarrow \mathbb{C}^2$ fully characterizes the operator. 
See~\fref{app:proof:eq:basic_form_SFFT} for a formal proof that~$\oper{S}_{\up{R}_\up{FR}}$ can be represented as in~\eqref{eq:basic_form_SFFT}. 
The \emph{receiving operator}~$\oper{S}_{\up{R}_{\up{R}\up{F}}}\hspace{-1mm}:L^2\rightarrow \mathbb{C}^M$ is, similarly to the transmitting operator, a finite-rank operator whose action on a function~$\phv{t}\in L^2$ can be represented as
\begin{align}
    &\big[\oper{S}_{\up{R}_{\up{R}\up{F}}}\phv{t}\big]_m 
     =\big\langle\phv{t},\overline{\vect{s}}_{\up{R}_{\up{R}\up{F}}}(m;\cdot)\big\rangle_{L^2}
    \\
    &\quad=
    \oiint\displaylimits_{\Omega}
    \big\langle\phv{t}(\theta,\varphi),\overline{\vect{s}}_{\up{R}_{\up{R}\up{F}}}(m;\theta,\varphi)\big\rangle_{\mathbb{C}^2}
    \,\sin(\theta)\,\up{d}(\theta,\varphi).
    \label{eq:basic_form_SFTF}
\end{align}
Here, we use (i) the \mbox{kernel~$\vect{s}_{\up{R}_{\up{R}\up{F}}}\hspace{-1mm}:[M]\times\Omega\rightarrow \mathbb{C}^2$} to fully characterize the operator, (ii) the notation~$\overline{\vect{s}}_{\up{R}_{\up{R}\up{F}}}(m;\cdot)$ to represent an element in the previously introduced~$L^2$-space, induced by the (complex conjugated) kernel and the respective index~$m$, and (iii) the inner product of the same \mbox{$L^2$-space.} See~\fref{app:proof:eq:basic_form_SFTF} for a formal proof that~$\oper{S}_{\up{R}_\up{RF}}$ can be represented as in~\eqref{eq:basic_form_SFTF}. 

\def\vertT at (#1,#2,#3) named (#4){

    \begin{scope}[shift={(#1,#2)}]
        \draw[white, fill=white, line width=1.15] (0,0) circle (.2);
        \draw[line width=1.15, pattern=diagonal pattern node ,pattern color=#3] (0,0) circle (.2);
        \node at (0,.5) {\normalsize #4};
    \end{scope}
}
\def\vertB at (#1,#2,#3) named (#4){

    \begin{scope}[shift={(#1,#2)}]
        \draw[white, fill=white, line width=1.15] (0,0) circle (.2);
        \draw[line width=1.15, pattern=diagonal pattern node, pattern color=#3] (0,0) circle (.2);
        \node at (0,-.5) {\normalsize #4};
    \end{scope}
}
\def\vertL at (#1,#2,#3) named (#4){

    \begin{scope}[shift={(#1,#2)}]
        \draw[white, fill=white, line width=1.15] (0,0) circle (.2);
        \draw[line width=1.15, pattern=diagonal pattern node, pattern color=#3] (0,0) circle (.2);
        \node at (-.5,0) {\normalsize #4};
    \end{scope}
}

\def\edgeL from (#1,#2) to (#3,#4) named (#5){
    \draw [line width=1pt, color=blue_120, arrows = {-Stealth[inset=0, length=8pt, angle'=35]}] (#1/2+#3/2-.4,#2/2+#4/2) -- (#1/2+#3/2-.4,#2/2+#4/2-4pt);
    \draw [line width=1pt, color=blue_120] plot[smooth, tension=1.2] coordinates {(#1,#2) (#1/2+#3/2-.4,#2/2+#4/2) (#3,#4)};
    \node[anchor=east] at (#1/2+#3/2-.4,#2/2+#4/2) {\color{blue_120}\normalsize #5};
}

\def\edgeR from (#1,#2) to (#3,#4) named (#5){
    \draw [line width=1pt, color=blue_120, arrows = {-Stealth[inset=0, length=8pt, angle'=35]}] (#1/2+#3/2+.4,#2/2+#4/2) -- (#1/2+#3/2+.4,#2/2+#4/2+4pt);
    \draw [line width=1pt, color=blue_120] plot[smooth, tension=1.2] coordinates {(#1,#2) (#1/2+#3/2+.4,#2/2+#4/2) (#3,#4)};
    \node[anchor=west] at (#1/2+#3/2+.45,#2/2+#4/2) {\color{blue_120}\normalsize #5};
}

\def\edgeT from (#1,#2) to (#3,#4) named (#5){
    \draw [line width=1pt, color=blue_120, arrows = {-Stealth[inset=0, length=8pt, angle'=35]}] (#1/2+#3/2,#2/2+#4/2-.4) -- (#1/2+#3/2-.1,#2/2+#4/2-.4);
    \draw [line width=1pt, color=blue_120] plot[smooth, tension=1.2] coordinates {(#1,#2) (#1/2+#3/2,#2/2+#4/2-.4) (#3,#4)};
    \node[anchor=north] at (#1/2+#3/2,#2/2+#4/2-.4) {\color{blue_120}\normalsize #5};
}

\def\edgeB from (#1,#2) to (#3,#4) named (#5){
    \draw [line width=1pt, color=blue_120, arrows = {-Stealth[inset=0, length=8pt, angle'=35]}] (#1/2+#3/2,#2/2+#4/2+.4) -- (#1/2+#3/2+.1,#2/2+#4/2+.4);
    \draw [line width=1pt, color=blue_120] plot[smooth, tension=1.2] coordinates {(#1,#2) (#1/2+#3/2,#2/2+#4/2+.4) (#3,#4)};
    \node[anchor=south] at (#1/2+#3/2,#2/2+#4/2+.4) {\color{blue_120}\normalsize #5};
}

\def\edgeRS from (#1,#2) to (#3,#4) named (#5){
    \draw [line width=1pt, color=blue_120, arrows = {-Stealth[inset=0, length=8pt, angle'=35]}] (.25,.55*\distV) -- (0.25,.6*\distV);
    \draw [line width=1pt, color=blue_120] plot[smooth, tension=1.2] coordinates {(#1,#2) (#1/2+#3/2+.25,#2/2+#4/2) (#3,#4)};
    \node[anchor=east] at (#1/2+#3/2+.2+.05,#2/2+#4/2-.05) {\color{blue_120}\normalsize #5};
}

\begin{figure}[tp]
    \centering
    {
    \small
    \begin{tikzpicture}
        \def\distH{2.6};
        \def\distV{2.7};

        \clip (-.8,-.9*\distV) rectangle (3.4*\distV,1.25*\distV);

        \draw [line width=1pt, color=blue_120] plot[smooth, tension=1.2] coordinates {(2.6*\distH,.7*\distV) (2.25*\distH,-.1*\distV) (2.01*\distH,-\distV/2)};
        \draw [line width=1pt, color=blue_120, arrows = {-Stealth[inset=0, length=8pt, angle'=35]}] (2.525*\distH,.512*\distV) -- (2.525*\distH-.04,.512*\distV-.1);
        \node[anchor=east] at (2.55*\distH,.515*\distV) {\color{blue_120}\small~$-\frac{1}{2\sqrt{R_0}}$};
        \draw[shift={(2.4*\distH,.35*\distV)}, white, fill=white] (-.2,-.4) rectangle (.4,.1);
        \draw[shift={(2.3*\distH,-.03*\distV)}, white, fill=white] (-.2,-.1) rectangle (.2,.2);

        \draw [line width=1pt, color=blue_120] plot[smooth, tension=1.2] coordinates {(0,.8*\distV) (\distH/2,.25*\distV) (.965*\distH,-\distV/2)};
        \draw [line width=1pt, color=blue_120, arrows = {-Stealth[inset=0, length=8pt, angle'=35]}] (\distH/2,.25*\distV) -- (\distH/2-.1,.25*\distV+.15);
        \node[anchor=south west] at (\distH/2-.2,.25*\distV-.1) {\color{blue_120}\normalsize~$-\mat{K}_{\phv{v}_\up{Rx}}$};
        \draw[shift={(.7*\distH,-.05*\distV)}, white, fill=white] (-.3,-.1) rectangle (.4,.35);

        \edgeL from (\distH,\distV/2) to (\distH,-\distV/2) named ($\mat{S}_\up{RF}$);
        \edgeL from (2*\distH,\distV/2) to (2*\distH,-\distV/2) named ($\mat{S}_{\up{T}_\up{RR}}$);
        \edgeL from (3*\distH,\distV/2) to (3*\distH,-\distV/2) named ($\oper{S}_{\up{R}_\up{FF}}$);

        \edgeR from (\distH,\distV/2) to (\distH,-\distV/2) named ($\mat{S}_{\up{T}_\up{TT}}$);
        \edgeR from (2*\distH,\distV/2) to (2*\distH,-\distV/2) named ($\oper{S}_{\up{R}_\up{RR}}$);
        \edgeRS from (0,\distV*.3) to (0,\distV*.8) named ($\mat{Z}_\up{Rx}$);

        \edgeT from (2*\distH,\distV/2) to (\distH,\distV/2) named ($\mat{S}_{\up{T}_\up{TR}}$);
        \edgeT from (3*\distH,\distV/2) to (2*\distH,\distV/2) named ($\oper{S}_{\up{R}_\up{RF}}$);

        \edgeB from (3*\distH,-\distV/2) to (2*\distH,-\distV/2) named ($\oper{S}_{\up{R}_\up{FR}}$);
        \edgeB from (2*\distH,-\distV/2) to (1*\distH,-\distV/2) named ($\mat{S}_{\up{T}_\up{RT}}$);

        \draw [line width=1pt, color=blue_120] plot[smooth, tension=1.2] coordinates {(0,-.8*\distV) (\distH/2,-.75*\distV) (\distH,-\distV/2)};
        \draw [line width=1pt, color=blue_120, arrows = {-Stealth[inset=0, length=8pt, angle'=35]}] (\distH/2,-.75*\distV) -- (\distH/2+.1,-.75*\distV+.04); 
        \node[anchor=south] at (\distH/2-.15,-.75*\distV) {\color{blue_120}\normalsize~$\mat{K}_{\phv{v}_\up{Tx}}$};

        \draw [line width=1pt, color=blue_120] plot[smooth, tension=1.2] coordinates {(0,-.2*\distV) (\distH/2,-.45*\distV) (\distH,-\distV/2)};
        \draw [line width=1pt, color=blue_120, arrows = {-Stealth[inset=0, length=8pt, angle'=35]}] (\distH/2,-.45*\distV) -- (\distH/2+.1,-.45*\distV-.02);
        \node[anchor=south] at (\distH/2,-.45*\distV) {\color{blue_120}\normalsize~$\mat{K}_{\phv{v}_\Gamma}$};

        \draw [line width=1pt, color=blue_120] plot[smooth, tension=1.2] coordinates {(0,.3*\distV) (\distH/2,-.15*\distV) (\distH,-\distV/2)};
        \draw [line width=1pt, color=blue_120, arrows = {-Stealth[inset=0, length=8pt, angle'=35]}] (\distH/2,-.15*\distV) -- (\distH/2+.1,-.15*\distV-.09);
        \node[anchor=south] at (\distH/2,-.135*\distV) {\color{blue_120}\normalsize~$\mat{K}_{\phv{i}_\Gamma}$};

        \draw [line width=1pt, color=blue_120] plot[smooth, tension=1.2] coordinates {(0,.8*\distV) (\distH/2,.7*\distV) (.98*\distH,\distV/2)};
        \draw [line width=1pt, color=blue_120, arrows = {-Stealth[inset=0, length=8pt, angle'=35]}] (\distH/2,.7*\distV) -- (\distH/2-.1,.7*\distV+.03);
        \node[anchor=north] at (\distH/2,.7*\distV) {\color{blue_120}\normalsize~$\mat{K}_{\phv{v}_\up{Rx}}$};

        \draw [line width=1pt, color=blue_120] plot[smooth, tension=1.2] coordinates {(2.6*\distH,.7*\distV) (2.25*\distH,.65*\distV) (2*\distH,\distV/2)};
        \draw [line width=1pt, color=blue_120, arrows = {-Stealth[inset=0, length=8pt, angle'=35]}] (2.25*\distH,.65*\distV) -- (2.25*\distH-.1,.65*\distV-.025);
        \node[anchor=south] at (2.25*\distH,.65*\distV+.1) {\color{blue_120}\normalsize~$\frac{1}{2\sqrt{R_0}}$};

        \vertL at (0,\distV*.8,white) named ($\phv{v}_\up{Rx}$);
        \vertL at (0,\distV*.3,black) named ($\phv{i}_\Gamma$);
        \vertL at (0,-\distV*.2,black) named ($\phv{v}_\Gamma$);
        \vertL at (0,-\distV*.8,black) named ($\phv{v}_\up{Tx}$);
    
        \vertT at (\distH,\distV/2,white) named ($\phv{b}_\up{T}$);
        \vertB at (\distH,-\distV/2,white) named ($\phv{a}_\up{T}$);

        \vertT at (\distH*2,\distV/2,white) named ($\phv{b}_\up{R}$);
        \vertB at (\distH*2,-\distV/2,white) named ($\phv{a}_\up{R}$);

        \vertT at (\distH*2.6,\distV*.7,black) named ($\phv{v}_\Upsilon$);

        \vertT at (\distH*3,\distV/2,black) named ($\phv{b}_\up{F}$);
        \vertB at (\distH*3,-\distV/2,white) named ($\phv{a}_\up{F}$);

        \node[shift={(\distH/2,0)}, anchor=north] at (0,1.2*\distV) {\normalsize RF frontend};
        \node[shift={(1.5*\distH,0)}, anchor=north] at (0,1.2*\distV) {\normalsize tuning network};
        \node[shift={(2.6*\distH,0)}, anchor=north] at (0,1.2*\distV) {\normalsize radiating structure};

        \draw[shift={(\distH,0)}, line width=1, dashed, dash pattern=on 6pt off 3pt] (0,1.2*\distV) -- (0,.8*\distV);
        \draw[shift={(2*\distH,0)}, line width=1, dashed, dash pattern=on 6pt off 3pt] (0,1.2*\distV) -- (0,.8*\distV);
        
    \end{tikzpicture}}
\caption{Signal-flow graph of our model: The nodes represent (i) complex vectors that correspond to voltages, currents, and circuit theoretic power waves, or (ii) elements in an~$L^2$-space, which represent the far-field radiation pattern of the incoming and outgoing waves. The edges of the graph represent bounded linear operators between the respective spaces. To enhance readability, the nodes representing inputs are hatched.}
\label{fig:signal_flow_graph}
\end{figure}

The \emph{scattering operator}~$\oper{S}_{\up{R}_{\up{F}\up{F}}}\hspace{-1mm}:L^2\rightarrow L^2$ can, according to Saxon~\cite[Eq.~4]{saxon_tensor_scattering_matrix_for_the_electromagnetic_field}, be represented by a surface-integral over the unit sphere~$S^2$ as
\begin{align}
    \big(\oper{S}_{\up{R}_{\up{F}\up{F}}}\phv{t}\big)(\hat{\vect{r}})
    &=
     \oiint\displaylimits_{S^2} 
    \mat{S}_{\up{R}_{\up{F}\up{F}}}(\hat{\vect{r}};\hat{\vect{r}}')\phv{t}(\hat{\vect{r}}')
    \,\up{d}\vect{S}'.
    \label{eq:basic_form_SFFF}
\end{align}
Here,~$\hat{\vect{r}}$ and~$\hat{\vect{r}}'$ are unit vectors corresponding to spherical coordinates~$(\theta,\varphi)$ and~$(\theta', \varphi')$, respectively.\footnote{To ensure correct notation, functions that map the unit vectors~$\hat{\vect{r}}$ and~$\hat{\vect{r}}'$ to their respective spherical coordinates would need to be added in~\eqref{eq:basic_form_SFFF}. \new{However, to simplify notation, we do not explicitly include these functions.}}
Furthermore,~$\up{d}\vect{S}'$ denotes the differential vector element of surface area normal to the unit sphere~$S^2$ at position~$\hat{\vect{r}}'$ and the kernel~$\mat{S}_{\up{R}_{\up{F}\up{F}}}$ fully characterizes this scattering operator.
This kernel~$\mat{S}_{\up{R}_{\up{F}\up{F}}}$ can, for example, be a proper function of the form~$\Omega\times\Omega\rightarrow\mathbb{C}^{2\times 2}$. In this special case one can directly parametrize the integral in~\eqref{eq:basic_form_SFFF} by~$(\theta,\varphi)$ and write 
\begin{align}
    &\!\!\!\big(\oper{S}_{\up{R}_{\up{F}\up{F}}}\phv{t}\big)(\theta,\varphi)
    \nonumber
    \\
    &=
    \oiint\displaylimits_{\Omega} 
    \mat{S}_{\up{R}_{\up{F}\up{F}}}(\theta,\varphi;\theta',\varphi')\phv{t}(\theta',\varphi')
    \,\sin(\theta')\,\up{d}(\theta',\varphi').
\end{align}
Note that more general mathematical tools, such as distributions, might be necessary to describe the kernel~$\mat{S}_{\up{R}_\up{FF}}$, e.g., if no scatterer is present and~$\oper{S}_{\up{R}_{\up{F}\up{F}}}$ acts as the identity operator.

\setcounter{equation}{36}
\begin{figure*}[b]
    \begingroup
    \addtolength{\jot}{-.34em}
    \hrule
    \begin{align}
        \label{eq:gain_operator_equation_first}
        \mathmakebox[\widthof{$\mathbb{G}_{\phv{v}_\up{Tx}}^{\phv{a}_\up{F}}$}][l]{
        \mathbb{G}_{\phv{v}_\up{Tx}}^{\phv{a}_\up{F}}}
        &=
        \oper{S}_{\up{R}_\up{FR}}
        \left(
            \mat{I}_{M}
            -\mat{L}_2
         \right)^{-1}
         \mat{S}_{\up{T}_\up{RT}}
         \left(
            \mat{I}_{N}
            -\mat{L}_1
            -\mat{L}_3
        \right)^{-1}
        \mat{K}_{\phv{v}_\up{Tx}}
        \\
        \mathmakebox[\widthof{$\mathbb{G}_{\phv{v}_\up{Tx}}^{\phv{a}_\up{F}}$}][l]{
        \mathbb{G}_{\phv{b}_\up{F}}^{\phv{a}_\up{F}}
        }
        &=
        \oper{S}_{\up{R}_\up{FR}}
        \Big(
            \mat{S}_{\up{T}_\up{RT}}
            \mat{S}_\up{RF}
            \big(
                \mat{I}_{N}
                -\mat{L}_5
            \big)^{-1}
            \mat{S}_{\up{T}_\up{TR}}
            +
            \mat{S}_{\up{T}_\up{RR}}
        \Big)
        \Big(
            \mat{I}_{M}
            -\mat{L}_6
            -\mat{L}_7
        \Big)^{-1}
        \oper{S}_{\up{R}_\up{RF}}
        \,+\,
        \oper{S}_{\up{R}_\up{FF}}
        \\
        \mathmakebox[\widthof{$\mathbb{G}_{\phv{v}_\up{Tx}}^{\phv{a}_\up{F}}$}][l]{
        \mathbb{G}_{\phv{v}_\up{Tx}}^{\phv{v}_\up{Rx}}
        }
        &=
        {\color{black}
            \mat{K}_{\phv{v}_\up{Rx}}
            \Big(
                \mat{S}_{\up{T}_\up{TR}}
                \oper{S}_{\up{R}_\up{RR}}
                \big(
                    \mat{I}_M
                    -
                    \mat{L}_2
                \big)^{-1}
                \mat{S}_{\up{T}_\up{RT}}
                +
                \mat{S}_{\up{T}_\up{TT}}
                -\mat{I}_N
            \Big)
            }{\color{black}
            \Big(
                \mat{I}_N
                -
                \mat{L}_1
                -
                \mat{L}_3
            \Big)^{-1}
        }
        \mat{K}_{\phv{v}_\up{Tx}}
        \\
        \mathmakebox[\widthof{$\mathbb{G}_{\phv{v}_\up{Tx}}^{\phv{a}_\up{F}}$}][l]{
        \mathbb{G}_{\phv{v}_\Gamma}^{\phv{v}_\up{Rx}}
        }
        &=
        {\color{black}
            \mat{K}_{\phv{v}_\up{Rx}}
            \Big(
                \mat{S}_{\up{T}_\up{TR}}
                \oper{S}_{\up{R}_\up{RR}}
                \big(
                    \mat{I}_M
                    -
                    \mat{L}_2
                \big)^{-1}
                \mat{S}_{\up{T}_\up{RT}}
                +
                \mat{S}_{\up{T}_\up{TT}}
                -\mat{I}_N
            \Big)
            }{\color{black}
            \Big(
                \mat{I}_N
                -
                \mat{L}_1
                -
                \mat{L}_3
            \Big)^{-1}
        }
        \mat{K}_{\phv{v}_\Gamma}
        \\
        \mathmakebox[\widthof{$\mathbb{G}_{\phv{v}_\up{Tx}}^{\phv{a}_\up{F}}$}][l]{
        \mathbb{G}_{\phv{i}_\Gamma}^{\phv{v}_\up{Rx}}
        }
        &=
        {\color{black}
            \mat{K}_{\phv{v}_\up{Rx}}
            \Big(
                \mat{S}_{\up{T}_\up{TR}}
                \oper{S}_{\up{R}_\up{RR}}
                \big(
                    \mat{I}_M
                    -
                    \mat{L}_2
                \big)^{-1}
                \mat{S}_{\up{T}_\up{RT}}
                +
                \mat{S}_{\up{T}_\up{TT}}
                -\mat{I}_N
            \Big)
            }{\color{black}
            \Big(
                \mat{I}_N
                -
                \mat{L}_1
                \mat{L}_3
            \Big)^{-1}
        }
        \mat{K}_{\phv{i}_\Gamma}
        +\mat{Z}_\up{Rx}
        \\
        \mathmakebox[\widthof{$\mathbb{G}_{\phv{v}_\up{Tx}}^{\phv{a}_\up{F}}$}][l]{
        \mathbb{G}_{\phv{b}_\up{F}}^{\phv{v}_\up{Rx}}
        }
        &=
        {\color{black}
            \mat{K}_{\phv{v}_\up{Rx}}
            \Big(
                \mat{I}_N
                -
                \mat{S}_\up{RF}
            \Big)
            \Big(
            \mat{I}_N-\mat{L}_5
            \Big)^{-1}
            }{\color{black}
            \mat{S}_{\up{T}_\up{TR}}
        \Big(
                \mat{I}_M
                -\mat{L}_6
                -\mat{L}_7
        \Big)^{-1}
        }
        \oper{S}_{\up{R}_\up{RF}}
        \\
        \mathmakebox[\widthof{$\mathbb{G}_{\phv{v}_\up{Tx}}^{\phv{a}_\up{F}}$}][l]{
        \mathbb{G}_{\phv{v}_\Upsilon}^{\phv{v}_\up{Rx}}
        }
        &=
        {\color{black}
            \mat{K}_{\phv{v}_\up{Rx}}
            \Big(
                \mat{I}_N
                -
                \mat{S}_\up{RF}
            \Big)
            \Big(
            \mat{I}_N-\mat{L}_5
            \Big)^{-1}
            }{\color{black}
            \mat{S}_{\up{T}_\up{TR}}
        \Big(
            \mat{I}_M
            -\mat{L}_6
            -\mat{L}_7
        \Big)^{-1}
        }
        \Big(
        \mat{I}_M
        -
        \oper{S}_{\up{R}_\up{RR}}
        \Big)
        \frac{1}{2\sqrt{R_0}}
        \label{eq:gain_operator_equation_last}
    \end{align}
    \hrule
    \subfloat{
    \begin{minipage}{0.49\textwidth}
        \begin{align}
            \label{eq:aux_L_equation_first}
            \mat{L}_1
            &\triangleq
            \mat{S}_\up{RF}
            \mat{S}_{\up{T}_\up{TT}}
            \\
            \mat{L}_2
            &\triangleq
            \mat{S}_{\up{T}_\up{RR}}
            \oper{S}_{\up{R}_\up{RR}}
            \\
            \mat{L}_3
            &\triangleq
            \mat{S}_\up{RF}
            \mat{S}_{\up{T}_\up{TR}}
            \oper{S}_{\up{R}_\up{RR}}
            \left(
            \mat{I}_{M}
            -\mat{L}_2
            \right)^{-1}
            \mat{S}_{\up{T}_\up{RT}}
            \\   
            \mat{L}_4
            &\triangleq
            \mat{S}_{\up{T}_\up{RT}}
            \mat{S}_\up{RF}
            \left(
            \mat{I}_{N}
            -\mat{L}_5
            \right)^{-1}
            \mat{S}_{\up{T}_\up{TR}}
            \oper{S}_{\up{R}_\up{RR}}     
        \end{align}
    \end{minipage}
    }
    \hfill
    \subfloat{
    \begin{minipage}{0.49\textwidth}
        \begin{align}
            \mat{L}_5
            &\triangleq
            \mat{S}_{\up{T}_\up{TT}}
            \mat{S}_\up{RF}
            \\
            \mat{L}_6
            &\triangleq
            \oper{S}_{\up{R}_\up{RR}}
            \mat{S}_{\up{T}_\up{RR}}
            \\
            \mat{L}_7
            &\triangleq
            \oper{S}_{\up{R}_\up{RR}}
            \mat{S}_{\up{T}_\up{RT}}
            \mat{S}_\up{RF}
            \left(
            \mat{I}_{N}
            -\mat{L}_5
            \right)^{-1}
            \mat{S}_{\up{T}_\up{TR}}.
            \label{eq:aux_L_equation_last}
            \\
            \nonumber
        \end{align}
    \end{minipage}
    }
    \vspace{3mm}
    \hrule
    \endgroup
\end{figure*}
\setcounter{equation}{22}

\subsection{Input-Output Relationships}\label{sec:input_output_relationships} 
We now derive several input-output relationships for a REMS. 
We begin by mathematically describing the behavior of the RF frontend.
From basic circuit theory and~\fref{defi:circuit_theoretic_power_waves}, it follows that the PAs behavior can be described by 
\begin{align}
    \phv{a}_{\up{T}}^\up{Tx}
    &=
    \underbrace{
    (\mat{Z}_{\up{Tx}}+R_0\mat{I}_{N_\up{Tx}})^{-1}(\mat{Z}_{\up{Tx}}-R_0\mat{I}_{N_\up{Tx}})
    }_{\triangleq\,\mat{S}_{\up{RF}}^\up{Tx}}
    \,
    \phv{b}_{\up{T}}^\up{Tx}
    \nonumber
    \\
    &\;\;\;+
    \underbrace{
    (\mat{Z}_{\up{Tx}}+R_0\mat{I}_{N_\up{Tx}})^{-1}\sqrt{R_0}
    }_{\triangleq\,\tilde{\mat{K}}_{\phv{v}_\up{Tx}}}
    \,
    \phv{v}_{\up{Tx}},
    \label{eq:PA_equation}
\end{align}
where we define~$\mat{S}_{\up{RF}}^\up{Tx}$ and~$\tilde{\mat{K}}_{\phv{v}_\up{Tx}}$ to simplify notation later. 
Analogously, one can show that for the LNAs it holds that
\begin{align}
\phv{a}_{\up{T}}^\up{Rx}
    &=
    \underbrace{
    (\mat{Z}_{\up{Rx}}+R_0\mat{I}_{N_\up{Rx}})^{-1}(\mat{Z}_{\up{Rx}}-R_0\mat{I}_{N_\up{Rx}})
    }_{\triangleq\,\mat{S}_{\up{RF}}^\up{Rx}}
    \,
    \phv{b}_{\up{T}}^\up{Rx}
    \nonumber
    \\
    &\;\;\;+
    \underbrace{
    (\mat{Z}_{\up{Rx}}+R_0\mat{I}_{N_\up{Rx}})^{-1}
    \sqrt{R_0}\mat{Z}_{\up{Rx}}
    }_{\triangleq\,\tilde{\mat{K}}_{\phv{i}_\Gamma}}
    \,
    \phv{i}_{\Gamma}
    \nonumber
    \\
    &\;\;\;+
    \underbrace{
    (\mat{Z}_{\up{Rx}}+R_0\mat{I}_{N_\up{Rx}})^{-1}
    \sqrt{R_0}
    }_{\triangleq\,\tilde{\mat{K}}_{\phv{v}_\Gamma}}
    \,
    \phv{v}_{\Gamma},
    \label{eq:LNA_equation_1}
    \\
    \phv{v}_{\up{Rx}}
    &=
    \underbrace{ 
    (\sqrt{R_0})^{-1}
    \mat{Z}_\up{Rx}
    }_{\triangleq\,\tilde{\mat{K}}_{\phv{v}_\up{Rx}}}
    \,
    \big(\phv{b}_{\up{T}}^\up{Rx}-\phv{a}_{\up{T}}^\up{Rx}\big)
    +
    \mat{Z}_\up{Rx}
    \phv{i}_{\Gamma},
    \label{eq:LNA_equation_2}
\end{align}
where we define~$\mat{S}_{\up{RF}}^\up{Rx}$,~$\tilde{\mat{K}}_{\phv{v}_\up{Rx}}$,~$\tilde{\mat{K}}_{\phv{i}_\Gamma}$, and~$\tilde{\mat{K}}_{\phv{v}_\Gamma}$ to simplify notation later.
\new{From the RF frontend structure in \fref{fig:model} and~\fref{asm:basic_assumption_PA_LNA}, the impedance matrices~$\mat{Z}_\up{Tx}$ and~$\mat{Z}_\up{Rx}$ are diagonal with non-negative real parts. 
Thus, the inverses in~\eqref{eq:PA_equation},~\eqref{eq:LNA_equation_1}, and~\eqref{eq:LNA_equation_2} exist for any strictly positive reference resistance (e.g.,~$R_0 = 50\,\Omega$).}

To further simplify notation, we introduce the following auxiliary matrices:
\begin{align}
    \label{eq:aux_matrices_1}
    \mat{K}_{\phv{v}_\up{Tx}}
    &\triangleq
    \begin{bmatrix}
        \tilde{\mat{K}}_{\phv{v}_\up{Tx}} \\ \mat{0}_{N_\up{Rx},N_\up{Tx}}
    \end{bmatrix},\!
    \hphantom{a}
    \mat{K}_{\phv{v}_\Gamma}
    \triangleq
    \begin{bmatrix}
        \mat{0}_{N_\up{Tx},N_\up{Rx}} \\ \tilde{\mat{K}}_{\phv{v}_\Gamma}
    \end{bmatrix},\!
    \hphantom{a}
    \mat{K}_{\phv{i}_\Gamma}
    \triangleq
    \begin{bmatrix}
        \mat{0}_{N_\up{Tx},N_\up{Rx}} \\ \tilde{\mat{K}}_{\phv{i}_\Gamma}
    \end{bmatrix}\!,
    \\
    \label{eq:aux_matrices_2}
    \mat{K}_{\phv{v}_\up{Rx}}
    &\triangleq
    \begin{bmatrix}
        \mat{0}_{N_\up{Rx},N_\up{Tx}} & \tilde{\mat{K}}_{\phv{v}_\up{Rx}}
    \end{bmatrix},
    \hphantom{a}
    \mat{S}_\up{RF}
    \triangleq
    \begin{bmatrix}
        \mat{S}_\up{RF}^\up{Tx}&\mat{0}_{N_\up{Tx},N_\up{Rx}}
        \\
        \mat{0}_{N_\up{Rx},N_\up{Tx}}&\mat{S}_\up{RF}^\up{Rx}
    \end{bmatrix}. 
\end{align}

The REMS's behavior can be described by the system of linear equations consisting of \eqref{eq:matrix_ST}, \eqref{eq:extrinsic_noise_equation}, \eqref{eq:matrix_SF},~\eqref{eq:PA_equation},~\eqref{eq:LNA_equation_1},~\eqref{eq:LNA_equation_2},~\eqref{eq:aux_matrices_1}, and \eqref{eq:aux_matrices_2}. 
Alternatively, this system of equations can be visualized as the signal-flow graph depicted in~\fref{fig:signal_flow_graph}.
Visualizing the REMS's behavior with such a signal-flow graph offers the advantage that the various input-output relationships can be derived quickly. 
For example, the return loop method introduced by Riegle and Lin in~\cite{riegle_lin_matrix_signal_flow_graphs_and_an_optimum_topological_method_for_evaluating_their_gains} can be used to derive input-output relationships.
The inputs to our REMS model include the signal and noise source voltages and \mbox{currents~$\phv{v}_\up{Tx}$,~$\phv{v}_\Gamma$,~$\phv{i}_\Gamma$,~$\phv{v}_\Upsilon$,} and the incoming far-field power wave pattern~$\phv{b}_\up{F}$, while the outputs of the model are the LNA receive voltage vector~$\phv{v}_\up{Rx}$ and the outgoing far-field power wave pattern~$\phv{a}_\up{F}$.
For an output~$\phv{s}_\up{out}$ and an input~$\phv{s}_\up{in}$, we denote the operator characterizing the linear dependence of~$\phv{s}_\up{out}$ on~$\phv{s}_\up{in}$ as~$\oper{G}_{\phv{s}_\up{in}}^{\phv{s}_\up{out}}$. 
The operator $\oper{G}_{\phv{s}_\up{in}}^{\phv{s}_\up{out}}$ is implicitly defined by 
\begin{align}
    \phv{s}_\up{out}
    \left\rvert{
        {
        \scriptscriptstyle
        \begin{subarray}{l}
            \hspace{1mm}        \\
            \hspace{1mm}        \\
            \text{inputs other than~$\phv{s}_\up{in}$ are disabled}  
        \end{subarray}
        }
    }
    \right.
    &=
    \oper{G}_{\phv{s}_\up{in}}^{\phv{s}_\up{out}}
    \phv{s}_\up{in}
    \label{eq:define_G_matrix},
\end{align}
where the vertical bar~$\rvert$ means ``evaluated at.'' 
Using this notation one can write
\begin{align}
            \phv{a}_\up{F}
            &=
            \mathmakebox[\widthof{$\mathbb{G}_{\phv{v}_\up{Tx}}^{\phv{v}_\up{Rx}}\phv{v}_\up{Tx}$}][c]{\mathbb{G}_{\phv{v}_\up{Tx}}^{\phv{a}_\up{F}}\phv{v}_\up{Tx}}
            +
            \mathmakebox[\widthof{$\mathbb{G}_{\phv{b}_\up{F}}^{\phv{v}_\up{Rx}}\phv{b}_\up{F}$}][c]{\mathbb{G}_{\phv{b}_\up{F}}^{\phv{a}_\up{F}}\phv{b}_\up{F}}
            +
            \mathmakebox[\widthof{$\mathbb{G}_{\phv{v}_\Gamma}^{\phv{v}_\up{Rx}}\phv{v}_\Gamma$}][c]{\mathbb{G}_{\phv{v}_\Gamma}^{\phv{a}_\up{F}}\phv{v}_\Gamma}
            +
            \mathmakebox[\widthof{$\mathbb{G}_{\phv{i}_\Gamma}^{\phv{v}_\up{Rx}}\phv{i}_\Gamma$}][c]{\mathbb{G}_{\phv{i}_\Gamma}^{\phv{a}_\up{F}}\phv{i}_\Gamma}
            +
            \mathmakebox[\widthof{$\mathbb{G}_{\phv{v}_\Upsilon}^{\phv{v}_\up{Rx}}\phv{v}_\Upsilon$}][c]{\mathbb{G}_{\phv{v}_\Upsilon}^{\phv{a}_\up{F}}\phv{v}_\Upsilon}
            \label{eq:input_output_relationship_bF}
            \\
            \phv{v}_\up{Rx}
            &=
            \mathbb{G}_{\phv{v}_\up{Tx}}^{\phv{v}_\up{Rx}}\phv{v}_\up{Tx}
            +
            \mathbb{G}_{\phv{b}_\up{F}}^{\phv{v}_\up{Rx}}\phv{b}_\up{F}
            +
            \mathbb{G}_{\phv{v}_\Gamma}^{\phv{v}_\up{Rx}}\phv{v}_\Gamma
            +
            \mathbb{G}_{\phv{i}_\Gamma}^{\phv{v}_\up{Rx}}\phv{i}_\Gamma
            +
            \mathbb{G}_{\phv{v}_\Upsilon}^{\phv{v}_\up{Rx}}\phv{v}_\Upsilon.
            \label{eq:input_output_relationship_rx}
\end{align}
In~\eqref{eq:gain_operator_equation_first}--\eqref{eq:gain_operator_equation_last}, we explicitly show how most of the linear operators appearing in~\eqref{eq:input_output_relationship_bF} and~\eqref{eq:input_output_relationship_rx} can be constructed from model parameters; the auxiliary matrices~$\mat{L}_1$ to~$\mat{L}_7$ are defined in~\eqref{eq:aux_L_equation_first} to~\eqref{eq:aux_L_equation_last}, respectively.
We refrain from giving constructing formulas for~$\mathbb{G}_{\phv{v}_\Gamma}^{\phv{a}_\up{F}}$,~$\mathbb{G}_{\phv{i}_\Gamma}^{\phv{a}_\up{F}}$, and~$\mathbb{G}_{\phv{v}_\Upsilon}^{\phv{a}_\up{F}}$, because the influence of noise on the outgoing far-field power wave pattern~$\vect{a}_\up{F}$ can typically be ignored.
\subsection{Wireless Channel Between Two REMSs}\label{sec:LoS_channel} 
\new{
In Section~\ref{sec:input_output_relationships}, we demonstrate how to calculate various input-output relationships for a single REMS. 
However, deriving input-output relationships for entire wireless systems is often necessary as well. 
To illustrate how our REMS modeling method can be applied to derive these relationships for entire systems, we present the following example scenario. 
}

\new{
Consider two REMSs placed far apart in empty space, with the first REMS equipped with~$M^{(1)}$ antennas and the second REMS equipped with~$M^{(2)}$ antennas. 
Our goal is to calculate the \emph{transmission coefficients} (see~\fref{defi:transmission_coefficient}) of the wireless channel between the two REMSs. 
Since the two REMSs are separated by a large distance, the transmission coefficients can be approximated utilizing~\fref{prop:calculate_channel}, provided below. 
A proof of~\fref{prop:calculate_channel} is provided in~\fref{app:proof:proposition_channel}. 
}

\begin{defi}[Transmission Coefficient  of a Wireless Channel]\label{defi:transmission_coefficient}
    Given a system consisting of~$U$ REMSs, for which~\fref{asm:basic_assumption_radiating_part} holds.
    One can treat the subsystem consisting of the radiating structures of these~$U$ REMSs and the wireless channel between them as a multiport with~$\sum_{u\in[U]}M^{(u)}$ ports, analogous to the multiport~$Z_\up{A}$ in~\cite[Fig.~2]{ivrlac_nossek_toward_a_circuit_theory_of_communication}).
    \mbox{For~$u^{(1)},u^{(2)}\in [U]$} and \mbox{$\ell^{(1)}\in [M^{(u^{(1)})}]$,}~$\ell^{(2)}\in [M^{(u^{(2)})}]$, we define the scattering parameter that characterizes the scattering from the incoming circuit-theoretic power wave~$(\phv{a}_{\tilde{R}}^{(u^{(1)})})_{\ell^{(1)}}$ at the~$\ell^{(1)}$th port of REMS~$u^{(1)}$ to the outgoing circuit-theoretic power wave~$(\phv{b}_{\tilde{R}}^{(u^{(2)})})_{\ell^{(2)}}$ at the~$\ell^{(2)}$th port of REMS~$u^{(2)}$ as the transmission coefficient~$S_{u^{(2)}_{\ell^{(2)}} u^{(1)}_{\ell^{(1)}}}$.
\end{defi}

\begin{prop}\label{prop:calculate_channel}
    Given two REMSs, each with finite spatial support, for which~\fref{asm:basic_assumption_radiating_part} hold. 
    Assume that they are placed in empty space and their radiating structures possess~$M^{(1)}$ and~$M^{(2)}$ ports, respectively. 
    Additionally, assume that the kernels~$\vect{s}_{\up{R}_{\up{F}\up{R}}}^{(i)}$,~$\vect{s}_{\up{R}_{\up{R}\up{F}}}^{(i)}$, and~$\mat{S}_{\up{R}_{\up{F}\up{F}}}^{(i)}$, where~$i\in\{1,2\}$, of both REMSs are continuous functions.
    Let~$\vect{d}=d\hat{\vect{d}}$ be the Cartesian vector pointing from the center of the first REMS to the center of the second REMS, where~$d$ can be interpreted as the distance between the REMSs' and the unit vector~$\hat{\vect{d}}$ characterizes their relative orientation to each other.
    Let~$\ell^{(1)}\in[M^{(1)}]$ and~$\ell^{(2)}\in[M^{(2)}]$. 
    Then, as~$d\rightarrow\infty$, the transmission coefficient from the~$\ell^{(1)}$th port of the first REMS to the~$\ell^{(2)}$th port of the second REMS is given by\footnote{For convenience, we abuse the notation of the kernel function arguments in~\eqref{eq:transmission_coefficient_construction_formula}--\eqref{eq:definition_matrix_c}. Specifically, instead of using spherical altitude and azimuth coordinates, we represent directions with unit vectors (e.g.,~$\hat{\vect{d}}$).}
    \begin{align}
        \lim_{d\rightarrow\infty}\!
        S_{2_{\ell^{(2)}}1_{\ell^{(1)}}}
        \!=\!
        \Big(
            \vect{s}_{\up{R}_{\up{R}\up{F}}}^{(2)}
            (\ell^{(2)};-\hat{\vect{d}})
        \Big)^{\!\!\T}
        \!\mat{C}
        \big(
            \mat{I}_2\!-\!\mat{M}
        \big)^{-1}
        \vect{s}_{\up{R}_{\up{F}\up{R}}}^{(1)}
        (\ell^{(1)};\hat{\vect{d}}),
        \label{eq:transmission_coefficient_construction_formula}
    \end{align}
    where the matrices~$\mat{C}$ and~$\mat{M}$ are given by 
    \begin{align}
        \mat{M}
        &\triangleq
        \mat{S}_{\up{R}_{\up{F}\up{F}}}^{(1)}(\hat{\vect{d}};\hat{\vect{d}})
        \mat{C}
        \mat{S}_{\up{R}_{\up{F}\up{F}}}^{(2)}(-\hat{\vect{d}};-\hat{\vect{d}})
        \mat{C}
        \label{eq:definition_matrix_m}
        \\
        \mat{C}
        &\triangleq
        \frac{2\pi}{j k}
        \frac{e^{-j k d}}{d}
        \begin{bmatrix}
            1 & 0 \\
            0 &  -1
        \end{bmatrix}.
        \label{eq:definition_matrix_c}
    \end{align}
\end{prop}
%

\def\powerNode at (#1,#2,#3){
    \begin{scope}[shift={(#1, #2)}]
        \draw  [line width=1pt, fill=white] (0,0) ellipse (0.4 and 0.4);
        \node[color=black] at (0,0) {#3};     
    \end{scope}
}

\def\powerNodeII at (#1,#2,#3){
    \begin{scope}[shift={(#1, #2)}]
        \draw[rounded corners = 11.5, line width=1pt, fill=white] (-.8, -.4) rectangle (.8, .4) {};
        \node[color=black] at (0,0) {#3};     
    \end{scope}
}

\def\causalRelation at (#1,#2,#3,#4,#5,#6,#7){

    \draw [line width=1pt, color=blue_120, arrows = {-Stealth[inset=0, length=6pt, angle'=35]}] (#1,#2) -- (#3*0.55+#1*0.45,#4*0.55+#2*0.45);
    \draw [line width=1pt, color=blue_120] (#3,#4) -- (#3/2+#1/2,#4/2+#2/2);

    \node[shift={(#3*0.5+#1*0.5,#4*0.5+#2*0.5)}, anchor=north] at (#6,#7) {\color{blue_120}#5};
} 

\def\gainArrow at (#1,#2,#3,#4,#5,#6,#7){
    \draw [densely dotted, line width=1pt, color=blue, arrows = {-Stealth[inset=0, length=6pt, angle'=35]}] (2*\distPowerNodes,1) -- (2.1*\distPowerNodes,1);

    \draw [densely dashed, line width=1pt, color=blue] plot[smooth, tension=.8] coordinates {(0,0) (2*\distPowerNodes,1) (4.2*\distPowerNodes,0)};
  
    \node[shift={(2.1*\distPowerNodes,1)}, anchor=south] at (#6,#7) {\color{blue}#5};
}

\begin{figure}[tp]
    \centering
    {
    \small
    \begin{tikzpicture}
        \def\distPowerNodes{1.85};

        \causalRelation at (0,0,\distPowerNodes,0,$\eta_\up{matching}$,0,0);
        \causalRelation at (\distPowerNodes,0,2*\distPowerNodes,0,~$\eta_\up{tuning}$ ,0,0);
        \causalRelation at (2*\distPowerNodes,0,3*\distPowerNodes,0,~$\eta_\up{radiating}$ ,0,0);
        \causalRelation at (3*\distPowerNodes,0,4*\distPowerNodes,0, {$D$ } ,0,0);

        \gainArrow at (0, 0, 3*\distPowerNodes, 0.8*\distPowerNodes, {$G_\up{REMS}$}, 0, 0);
       
        \powerNode at (0,0,$P_\up{A}$);
        \powerNode at (\distPowerNodes, 0,$P_\up{T}$);
        \powerNode at (2*\distPowerNodes, 0,$P_\up{R}$);
        \powerNode at (3*\distPowerNodes, 0,$P_\up{F}$);
        \powerNode at (4*\distPowerNodes, 0,{$4\pi \,I$});

    \end{tikzpicture}}
\caption{The relationship between various REMS power metrics is illustrated with solid lines. The REMS gain, which relates the radiation intensity to the available power from the power amplifiers, is represented by a dashed line. In general, the matching efficiency~$\eta_\up{matching}$, the tuning network efficiency~$\eta_\up{tuning}$, the radiation efficiency~$\eta_\up{radiation}$, and the directivity~$D(\theta,\varphi)$ are all influenced by the voltage applied by the REMS power amplifiers and the configuration of the reconfigurable elements.}
\label{fig:power_and_gain_terms}
\end{figure}
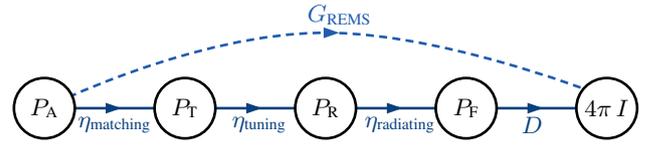
\subsection{Power and Gain Metrics}

Besides the input-output relationships of a REMS and the wireless channels between multiple REMSs, one is often also interested in the total power output by the PAs, the radiated power of the REMS, or other related power metrics.
We now define a range of potentially useful power metrics.
We start with the following three interface-related power metrics.

\begin{defi}[Interface-Related Power Metrics]\label{defi:basic_power_metrics}
    We define the \emph{tuning network accepted power} as
    \begin{align}
        \label{eq:defi_PT}
        \hspace{.5cm}P_\up{T}
        &\triangleq
        \mathmakebox[2cm][c]{\;\Re\{\phv{v}_\up{T}^\He\phv{i}_\up{T}\}}
        =
        \|\phv{a}_\up{T}\|_2^2
        -
        \|\phv{b}_\up{T}\|_2^2,
        \intertext{
        the \emph{radiating structure accepted power} as
        }
        \label{eq:defi_PR}
        P_\up{R}
        &\triangleq
        \mathmakebox[2cm][c]{\;
        \Re\{\phv{v}_\up{R}^\He\phv{i}_\up{R}\}
        }
        =
        \|\phv{a}_\up{R}\|_2^2
        -
        \|\phv{b}_\up{R}\|_2^2,
        \intertext{
        and the \emph{total far field radiated power} as 
        }
        \label{eq:defi_PF}
        P_\up{F}
        &\triangleq
        \mathmakebox[2cm][c]{\;
        P^\nearrow 
        -
        P^\swarrow
        }
        =
        \|\phv{a}_\up{F}\|_{L^2}^2
        -
        \|\phv{b}_\up{F}\|_{L^2}^2.
    \end{align}
    The right-hand side quantities in~\eqref{eq:defi_PT}--\eqref{eq:defi_PF} are derived  in~\fref{app:proof_power_metrics_basic}.
\end{defi}
In contrast to these interface-related power metrics, which specify the actual power flowing into a subsystem, the next power metric specifies a potentially available power.

\setcounter{equation}{50}
\begin{defi}[Power Amplifiers Available Power]\label{defi:power_metrics_PA}
    For a fixed PA output voltage vector~$\phv{v}_\up{Tx}\in\mathbb{C}^{N_\up{Tx}}$, we define the \emph{power amplifiers available power} as 
    \begin{align}
        \label{eq:defi_PA}
        P_\up{A}
        &\triangleq
        \max_{
            \mat{S}_{\up{R}_\up{RR}},\,\mat{S}_\up{T} 
        }
        P_\up{T}
        \left\rvert{
            {
            \scriptscriptstyle
            \begin{subarray}{l}
                \hspace{1mm}        \\
                \hspace{1mm}        \\
                \mathmakebox[\widthof{$\phv{v}_\Upsilon$}][l]{\phv{v}_\Gamma}=\,\mat{0}    \\
                \mathmakebox[\widthof{$\phv{v}_\Upsilon$}][l]{\phv{i}_\Gamma}=\,\mat{0}    \\
                \mathmakebox[\widthof{$\phv{v}_\Upsilon$}][l]{\phv{v}_\Upsilon}=\,\mat{0}  \\
                \mathmakebox[\widthof{$\phv{v}_\Upsilon$}][l]{\phv{b}_\up{F}}=\,\mat{0}    
            \end{subarray}
            }
        }
        \right.
        =
        \frac{1}{4}\phv{v}_\up{Tx}^\He
        \Re\{\mat{Z}_\up{Tx}\}^{-1}
        \phv{v}_\up{Tx},
    \end{align}
    where the vertical bar~$\rvert$ means ``evaluated at.'' 
    The right-hand side quantity in~\eqref{eq:defi_PA} is derived in~\fref{app:proof_power_metrics_PA}.
\end{defi}

The PAs' available power can now be used to define a new gain metric for REMSs acting as a transmitter.
\begin{defi}[REMS Gain]\label{defi:REMS_transmitting_gain}
    Given a REMS acting as a transmitter. 
    Let~$\phv{v}_\up{Tx}\in\mathbb{C}^{N_\up{Tx}}$ be the PA output voltage vector and~$(\theta,\varphi)\in\Omega$ specify a direction.
    We define the \emph{REMS gain} corresponding to~$\phv{v}_\up{Tx}$ and~$(\theta,\varphi)$ as 
    \begin{align}
        &G_\up{REMS}(\phv{v}_\up{Tx};\,\theta,\varphi)
        \triangleq
        \frac{1}{P_A}
        4\pi 
        I(\theta,\varphi)
        \left\rvert{
            {
            \scriptscriptstyle
            \begin{subarray}{l}
                \hspace{1mm}        \\
                \hspace{1mm}        \\
                \mathmakebox[\widthof{$\phv{v}_\Upsilon$}][l]{\phv{v}_\Gamma}=\,\mat{0}    \\
                \mathmakebox[\widthof{$\phv{v}_\Upsilon$}][l]{\phv{i}_\Gamma}=\,\mat{0}    \\
                \mathmakebox[\widthof{$\phv{v}_\Upsilon$}][l]{\phv{v}_\Upsilon}=\,\mat{0}  \\
                \mathmakebox[\widthof{$\phv{v}_\Upsilon$}][l]{\phv{b}_\up{F}}=\,\mat{0}    
            \end{subarray}
            }
        }
        \right.
        \\
        &\quad =
        16\pi
        \frac{
        \|
        \big(
            \mathbb{G}_{\phv{v}_\up{Tx}}^{\phv{a}_\up{F}}
            \phv{v}_\up{Tx}
        \big)
        (\theta,\varphi)
        \|_2^2}
        {\phv{v}_\up{Tx}^\He
        \Re\{\mat{Z}_\up{Tx}\}^{-1}
        \phv{v}_\up{Tx}}.
        \label{eq:calculate_GA}
    \end{align}
\end{defi}
The REMS gain~$G_\up{REMS}$ characterizes the combined effect of (i) the \emph{REMS matching efficiency} \mbox{$\eta_\up{matching}\triangleq P_\up{T}/P_\up{A}$}, (ii) the \emph{tuning network efficiency} \mbox{$\eta_\up{tuning}\triangleq P_\up{R}/P_\up{T}$}, (iii) the radiating structure's \emph{radiation efficiency} \mbox{$\eta_\up{radiating}\triangleq P_\up{F}/P_\up{R}$}, and (iv) the radiation structure's directivity \mbox{$D(\theta,\varphi)\triangleq (4\pi\,I(\theta,\varphi))/P_\up{F}$} as depicted in~\fref{fig:power_and_gain_terms}; where \mbox{$I(\theta,\varphi)=\|\phv{a}_\up{F}(\theta,\varphi)\|_2^2$} denotes the radiation intensity (cf.~\fref{rem:Intensity}).

\begin{rem}
    The REMS gain~$G_\up{REMS}$ can be interpreted as the radiation intensity gain of the REMS relative to a reference system containing a single lossless, isotropically radiating antenna, whose feed is perfectly matched to the reference system PA's output stage, conditioned on that~$P_\up{A}$ of the REMS is equal to~$P_\up{A}$ of the reference system. 
\end{rem}
\begin{rem}
    \new{
    The REMS gain~$G_\up{REMS}$ is, in some sense, defined with respect to~$P_\up{A}$.} 
    Analogous, one can also define gain metrics \new{with respect to}~$P_\up{T}$ and~$P_\up{R}$. 
    Moreover, for nonreciprocal REMS, it might be useful to distinguish between transmitting and receiving gain. 
\end{rem}

\def\wavesT at (#1,#2,#3,#4,#5){
    \draw [shift={(#1,#2)}, line width=1.5pt, color=#5, arrows = {-Stealth[inset=0, length=8pt, angle'=35]}] (.5,.3) -- (-.5,.3);

    \node[shift={(#1,#2)}, anchor=north] at (0.25,.2) {\small #3};

    \draw [shift={(#1,#2)}, line width=1.5pt, color=#5, arrows = {-Stealth[inset=0, length=8pt, angle'=35]}] (-.5,-.3) -- (.5,-.3);

    \node[shift={(#1,#2)}, anchor=north] at (0.25,-.4) {\small #4};
   
}

\def\wavesIII at (#1,#2,#3,#4,#5){
    \draw [shift={(#1,#2)}, line width=1.5pt, color=#5, arrows = {-Stealth[inset=0, length=8pt, angle'=35]}] (.5,.3) -- (-.5,.3);

    \node[shift={(#1,#2)}, anchor=north] at (0.25,.2) {\small #3};

    \draw [shift={(#1,#2)}, line width=1.5pt, color=#5, arrows = {-Stealth[inset=0, length=8pt, angle'=35]}] (-.5,-.4) -- (.5,-.4);

    \node[shift={(#1,#2)}, anchor=north] at (0.25,-.5) {\small #4};
   
}

\begin{figure}[tp]
    \centering
    {
    \small
    \begin{tikzpicture}
        \clip (0,1.3) rectangle (8,6.1);

        \draw[black_light, line width=1.5pt, dash pattern=on 12pt off 10pt, fill=grey_10] (2.5,0) circle (6);

        \begin{scope}[shift={(5.5,2.7)}]
            \draw[rounded corners=18, pattern color=black_60, grey_10, pattern=diagonal behind text] (-1,-.7) rectangle (1,.7);
            \node[anchor=center, color=black, fill=grey_10] at (0,0) {\color{black_60} \normalsize~$\,\,V_R\,\,$};
        \end{scope}

        \begin{scope}[shift={(1.5,3.2)}]
            \begin{scope}[shift={(30:.5)}]
                \draw[black_light, line width=1.5pt, dash pattern=on 6pt off 4pt, fill=white, rotate=30] (-.9,-.7) rectangle (.6,.7);
            \end{scope}
            \draw[white, fill=white] (0,0) circle (1);
            \draw[black_light, line width=1.5pt, dash pattern=on 6.08pt off 4pt, fill=white] (72:1) arc (72:348:1);
            \begin{scope}[shift={(30:.5)}]
                \draw[white, line width=0pt, fill=white, rotate=30] (-.872,-.672) rectangle (.572,.672);
            \end{scope}
            \begin{scope}[shift={(30:.5)}]
                \draw[black, line width=2pt, fill=white, rotate=30] (-.9,-.55) rectangle (.8,.55);
                \draw[black, line width=2pt, fill=white, rotate=30, dash pattern=on 6pt off 4pt] (0,-.55) -- (1.3,-.55);
                \draw[black, line width=2pt, fill=white, rotate=30, dash pattern=on 6pt off 4pt] (0,.55) -- (1.3,.55);
            \end{scope}
            \draw[black, line width=2pt, fill=white] (0,0) circle (.85);
            \begin{scope}[shift={(30:.5)}]
                \draw[grey_10, line width=2pt, fill=grey_10, rotate=30] (-.9,-.48) rectangle (1.1,.48);
                \draw[white, line width=2pt, fill=white, rotate=30] (-.9,-.48) rectangle (.56,.48);
            \end{scope}
            \begin{scope}[shift={(30:.5)}]
                \draw[black_light, line width=1.5pt, dash pattern=on 6pt off 4pt, fill=white, rotate=30] (.6,-.47) -- (.6,.47);
            \end{scope}

            \node [shift={(.5,-1.5)}] at (0,0) {\color{black_light}\normalsize~$\tilde{S}_{\up{port},m}$};
            \draw [color=black_light, shift={(.5,-1.5)}, line width=.8pt] plot[smooth, tension=1.4] coordinates {(0,0) (-.1,.25) (-.2,.4)};

            \node [shift={(.8,1.8)}] at (0,0) {\color{black_light}\normalsize~$S_{\up{port},m}$};
            \draw [color=black_light, shift={(.8,1.5)}, line width=.8pt] plot[smooth, tension=1.4] coordinates {(0.1,.05) (.2,.-.3) (.18,-.85)};

            \begin{scope}[shift={(1.52,-.63)}]
                \begin{scope}[rotate=30]
                    \wavesIII at (0,0,$\phs{b}_{\tilde{\up{R}},m}$,$\phs{a}_{\tilde{\up{R}},m}$,black)
                \end{scope}
            \end{scope}

            \begin{scope}[shift={(30:.5)}]
                \draw [shift={(0,0)}, rotate=30, line width=1pt, color=black_light, arrows = {-Stealth[inset=0, length=4pt, angle'=35]}] (.6,0) -- (-.1,0);

                \node[shift={(0,0)}] at (-.1,.25) {\small \color{black_light}~$\hat{\vect{n}}_m$};
            \end{scope}

        \end{scope}

        \begin{scope}[shift={(3.45,1.6)}]
            \node [shift={(.7,3.3)}] at (0,0) {\color{black_light}\normalsize~$S_R$};
            \draw [color=black_light, shift={(.7,3.3)}, line width=.8pt] plot[smooth, tension=1.4] coordinates {(.1,.1) (.3,.24) (.5,.55)};
        \end{scope}

        \begin{scope}[shift={(7,5)}]

                \draw [shift={(-.95,-.15)}, rotate=53, line width=1pt, color=black_light, arrows = {-Stealth[inset=0, length=4pt, angle'=35]}] (0,0) -- (.7,0);

                \node[shift={(-.95,-.15)}] at (75:0.4) {\small \color{black_light}~$\hat{\vect{r}}$};

                \begin{scope}[shift={(.2,0)}]
                    \begin{scope}[rotate=53]
                        \wavesT at (0,0,$\phv{b}_\up{F}$,$\phv{a}_\up{F}$,black)
                    \end{scope}
                \end{scope}
                
        \end{scope}

    \end{tikzpicture}}
\caption{
Qualitative sketch of the volume~$V_R$ and its surface~$\partial V_R$. 
$V_R$ is defined as the complement of a unit ball with radius~$R$, minus the volume associated with the~$M$ (circuit-theoretic) ports that connect the REMS radiating part to the rest of the REMS. 
$\partial V_R$ can be divided into $S_R$, $S_{\up{port},m}$, and~$\tilde{S}_{\up{port},m}$ for~$m\in[M]$. For each~$m$, the surface~$S_{\up{port},m}$ is the actual terminal surface of the~$m$th port, while $\tilde{S}_{\up{port},m}$ denotes a surface inside the port's perfectly conducting shielding, so that~$S_{\up{port},m}\cup\tilde{S}_{\up{port},m}$ forms a closed surface.
}
\label{fig:scetch_reciprocity}
\end{figure}
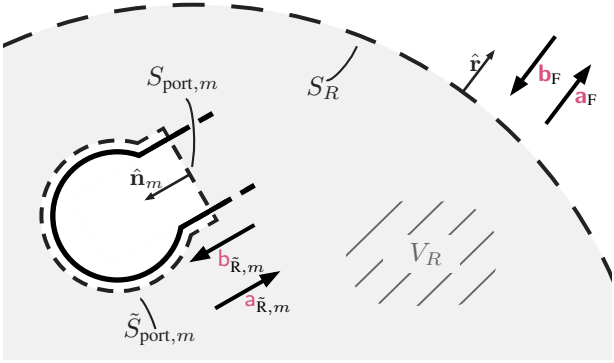

\subsection{Reciprocity Relations for the Radiating Structure}
It is often assumed that the wireless channel, including the radiating structure of all participating systems, behaves in reciprocal manner.
Whether a passive linear time-invariant wireless system behaves reciprocally or not is determined by the permittivity and permeability tensors describing the macroscopic material characteristics of the objects within the system.
A sufficient condition for reciprocity is that all relevant materials are isotropic, meaning their properties do not change when tested in different directions.
In this case, the permittivity and permeability tensors reduce to (position-dependent) scalar quantities.
As a consequence of this, Lorenz reciprocity (cf.~\cite[Eq.~3.61]{balanis_antenna_theory}) holds. 
I.e., in a source-free reciprocal system, it holds that 
\begin{align}
    \oiint_{\partial V}
    \hat{\vect{n}}^\T
    \big(
    \phv{E}_1 \times \phv{H}_2 
    -
    \phv{E}_2 \times \phv{H}_1 
    \big)
    \up{d} S
    &=\SI{0}{\watt}.
    \label{eq:lorenz_reciprocity_definition}
\end{align}
Here,~$V$ denotes an arbitrary volume within the source-free reciprocal system, 
$\up{d} S$ denotes the area of the differential element on the surface of the volume~$V$,
$\phv{E}_\ell$ and~$\phv{H}_\ell$ are the electric field and the magnetic field strength phasors, respectively, induced by one of two sources indexed by~\mbox{$\ell\in\{1,2\}$}, 
and~$\hat{\vect{n}}$ is the outward pointing unit vector perpendicular to the surface~$\partial V$. 

In the following proposition, the impact of Lorentz reciprocity on the parameters of the radiation structure is explained. 
A proof of~\fref{prop:reciprocity} can be found in~\fref{app:proof_reciprocity}.

\begin{prop}\label{prop:reciprocity}
Given a REMS for which~\fref{asm:basic_assumption_radiating_part} holds and whose~$M$ ports connecting the radiating part to the rest of the REMS support exactly one propagation mode\footnote{With propagation mode, we refer to the specific way in which signals travel through a port.} each.
For~$R>0$, we introduce the volume~$V_R$ as the complement of the unit ball with radius~$R$ (the REMS at its center) minus the volume occupied by the~$M$ ports, as sketched in \fref{fig:scetch_reciprocity}.
Assume Lorentz reciprocity holds for the volume~$V_R$ as~\mbox{$R\rightarrow\infty$}.
Then, the parameters describing the REMS radiating structure display symmetrical properties.
Specifically, it holds that 
\begin{align}
    \mat{S}_{\up{R}_{\up{R}\up{R}}}
    &=
    \mat{S}_{\up{R}_{\up{R}\up{R}}}^\T
    \label{eq:symmetry_TT}
    \\
    \vect{s}_{\up{R}_{\up{R}\up{F}}}
    &=
    \vect{s}_{\up{R}_{\up{F}\up{R}}},
    \label{eq:symmetry_TF}
\end{align}
and for all~$(\theta,\varphi),(\theta',\varphi')\in\Omega$, it holds that 
\begin{align}
    \mat{S}_{\up{R}_{\up{F}\up{F}}}(\hat{\vect{r}};\hat{\vect{r}}')
    &=
    \mat{S}_{\up{R}_{\up{F}\up{F}}}^\T(\hat{\vect{r}}';\hat{\vect{r}}),
    \label{eq:symmetry_FF}
\end{align}
where~$\hat{\vect{r}}$, and~$\hat{\vect{r}}'$ are the unit vectors corresponding to~$(\theta,\varphi)$ and~$(\theta',\varphi')$, respectively.
\end{prop}
\begin{rem}
It has been shown before that equations~\eqref{eq:symmetry_TT} and~\eqref{eq:symmetry_FF} are true for reciprocal systems. A proof of~\eqref{eq:symmetry_FF} can be found in~\cite[Eq.~7]{saxon_tensor_scattering_matrix_for_the_electromagnetic_field}, for example. 
Furthermore, results similar to equation~\eqref{eq:symmetry_TF} can be found in the literature (e.g., in~\cite[Ch.~II,~Sec.~1.5]{kerns_plane_wave_scattering_matrix_theory_of_antennas_and_antenna_antenna_interactions}).
However, to our knowledge, we are the first to prove the symmetry behavior of reciprocal systems in the form of~\eqref{eq:symmetry_TF}, i.e., using the mathematical formalism based on spherical waves as introduced in~Section~\ref{sec:radiating_structure_and_far_field}.
\end{rem}

\subsection{Obtaining the Model Parameters}\label{sec:obtaining_model_params}
In order to obtain the model parameters describing the RF frontend and the tuning network, one can, for example, use simulations, measurements, or datasheets of the respective components.
Which method is best suited depends on the application and is therefore not discussed further.
Instead, we now explain how to extract the model parameters characterizing the radiating structure operator~$\oper{S}_\up{R}$ from a \emph{single} full-wave EM simulation.
A full-wave EM simulation can, for example, be conducted with Ansys HFSS~\cite{ansys_hffs}.
We demonstrate how to obtain the model parameters for both reciprocal and non-reciprocal radiating structures.
As derived in~Section~\ref{sec:radiating_structure_and_far_field}, the operator~$\oper{S}_\up{R}$ from \fref{eq:matrix_SF} is fully characterized by the matrix~$\mat{S}_{\up{R}_\up{RR}}\in\mathbb{C}^{M\times M}$ and the three kernels~$\vect{s}_{\up{R}_\up{FR}}:[M]\times\Omega\rightarrow\mathbb{C}^2$,~$\vect{s}_{\up{R}_\up{RF}}:[M]\times\Omega\rightarrow\mathbb{C}^2$, and~$\mat{S}_{\up{R}_\up{FF}}:\Omega\times\Omega\rightarrow\mathbb{C}^{2\times 2}$. 
Unfortunately, the three kernels are, in general, infinite-dimensional and therefore cannot be represented by a computer. 
Consequently, we show how to obtain the matrix~$\mat{S}_{\up{R}_\up{RR}}$ and the corresponding \emph{sampled kernels}, which can be stored using a finite number of complex values. 

\begin{defi}[Sampled Kernels]\label{defi:sampled_kernels}
    For a finite subset~\mbox{$\tilde{\Omega}\subset \Omega$} we define the sampled kernels as the following functions:
    \begin{align}
        \vect{s}_{\up{R}_{\up{F}\up{R}}}^\up{sampled}&: 
        [M]\times\tilde{\Omega}\rightarrow\mathbb{C}^2
        ,\,
        (m;\theta,\varphi)\mapsto 
        \vect{s}_{\up{R}_{\up{F}\up{R}}}(m;\theta,\varphi)
        \\
        \vect{s}_{\up{R}_{\up{R}\up{F}}}^\up{sampled}&: 
        [M]\times\tilde{\Omega}\rightarrow\mathbb{C}^2
        ,\,
        (m;\theta,\varphi)\mapsto 
        \vect{s}_{\up{R}_{\up{R}\up{F}}}(m;\theta,\varphi)
        \\
        \mat{S}_{\up{R}_{\up{F}\up{F}}}^\up{sampled}&: 
        \tilde{\Omega}\times\tilde{\Omega}\rightarrow\mathbb{C}^{2\times 2}
        ,\,
        (\theta,\varphi;\theta',\varphi')\mapsto 
        \mat{S}_{\up{R}_{\up{F}\up{F}}}(\theta,\varphi;\theta',\varphi').
    \end{align}
\end{defi}
\begin{rem}\label{rem:sampling}
    Bucci and Franceschetti demonstrated in~\cite[Sec.~2-4]{bucci_franceschetti_on_the_spatial_bandwidth_of_scattered_fields} that the electric fields in the far-field region scattered by a spatially bounded object can be described by almost space-bandlimited functions.
    \new{
    Motivated by this result, we assume that for each REMS, the kernels~$\vect{s}_{\up{R}_\up{FR}}$,~$\vect{s}_{\up{R}_\up{RF}}$, and~$\mat{S}_{\up{R}_\up{FF}}$ can be efficiently represented to an arbitrary degree of accuracy using the sampled kernels introduced in~\fref{defi:sampled_kernels}.}
\end{rem}
The following paragraphs are organized such that each discusses how to obtain the model parameters associated with one of the four operators that constitute~$\oper{S}_\up{R}$ in~\fref{eq:matrix_SF}. 
\subsubsection{Inter-Element Coupling Operator~$\oper{S}_{\up{R}_\up{RR}}$}
Extracting~$\mat{S}_{\up{R}_{\up{R}\up{R}}}$ through simulation software is straightforward and is therefore not discussed in detail.
In general,~$M^2$ complex values must be stored to fully characterize this operator.
If the radiating structure behaves reciprocally, it follows from~\fref{prop:reciprocity} that~$\mat{S}_{\up{R}_{\up{R}\up{R}}}=\mat{S}_{\up{R}_{\up{R}\up{R}}}^\T$. 
Consequently, in this case, roughly half of the complex values are sufficient to describe this operator. 
\subsubsection{Transmitting Operator~$\oper{S}_{\up{R}_\up{FR}}$}
The sampled kernel~$\vect{s}_{\up{R}_{\up{F}\up{R}}}^\up{sampled}$ can be obtained by sequentially exciting each of the REMS's~$M$ ports, while terminating all other ports with the reference impedance~$R_0$.
Consequently, when the~$m$th port is excited with the circuit-theoretic power wave~$1\sqrt{\si{\watt}}$, it follows \mbox{that~$\phv{a}_{\tilde{\up{R}},m}=1\sqrt{\si{\watt}}$,} and for~$n \neq m$ it holds that~$\phv{a}_{\tilde{\up{R}},n}=0\sqrt{\si{\watt}}$; where~$\sqrt{\si{\watt}}$ is the respective physical unit. 
%
By substituting~$\phv{a}_{\tilde{\up{R}}}$ into~\eqref{eq:basic_form_SFFT}, it can be verified that for~$(\theta,\varphi)\in\tilde{\Omega}$,~$\vect{s}_{\up{R}_{\up{F}\up{R}}}^\up{sampled}(m;\theta,\varphi)$ can be extracted from the outgoing far-field power wave pattern in the direction~$(\theta,\varphi)$.
Storing the sampled kernel~$\vect{s}_{\up{R}_{\up{F}\up{R}}}^\up{sampled}$ necessitates storing~$2 M |\tilde{\Omega}|$ complex values.

\subsubsection{Receiving Operator~$\oper{S}_{\up{R}_\up{RF}}$}\label{sec:obtain_s_R_RF}
For reciprocal systems, it holds that~$\vect{s}_{\up{R}_{\up{R}\up{F}}}=\vect{s}_{\up{R}_{\up{F}\up{R}}}$ (see~\fref{prop:reciprocity}).
Therefore, it is sufficient to extract only~$\vect{s}_{\up{R}_{\up{F}\up{R}}}^\up{sampled}$. 
For non-reciprocal systems,~$\vect{s}_{\up{R}_{\up{R}\up{F}}}^\up{sampled}$ must be extracted separately.
Extracting the sampled kernel~$\vect{s}_{\up{R}_{\up{R}\up{F}}}^\up{sampled}$ is a less straightforward task than obtaining~$\mat{S}_{\up{R}_{\up{R}\up{R}}}$ and~$\vect{s}_{\up{R}_{\up{F}\up{R}}}^\up{sampled}$.
The kernel~$\vect{s}_{\up{R}_{\up{R}\up{F}}}^\up{sampled}$ characterizes the REMS's response to incoming converging spherical waves, which currently are not typically simulated directly by simulation software.
Instead, simulation software, such as Ansys HFSS~\cite{ansys_hffs}, supports the simulation of scattering of an incoming plane wave with an electric field of the form
\begin{align}
    \phv{E}_\up{p} (r\hat{\vect{r}}) 
    &=
    \phv{q} e^{j k r \hat{\vect{r}}^\T\hat{\vect{n}}_\up{p}}.
    \label{eq:plane_wave}
\end{align}
Here, the unit vector~$\hat{\vect{n}}_\up{p}$ specifies the direction from which the wave is incoming and~$\phv{q}\in\mathbb{C}^3$ specifies the amplitude and polarization of the plane wave. 
Fortunately, according to Saxon~\cite[Eq.~16]{saxon_tensor_scattering_matrix_for_the_electromagnetic_field} and Nieto-Vesperinas~\cite[Eq.~5.45]{nieto_vesperinas_scattering_and_diffraction_in_physical_optics}, in the far field of the scattering object, such a plane wave can be decomposed into an incoming and an outgoing spherical wave. 
Specifically, it holds that 
\begin{align}
    &\lim_{r\rightarrow\infty}
    \phv{E}_\up{p} (r\hat{\vect{r}}) 
    \nonumber
    \\
    &\quad =
    \frac{2\pi}{j k}\phv{q}
    \delta(\hat{\vect{r}}-\hat{\vect{n}}_\up{p})
    \frac{e^{+jkr}}{r}
    -
    \frac{2\pi}{j k}\phv{q}
    \delta(\hat{\vect{r}}+\hat{\vect{n}}_\up{p})
    \frac{e^{-jkr}}{r}.
\end{align}
From this decomposition and~\fref{defi:far_field_power_wave_pattern} it follows that by generating a plane wave incoming from the direction indicated by the unit vector~$\hat{\vect{n}}_\up{p}$, with RMS magnitude~\mbox{$\|\phv{E}_\up{p}\|_2=\|\phv{q}\|_2$} and polarization specified by~$\phv{q}$, one can determine the REMS's response to the incoming spherical wave characterized \\by~\mbox{$\phv{b}_\up{F}=\frac{2\pi}{jk\sqrt{Z_0}}\phv{q}\delta(\hat{\vect{r}}-\hat{\vect{n}}_\up{p})$.}
Consequently, it follows from~\eqref{eq:basic_form_SFTF} that for each~$m\in[M]$,~$(\theta,\varphi)\in\tilde{\Omega}$, the sampled kernel can be calculated as follows:
\begin{align}
    \vect{s}_{\up{R}_{\up{R}\up{F}}}^\up{sampled}(m;\theta,\varphi)
    &=
    \frac{jk\sqrt{Z_0}}{2\pi}
    \begin{bmatrix}
        \phv{b}_{\tilde{\up{R}},m}
        \big|_{\phv{q}=\hat{\vect{\theta}};\;\hat{\vect{n}}_\up{p}=\hat{\vect{r}}}
        \\
        \phv{b}_{\tilde{\up{R}},m}
        \big|_{\phv{q}=\hat{\vect{\varphi}};\;\hat{\vect{n}}_\up{p}=\hat{\vect{r}}}
    \end{bmatrix}\!.
\end{align}
Here,~$\phv{b}_{\tilde{\up{R}},m}\big|_{\phv{q}=\cdot;\;\hat{\vect{n}}_\up{p}=\hat{\vect{r}}}$ is the circuit-theoretic power wave that is induced in port~$m$ by the plane wave incoming from~$(\theta,\varphi)$, with an (RMS) magnitude~\mbox{$\|\phv{E}_\up{p}\|_2=\SI[per-mode = symbol]{1}{\volt\per\meter}$} and the corresponding polarization.
Storing the sampled kernel~$\vect{s}_{\up{R}_{\up{R}\up{F}}}^\up{sampled}$ requires~$2 M |\tilde{\Omega}|$ complex values.
Unfortunately, the sampled kernel cannot be directly inserted into~\eqref{eq:basic_form_SFTF}. 
For this reason, we propose the following approximation formula, which utilizes the sampled kernel on~$\vect{s}_{\up{R}_{\up{R}\up{F}}}^\up{sampled}$ and can be implemented in a practical system:
\begin{align}
    &\!\!\!
    \big[\oper{S}_{\up{R}_{\up{R}\up{F}}}\phv{b}_\up{F}\big]_m
    =
    \oiint\displaylimits_{\Omega} 
    \big\langle\phv{b}_\up{F}(\theta,\varphi),\overline{\vect{s}}_{\up{R}_{\up{R}\up{F}}}(m;\theta,\varphi)\big\rangle_{\mathbb{C}^2}
    \sin(\theta)\,\up{d}(\theta,\varphi)
    \\
    &\approx
    \sum\displaylimits_{(\theta,\varphi)\in\tilde{\Omega}}
    \big\langle\phv{b}_\up{F}(\theta,\varphi),\overline{\vect{s}}_{\up{R}_{\up{R}\up{F}}}^\up{sampled}(m;\theta,\varphi)\big\rangle_{\mathbb{C}^2}
    \,
    A_{\tilde{\Omega}}(\theta,\varphi),
    \label{eq:SFTF_approximation}
\end{align}
where~$A_{\tilde{\Omega}}$ is a positive area weight function\footnote{Such a choice of~$A_{\tilde{\Omega}}>0$ can, for example, be constructed using a Voronoi diagram on the unit sphere.} that only depends on the choice of~$\tilde{\Omega}$.
Note that the accuracy of the approximation in~\eqref{eq:SFTF_approximation} depends on the kernel~$\vect{s}_{\up{R}_\up{RF}}$ and the number of samples~$|\tilde{\Omega}|$; see also the discussion in~\fref{sec:limitations} on the limitations of our approach.

\input{tikzfigs/fig-sketch_experiments}
\begin{figure}[t]
    \centering
    {
    \small
    \begin{tikzpicture}
        \begin{axis}[%
            width=6cm,
            height=5cm,
            scale only axis,
            xmin=0,
            xmax=90,
            xtick={0,15,30,45,60,75,90},
            xlabel={$\alpha$ (\si{\deg})},
            ymin=-6e-4,
            ymax=6e-4,
            ytick={-8e-4, -6e-4, -4e-4, -2e-4, 0e-4, 2e-4, 4e-4, 6e-4, 8e-4},
            scaled ticks=false,
            ylabel ={transmission coefficient~$S_{{2_1}{1_1}}$ ($1$)},
            ylabel near ticks,
            xmajorgrids,
            ymajorgrids,
            axis background/.style={fill=white},
            every outer y axis line/.append style={black,very thick},
            legend style={at={(0.97,0.03)},anchor=south east,legend cell align=left, align=right, draw=black,thick, fill opacity=0.9}
            ]
    
            \addplot [color=blue_120, line width=1pt] table [col sep=comma] {tikzfigs/data/exp_1_S_our_real.dat};
            \addlegendentry{\footnotesize proposed (real part)}
            
             \addplot [only marks, color=blue_60, line width=1pt, mark=square*, mark size=1.5pt] table [col sep=comma] {tikzfigs/data/exp_1_S_ref_real.dat};
            \addlegendentry{\footnotesize conventional (real part)}
            
            \addplot [color=green_120, line width=1pt, dash pattern=on 3.5pt off 1pt] table [col sep=comma] {tikzfigs/data/exp_1_S_our_imag.dat};
            \addlegendentry{\footnotesize proposed (imag. part)}

            \addplot [only marks, color=green_60, line width=1pt, mark=square*, mark size=1.5pt] table [col sep=comma] {tikzfigs/data/exp_1_S_ref_imag.dat};
            \addlegendentry{\footnotesize conventional (imag. part)}
            
        \end{axis}
    \end{tikzpicture}
    }
    \caption{Results for \new{Validation~Scenario~1}. The transmission coefficients obtained by (i) the proposed method and (ii) the conventional method are plotted for various angles~$\alpha$. The results of the two methods agree well over the entire analyzed range of~$\alpha$.}
    \label{fig:plot_experiment_1}
    \vspace{.5cm}
    \centering
    {
    \small
    \begin{tikzpicture}
        \begin{axis}[%
            width=6cm,
            height=5cm,
            scale only axis,
            xmin=15,
            xmax=45,
            xtick={15,20,25,30,35,40,45},
            xlabel={$\alpha$ (\si{\deg})},
            ymin=-4e-5,
            ymax=2e-5,
            ytick={-4e-5, -3e-5, -2e-5, -1e-5, 0, 1e-5, 2e-5},
            scaled ticks=false,
            ylabel ={transmission coefficient~$S_{{3_1}{1_1}}$ ($1$)},
            ylabel near ticks,
            xmajorgrids,
            ymajorgrids,
            axis background/.style={fill=white},
            every outer y axis line/.append style={black,very thick},
            legend style={at={(0.97,0.97)},anchor=north east,legend cell align=left, align=left, draw=black,thick, fill opacity=0.9}
            ]
    
            \addplot [color=blue_120, line width=1pt] table [col sep=comma] {tikzfigs/data/exp_2_S_our_real.dat};
            \addlegendentry{\footnotesize proposed (real part)}
            
             \addplot [only marks, color=blue_60, line width=1pt, mark=square*, mark size=1.5pt] table [col sep=comma] {tikzfigs/data/exp_2_S_ref_real.dat};
            \addlegendentry{\footnotesize conventional (real part)}
            
            \addplot [color=green_120, line width=1pt, dash pattern=on 3.5pt off 1pt] table [col sep=comma] {tikzfigs/data/exp_2_S_our_imag.dat};
            \addlegendentry{\footnotesize proposed (imag. part)}

            \addplot [only marks, color=green_60, line width=1pt, mark=square*, mark size=1.5pt] table [col sep=comma] {tikzfigs/data/exp_2_S_ref_imag.dat};
            \addlegendentry{\footnotesize conventional (imag. part)}
            
        \end{axis}
    \end{tikzpicture}
    }
    \caption{Results for \new{Validation~Scenario~2}. The transmission coefficients obtained by (i) the proposed method and (ii) the conventional method are plotted for various angles~$\alpha$. The results of the two methods generally agree. However, the results from the conventional method  exhibit noise, which we attribute to numerical stability issues.}
    \label{fig:plot_experiment_2}
\end{figure}

\subsubsection{Scattering Operator~$\oper{S}_{\up{R}_\up{FF}}$}
Similar to the receiving operator~$\oper{S}_{\up{R}_\up{RF}}$, one cannot directly insert the sampled kernel~$\mat{S}_{\up{R}_{\up{F}\up{F}}}^\up{sampled}$ into~\eqref{eq:basic_form_SFFF}. 
Therefore, we propose the approximation 
\begin{align}
    &\!\!\!\big(\oper{S}_{\up{R}_{\up{F}\up{F}}}\phv{b}_\up{F}\big)(\theta,\varphi)
    =
    \oiint\displaylimits_{\Omega} 
    \mat{S}_{\up{R}_{\up{F}\up{F}}}(\theta,\varphi;\theta',\varphi')\phv{b}_\up{F}(\theta',\varphi')
    \,\up{d}\Omega
    \label{eq:SFFF_approx_intermediate}
    \\
    &\approx
    \hspace{-3mm}\sum\displaylimits_{(\theta',\varphi')\in\tilde{\Omega}}
    \mat{S}^\up{sampled}_{\up{R}_{\up{F}\up{F}}}(\theta,\varphi;\theta',\varphi')\phv{b}_\up{F}(\theta',\varphi')
    \,
    A_{\tilde{\Omega}}(\theta',\varphi').
    \label{eq:SFFF_approximation}
\end{align}
Similar as in~\eqref{eq:SFTF_approximation}, the accuracy of the approximation in~\eqref{eq:SFFF_approximation} depends on the kernel~$\mat{S}_{\up{R}_\up{FF}}$ and the number of samples~$|\tilde{\Omega}|$.
As with the receiving operator~$\oper{S}_{\up{R}_\up{RF}}$, when attempting to obtain~$\mat{S}_{\up{R}_{\up{F}\up{F}}}^\up{sampled}$ from simulation software, one encounters the problem that generating incoming converging spherical waves is typically not directly supported by full-wave EM simulation software.
By applying the same plane-wave approach proposed in~Section~\ref{sec:obtain_s_R_RF}, the total electric far field can be expressed as 
\begin{align}
    &\lim_{r\rightarrow\infty}
    \phv{E} (r\hat{\vect{r}}) 
    =
    \phv{q} e^{+j k r \hat{\vect{r}}^\T\hat{\vect{n}}_\up{p}}
    +
    \phv{s}_\phv{q}(\hat{\vect{r}},\hat{\vect{n}}_\up{p})\frac{e^{-jkr}}{r}
    \\
    &=
    \frac{2\pi}{j k}\phv{q}
    \delta(\hat{\vect{r}}\!-\!\hat{\vect{n}}_\up{p})
    \frac{e^{+jkr}}{r}
    \!+\!
    \Big(
    \phv{s}_\phv{q}(\hat{\vect{r}},\hat{\vect{n}}_\up{p})
    \!-\!
    \frac{2\pi}{j k}\phv{q}
    \delta(\hat{\vect{r}}\!+\!\hat{\vect{n}}_\up{p})\!
    \Big)
    \frac{e^{-jkr}}{r},
    \label{eq:plane_wave_scattering_as_spherical_waves}
\end{align}
where~$\phv{s}_\phv{q}(\hat{\vect{r}},\hat{\vect{n}}_\up{p})$ is the complex amplitude characterizing the scattered electric field traveling in the direction~$\hat{\vect{r}}$, when the REMS is excited by a plane wave incoming from~$\hat{\vect{n}}_\up{p}$ with polarization~$\phv{q}$ and an (RMS) electric field magnitude of \mbox{$\|\phv{E}\|_2=\SI[per-mode = symbol]{1}{\volt\per\meter}$}.
Continuing from~\eqref{eq:plane_wave_scattering_as_spherical_waves}, it was deduced by Saxon how the kernel~$\mat{S}_{\up{R}_{\up{F}\up{F}}}$ can be constructed explicitly. 
We adapt Saxon's construction formula~\cite[Eq.~21]{saxon_tensor_scattering_matrix_for_the_electromagnetic_field} to our framework to obtain 
{
    \renewcommand\arraystretch{1.2}
    \begin{align}
        &\!\!\!\mat{S}_{\up{R}_{\up{F}\up{F}}}^\up{sampled}(\theta,\varphi;\theta',\varphi')
        =\!
        \begin{bmatrix}
            -1 & 0 \\ \hphantom{-}0 & 1
        \end{bmatrix}
        \frac{1}{A_{\tilde{\Omega}}(-\theta',-\varphi')}
        \delta_{\theta,-\theta'}
        \delta_{\varphi,-\varphi'}
        \nonumber
        \\
        &\;\;\;+
        \underbrace{
        \frac{j k}{2\pi}
        \begin{bmatrix}
            \big[\phv{s}_{\hat{\vect{\theta}}'}(\theta,\varphi;\theta',\varphi')\big]_{\hat{\vect{\theta}}}
            &
            \big[\phv{s}_{\hat{\vect{\varphi}}'}(\theta,\varphi;\theta',\varphi')\big]_{\hat{\vect{\theta}}}
            \\
            \big[\phv{s}_{\hat{\vect{\theta}}'}(\theta,\varphi;\theta',\varphi')\big]_{\hat{\vect{\varphi}}}
            &
            \big[\phv{s}_{\hat{\vect{\varphi}}'}(\theta,\varphi;\theta',\varphi')\big]_{\hat{\vect{\varphi}}}
        \end{bmatrix}
        }_{\triangleq\, \tilde{\mat{S}}_{\up{R}_{\up{F}\up{F}}}^\up{sampled}}
        ,
        \label{eq:construction_formula_of_SFFF}
    \end{align}
where the appearance of the Pauli matrix is a consequence of the spherical coordinate system we use in~\fref{defi:far_field_power_wave_pattern}; 
and where we introduced the \emph{reduced sampled kernel of the scattering operator}~$\tilde{\mat{S}}_{\up{R}_{\up{F}\up{F}}}^\up{sampled}$.
}
Inserting~\eqref{eq:construction_formula_of_SFFF} into~\eqref{eq:SFFF_approximation} leads to 
\begin{align}
    &\!\!\!\big(\oper{S}_{\up{R}_{\up{F}\up{F}}}\phv{b}_\up{F}\big)(\theta,\varphi)
    \approx
    \begin{bmatrix}
        -1 & 0 \\ \hphantom{-}0 & 1
    \end{bmatrix}
    \phv{b}_\up{F}(\pi-\theta,\pi+\varphi)
    \label{eq:SFFF_approximation_alternative_form}
    \\
    &\;\;\;
    +\hspace{-3mm}\sum\displaylimits_{(\theta',\varphi')\in\tilde{\Omega}}
    \tilde{\mat{S}}^\up{sampled}_{\up{F}_{\up{F}\up{F}}}(\theta,\varphi;\theta',\varphi')\phv{b}_\up{F}(\theta',\varphi')
    \,
    A_{\tilde{\Omega}}(\theta',\varphi')
    \nonumber.
\end{align}
It is evident that, in a practical implementation, it is sufficient to store the \emph{reduced} sampled kernel of the scattering operator~$\tilde{\mat{S}}_{\up{R}_{\up{F}\up{F}}}^\up{sampled}$.
Storing~$\tilde{\mat{S}}_{\up{R}_{\up{F}\up{F}}}^\up{sampled}$ instead of~$\mat{S}_{\up{R}_{\up{F}\up{F}}}^\up{sampled}$ offers numerical advantages, as the first term in~\eqref{eq:construction_formula_of_SFFF} includes~$\left( A_{\tilde{\Omega}}(\cdot,\cdot) \right)^{-1}$, whose magnitude tends to grow linearly with the number of samples~$|\tilde{\Omega}|$.
%


\section{Experimental Validation}\label{sec:experimental_validation}

We now use numerical experiments to validate the proposed REMS-far-field-interaction formalism from~Section~\ref{sec:radiating_structure_and_far_field} and the proposed method for obtaining the radiating structure parameters from~Section~\ref{sec:obtaining_model_params}.
In particular, we compare predictions made using our REMS far-field interaction formalism with Ansys HFSS simulation results in two scenarios. 
\subsection{\new{Validation Scenario~1}}
In the first scenario, there are two REMSs placed in empty space, as illustrated in~\fref{fig:skech_experiment_1}. 
The REMS types involved are shown in~\fref{fig:experiment_type_A} and~\fref{fig:experiment_type_B}: a patch antenna (type~A) and a half-wave dipole antenna (type~B), each with one antenna port ($M=1$).
The wireless channel between the two REMSs can be manipulated by varying the rotation angle~$\alpha$ of REMS~1.
We analyze the system at a frequency of $\SI{5.4}{\giga\hertz}$ and compare the following two methods.
For the \emph{proposed} method, we first use the approach described in~Section~\ref{sec:obtaining_model_params} to obtain the parameters characterizing the radiating structures of both REMS types, relying on a single full-wave EM simulation for each REMS type.
Second, for each rotation angle~$\alpha$, we utilize~\fref{prop:calculate_channel} to calculate the transmission coefficient (cf.~\fref{defi:transmission_coefficient}) from REMS~1 to REMS~2 ($S_{{2_1}{1_1}}$).
\new{
For the \emph{conventional} method, we use Ansys HFSS to simulate the complete scenario for each rotation angle~$\alpha$, including both REMSs, and then directly export~$S_{{2_1}{1_1}}$ from the simulation software.
}
\fref{fig:plot_experiment_1} shows the results of the proposed and the conventional method.
The two methods agree well over the analyzed range of~$\alpha$. 
We attribute the small deviations between the two methods to (i) the finite separation distance between the two REMSs in this scenario, whereas in~\fref{prop:calculate_channel} we assume that the separation distance approaches infinity, and (ii) numerical inaccuracies in the \emph{conventional} method.
We refer to~Section~\ref{sec:experiment_discussion_comuputional} for a discussion on the numerical stability and required computational resources of the proposed and conventional methods.
The good agreement between the proposed and conventional methods validates (i) the proposed REMS-far-field-interaction formalism and (ii) the proposed method for obtaining the model parameters of the REMS radiating structure.
Specifically, the results of \new{Validation~Scenario~1} validate our \emph{modeling} of the transmitting operator~$\oper{S}_{\up{R}_{\up{F}\up{R}}}$ in~\eqref{eq:basic_form_SFFT}, and the receiving operator~$\oper{S}_{\up{R}_{\up{R}\up{F}}}$ in~\eqref{eq:basic_form_SFTF} as well as the \emph{extraction} of the respective kernels~$\vect{s}_{\up{R}_{\up{F}\up{R}}}$ and~$\vect{s}_{\up{R}_{\up{R}\up{F}}}$.
Because we presume that the influence of the scattering operator~$\oper{S}_{\up{R}_{\up{F}\up{F}}}$ on the transmission coefficient is negligibly small in this scenario, additional experiments, which are shown next, are required to validate the respective modeling approach in~\eqref{eq:basic_form_SFFF} and the method for obtaining the kernel~$\mat{S}_{\up{R}_{\up{F}\up{F}}}$ in~\eqref{eq:construction_formula_of_SFFF}.

\subsection{\new{Validation Scenario~2}}
In this scenario, there are three REMSs placed in empty space, as illustrated in~\fref{fig:skech_experiment_2}. 
We use the same REMS types as in \new{Validation~Scenario~1} (see~\fref{fig:experiment_type_A} and~\fref{fig:experiment_type_B}).
The wireless channel between the three REMSs can be manipulated by varying the rotation angle~$\alpha$ of REMS~2.
The absorber\footnote{For numerical stability reasons, we did not include an absorber in the Ansys HFSS simulation of the conventional method. Instead, we conducted an additional simulation without REMS~2 to isolate the contribution of the direct path between REMS~1 and REMS~3 to the transmission coefficient~$S_{{3_1}{1_1}}$. We then subtracted this contribution from the main simulation result.} between REMS~1 and REMS~3 ensures that the scattering operator~$\oper{S}^{(\up{A})}_{\up{R}_{\up{F}\up{F}}}$ of REMS~2 significantly impacts the result.
Again, we analyze the system \new{at~$\SI{5.4}{\giga\hertz}$}.
We mainly follow the same methodology used in \new{Validation~Scenario~1} to determine the transmission coefficient~$S_{{3_1}{1_1}}$ (cf.~\fref{defi:transmission_coefficient}) from REMS~1 to REMS~3 using both the \emph{proposed} and the \emph{conventional} methods.
In contrast to \new{Validation~Scenario~1}, for the proposed method, we now use the approximation
\begin{align}
    S_{{3_1}{1_1}}
    \approx&\,
    \Big(
        \vect{s}_{\up{R}_{\up{R}\up{F}}}^{(\up{B})}
        (1;90^\circ,0^\circ)
    \Big)^\T
    \mat{C}_{23}
    \label{eq:approximation_for_experiment_2}
    \\
    \nonumber
    &\cdot\mat{S}_{\up{R}_{\up{F}\up{F}}}^{(\up{A})}(90^\circ,90^\circ\!-\alpha;90^\circ,\!-\alpha)
    \mat{C}_{12}
    \vect{s}_{\up{R}_{\up{F}\up{R}}}^{(\up{A})}
    (1;90^\circ,0^\circ)
\end{align} 
\new{instead of~\fref{eq:transmission_coefficient_construction_formula} from~\fref{prop:calculate_channel}} to calculate the transmission coefficient\footnote{Eq.~\eqref{eq:approximation_for_experiment_2} is derived by adapting the proof of~\fref{app:proof:proposition_channel} to \new{Validation~Scenario 2} and applying the unilateral approximation from~\cite[Sec.~II-F]{ivrlac_nossek_toward_a_circuit_theory_of_communication}.}, where the matrices~$\mat{C}_{12}$ and~$\mat{C}_{23}$ are defined as follows:
\begin{align}
        \mat{C}_{12}
        &\triangleq
        \frac{2\pi}{j k}
        \frac{e^{-j k\cdot\SI{1.5}{\meter}}}{\SI{1.5}{\meter}}
        \begin{bmatrix}
            1 & 0 \\
            0 &  -1
        \end{bmatrix}
    \\
        \mat{C}_{23}
        &\triangleq
        \frac{2\pi}{j k}
        \frac{e^{-j k\cdot\SI{3}{\meter}}}{\SI{3}{\meter}}
        \,\,\,\,
        \begin{bmatrix}
            1 & 0 \\
            0 &  -1
        \end{bmatrix}. 
\end{align}
\fref{fig:plot_experiment_2} shows the results of the proposed and conventional method.
The two methods generally agree over the analyzed range of~$\alpha$, but the relative deviations between them are more pronounced than in \new{Validation~Scenario~1}.
We attribute these deviations primarily to errors resulting from numerical inaccuracies of the \emph{conventional} method (see~Section~\ref{sec:experiment_discussion_comuputional}).
Together with the results from \new{Validation~Scenario~1}, the agreement between the methods shown in~\fref{fig:plot_experiment_2} validates the REMS scattering operator approach in~\eqref{eq:basic_form_SFFF} and the method for obtaining the kernel~$\mat{S}_{\up{R}_{\up{F}\up{F}}}$.
Thus, we have validated that our proposed modeling approach for the transmitting operators~$\mathbb{S}_{\up{R}_\up{FR}}$, the receiving operators~$\mathbb{S}_{\up{R}_\up{RF}}$, and the scattering operators~$\mathbb{S}_{\up{R}_\up{FF}}$ is correct.
Furthermore, we have validated that the proposed method for obtaining the respective kernels for these three operators is also correct.
%

\subsection{Numerical Stability and Computational Resources}\label{sec:experiment_discussion_comuputional}
The transmission coefficients obtained using the conventional method exhibit noise, as can be seen particularly in~\fref{fig:plot_experiment_2}. 
We observed that the variance of this noise-like behavior reduced when investing more computational resources, i.e., additional CPU time and memory.
As a comparison, to generate~\fref{fig:plot_experiment_1} and~\fref{fig:plot_experiment_2}, for the conventional method, we invested~$120$~CPU-hours to obtain~$30$ transmission coefficients, while our proposed method required only about~$0.5$~CPU-hours to generate~$132$ transmission coefficients.\footnote{We did not optimize the simulation settings for either method to achieve best performance. Therefore, the CPU hours provided here only serve as a rough comparison of the computational effort.}
We have the following explanations for these~observations.

For the conventional method, we simulate the complete scenario at once. 
In doing so, the simulation software must describe both the EM fields in the transmitting REMS and the much weaker EM fields around the receiving REMS.
We presume that the need to handle both strong and comparably very weak signals makes the conventional method susceptible to numerical issues.
Consequently, extensive computational resources are required for the conventional method to accurately predict the transmission coefficients.
In contrast, our proposed method, where the two REMSs types are simulated separately, is significantly more numerically stable.
Moreover, the conventional method requires a completely new simulation for every physical change in the system (e.g., for \new{each} different angle~$\alpha$).
In contrast, our proposed method requires only one Ansys HFSS simulation for each REMS type to obtain the radiating structure~parameters (cf.~Section~\ref{sec:obtaining_model_params}).
As a consequence, the proposed method can predict transmission coefficients orders of magnitude faster than the conventional method, while producing more accurate results.
Because of this behavior, in both scenarios, we generated results for many more angles~$\alpha$ using the proposed method than with the conventional method. 
Consequently, in~\fref{fig:plot_experiment_1} and~\fref{fig:plot_experiment_2}, we illustrate the results of the proposed method with a continuous curve and those of the conventional method with discrete samples.


\section{Case Study: Physically Accurate Beam- and Null-Forming with an RRA}\label{sec:case_study}
\new{To demonstrate the efficiency of our REMS modeling method in the context of beamforming design, we now present a case study focusing on the RRA depicted in~\fref{fig:case_study_RRA}.\footnote{An Ansys HFSS project file containing the modeled RRA can be downloaded from \url{https://github.com/IIP-Group/RRA-case-study}.}}$^{,}$\footnote{\new{Our RRA is a basic example of a typical RRA system. More advanced designs for the overall structure and the patch antennas exist; see, e.g.,~\cite{yang_fan_chen_a_novel_compact_electromagnetic_bandgap_structure_and_its_applications_for_microwave_circuits}.}}
This RRA is a hybrid analog-digital beamforming system, and we analyze its behavior when it acts as a transmitter.
The goal of the RRA is to steer its transmit beams \new{toward}~$U\leq N$ \emph{primary users} (e.g., communication partners) while minimizing the power directed \new{toward}~$I\in\mathbb{Z}_{\geq 0}$ \emph{secondary users} (e.g., other receiving devices that are not part of the communication system).
Our RRA's \new{basic structure is} illustrated in~\fref{fig:basic_structure_RRA}, with~$N=2$ actively fed patch antennas and a large number~($R=100$) of passive and configurable reflective elements arranged as a~$10\times10$ rectangular array. 
For~$r\in [R]$, the~$r$th passive reflective element features a tunable impedance~$Z_{\up{R},r}$.

\subsection{Proposed Algorithm}\label{sec:algorithm}
\new{To enable simultaneous beam- and null-forming, we propose an efficient and physics-aware algorithm that jointly optimizes the configuration of the RRA's reconfigurable elements and the digital precoding matrix. }
We represent the values of the~$R$ impedances in the tuning network with the tuple~\mbox{$\mat{Z}_\up{R}\triangleq(Z_{\up{R},1},\ldots,Z_{\up{R},R})$}.
Furthermore, we denote the finite set of potential impedance values that can be chosen for each reconfigurable impedance \mbox{as~$\mathcal{Z}_\up{R}\subset \big\{c\in\mathbb{C}\big|\Re\{c\}\geq 0\big\}$}\footnote{\new{The impedance values of the reconfigurable elements are restricted to have a non-negative real part as the tuning network is assumed to be passive.}}; i.e., it holds that~$\mat{Z}_\up{R}\in\mathcal{Z}_\up{R}^R$.
For each primary user indexed by~\mbox{$u\in[U]$}, let~\mbox{$s_u\in\mathbb{C}$} be the symbol to be transmitted to primary user~$u$, and define the symbol vector \mbox{as~$\vect{s}\triangleq [s_1,\ldots,s_U]^\T$.}
Let~$\mat{T}\in\mathbb{C}^{N\times U}$ denote the precoding matrix, such \\that~\mbox{$\phv{v}_\up{Tx}\triangleq \mat{T}\vect{s}$} is the precoded PA output voltage vector. 
To steer the RRA's beams \new{toward} the primary users and minimize the energy \new{toward} the secondary users, the algorithm \emph{jointly optimizes} the analog reconfigurable impe-\\dances~$\mat{Z}_\up{R}$ (equal to analog beam- and null-forming) and the precoding matrix~$\mat{T}$ (equal to digital beam- and null-forming).
We assume that the algorithm is provided with the directions of all primary users~$\left\{(\theta_{\up{U},1}, \varphi_{\up{U},1}),\ldots,(\theta_{\up{U},U}, \varphi_{\up{U},U})\right\}$ and all secondary users~$\left\{(\theta_{\up{I},1}, \varphi_{\up{I},1}),\ldots,(\theta_{\up{I},I}, \varphi_{\up{I},I})\right\}$.

\subsubsection{Underlying Optimization Problem}
For~$u\in[U]$, let~$\vect{e}_u$ be the~$u$th column of the identity matrix~$\mat{I}_U$.
We now introduce the following quasi-power\footnote{We use ``quasi-power'' because one can informally interpret the \mbox{in~\eqref{eq:defi_P_signal}--\eqref{eq:defi_P_3_3_party}} defined terms as power metrics, even though they do not represent a physical power.} terms:
\begin{align}
    \label{eq:defi_P_signal}
    \mathmakebox[\widthof{$\tilde{P}_\up{second.}
    (\mat{Z}_\up{R},\mat{T})$}][l]{
        \tilde{P}_\up{signal}
        (\mat{Z}_\up{R},\mat{T})
    }
    &\triangleq
    \mathmakebox[\widthof{$\displaystyle\sum_{u\in[U],\iota\in[I]}$}][c]{
    \displaystyle\min_{u\in[U]} }
    G_\up{REMS}
    \left(
    \mat{T}\vect{e}_u
    ;
    \theta_{\up{U},u}
    ,
    \varphi_{\up{U},u}
    \right)
    \\
    \mathmakebox[\widthof{$\tilde{P}_\up{second.}
    (\mat{Z}_\up{R},\mat{T})$}][l]{
        \tilde{P}_\up{interf.}
        (\mat{Z}_\up{R},\mat{T})
    }
    \label{eq:defi_P_inter}
    &\triangleq
    \mathmakebox[\widthof{$\displaystyle\sum_{u\in[U],\iota\in[I]}$}][c]{
    \displaystyle\max_{\substack{u,u'\in[U]: \\ u'\neq u}} }
    G_\up{REMS}
    \left(
    \mat{T}\vect{e}_{u'}
    ;
    \theta_{\up{U},u}
    ,
    \varphi_{\up{U},u}
    \right)
    \\
    \label{eq:defi_P_3_3_party}
    \mathmakebox[\widthof{$\tilde{P}_\up{second.}
    (\mat{Z}_\up{R},\mat{T})$}][l]{
        \tilde{P}_\up{second.}
        (\mat{Z}_\up{R},\mat{T})
    }
    &\triangleq
    \mathmakebox[\widthof{$\displaystyle\sum_{u\in[U],\iota\in[I]}$}][c]{
    \displaystyle\max_{u\in[U],\,\iota\in[I]} }
    G_\up{REMS}
    \left(
    \mat{T}\vect{e}_{u}
    ;
    \theta_{\up{I},\iota}
    ,
    \varphi_{\up{I},\iota}
    \right)\!.
\end{align}
Here,~$G_\up{REMS}$ is the REMS gain (see~\fref{defi:REMS_transmitting_gain}) of the RRA when the impedance values~$\mat{Z}_\up{R}$ in the argument of the quasi-power terms are used for the reconfigurable impedances\footnote{\new{The effect of the argument~$\mat{Z}_\up{R}$ onto the REMS gain is not shown explicitly to simplify notation.}}.
The optimization problem underlying our algorithm can now be stated as follows: 
\begin{align}
\label{eq:optimization_problem}
    \maximize_{\substack{\mat{Z}_\up{R}\in\mathcal{Z}_\up{R}^R\\\,\,\,\,\,\,\,\,\,\,\mat{T}\in\mathbb{C}^{N\times U}}}
    \frac{
        \tilde{P}_\up{signal}(\mat{Z}_\up{R},\mat{T})
    }{
        \tilde{P}_\up{interf.}(\mat{Z}_\up{R},\mat{T})
        +
        \tilde{P}_\up{second.}(\mat{Z}_\up{R},\mat{T})
    }.
\end{align}
Informally,~\eqref{eq:defi_P_signal}--\eqref{eq:optimization_problem} can be interpreted as the algorithm jointly optimizing the impedances~$\mat{Z}_\up{R}$ and the precoding matrix~$\mat{T}$ to minimize the ratio of the minimum received signal power at the~$U$ primary users to the sum of the maximal interference power at the~$U$ primary users and the maximal power radiated in the direction of the~$I$ secondary users.

\input{tikzfigs/fig-case_study_RRA}
\input{tikzfigs/fig-algorithm_REMS}

\subsubsection{Algorithm  Overview}
The optimization problem in~\eqref{eq:optimization_problem} can be reformulated as
\begin{align}
\label{eq:optimization_problem_rewritten}
    \maximize_{\substack{\mat{Z}_\up{R}\in\mathcal{Z}_\up{R}^R}}
    \max_{\substack{\mat{T}\in\mathbb{C}^{N\times U}}}
    \frac{
        \tilde{P}_\up{signal}(\mat{Z}_\up{R},\mat{T})
    }{
        \tilde{P}_\up{interf.}(\mat{Z}_\up{R},\mat{T})
        +
        \tilde{P}_\up{second.}(\mat{Z}_\up{R},\mat{T})
    }.
\end{align}
Our proposed algorithm heuristically addresses the inner maximization problem in~\eqref{eq:optimization_problem_rewritten} by first approximating the wireless channel matrix between the~$N$ PAs of the RRA and the~$U$ primary users. 
Second, the algorithm applies this channel matrix and the zero-forcing precoding scheme (see~\cite[Sec.~4]{bjoernson_massive_mimo_networks}) to compute the precoding matrix~$\mat{T}$.
Since the set~$\mathcal{Z}_\up{R}^R$ is finite, the outer maximization problem in~\eqref{eq:optimization_problem_rewritten} is of combinatorial nature. 
As the number of choices for~$\mat{Z}_\up{R}$ grows exponentially with~$R$, an exhaustive search is infeasible.
We, therefore, propose a heuristic optimization approach based on coordinate ascent~\cite[Sec.~2.3.1]{bertsekas_nonlinear_programming}, which is computationally efficient. 
Specifically, the algorithm operates iteratively, optimizing the reconfigurable impedance tuple~$\mat{Z}_\up{R}$. 
During each iteration, all but one entry in~$\mat{Z}_\up{R}$ are held fixed. 
For the non-fixed entry, the algorithm tests all values from~$\mathcal{Z}_\up{R}$ to maximize an objective function, which we define below. 

\subsubsection{Algorithm  Details}
The proposed approach is detailed in~\fref{alg:1}. 
The main iteration loop (see lines~\mbox{\ref{al:main_loop_1}--\ref{al:main_loop_2}}) is executed~$I_\up{max}$ times.
In each iteration of the main loop, a new pseudo-random permutation (denoted by the tuple~$\tilde{\mat{R}}$) of the elements in the set~$[R]$ is generated (see line~\ref{al:pseudo_random}). 
Based on this permutation, the algorithm iterates over all~$R$ reconfigurable impedances (see lines~\mbox{\ref{al:r_loop_1}--\ref{al:r_loop_2}}).
During each iteration of this inner loop, all reconfigurable impedances are fixed except one, which is indexed by~$r$.
With an additional innermost loop, the best possible value for the non-fixed impedance is searched over the set~$\mathcal{Z}_\up{R}$ (see lines~\mbox{\ref{al:z_loop_1}--\ref{al:z_loop_2}}). 
Specifically, the reconfigurable impedance tuple~$\mat{Z}_\up{R}^\up{eval}$ used for evaluation in the corresponding coordinate ascent step is first constructed on line~\ref{al:set_in_z}.
Then, the channel to the~$U$ primary users is calculated using the function~$\mat{H}_\up{co}$ on line~\ref{al:channel_H}. 
The function~$\mat{H}_\up{co}$ is defined as follows:
\begin{align}
     \mat{H}_\up{co} ( \mat{Z}_\up{R})
     &\triangleq
        \begin{bmatrix}
            \hat{\vect{q}}_\up{co}^\T(\theta_{\up{U},1}, \varphi_{\up{U},1}) 
            \Big(
            \oper{G}_{\phv{v}_\up{Tx}}^{\phv{a}_\up{F}}
            (\theta_{\up{U},1}, \varphi_{\up{U},1})
            \Big)
            \\
            \vdots
            \\
            \hat{\vect{q}}_\up{co}^\T(\theta_{\up{U},U}, \varphi_{\up{U},U}) 
            \Big(
            \oper{G}_{\phv{v}_\up{Tx}}^{\phv{a}_\up{F}}
            (\theta_{\up{U},U}, \varphi_{\up{U},U})
            \Big)
        \end{bmatrix}\!.
\end{align}
\new{The matrix~$\big(\oper{G}_{\phv{v}_\up{Tx}}^{\phv{a}_\up{F}} (\theta, \varphi)\big)$ corresponds to} the linear map defined by~\mbox{$\mathbb{C}^{N_\up{Tx}}\rightarrow\mathbb{C}^2,\,\vect{v}\mapsto \big(\oper{G}_{\phv{v}_\up{Tx}}\vect{v}\big)(\theta,\phi)$ and~$\hat{\vect{q}}_\up{co}:\Omega\rightarrow\mathbb{C}^2$} \new{is an RRA-dependent function that is intended to approximate the polarization of the transmitted EM waves} as accurately as possible.\footnote{Informally, if all radiating components (e.g., antennas or reflective elements) of the REMS have the same polarization, then~$\hat{\vect{q}}_\up{co}$ should approximate the co-polarization of these radiating components in the far field of the REMS.}
\begin{algorithm}[tp]
    \caption{Physically Accurate Beam- and Null-forming}
    \label{alg:1}
    \begin{algorithmic}[1]
        \renewcommand{\algorithmicrequire}{\textbf{Input:}}
        \renewcommand{\algorithmicensure}{\textbf{Output:}}
        \newcommand{\algrule}[1][.2pt]{\par\vskip.5\baselineskip\hrule height #1\par\vskip.5\baselineskip}
        \REQUIRE~$R_0\!\in\!\mathbb{R}_{>0}$, 
       ~$N\!\in\!\mathbb{N}$, 
       ~$R\!\in\!\mathbb{N}$, 
       ~$\mathcal{Z}_\up{R}\!\subset\! \big\{c\!\in\!\mathbb{C}\big|\Re\{c\}\geq 0\big\}$,
        \\
       ~$Z_\up{R}^\up{init}\in\mathcal{Z}_\up{R}$, 
       ~$I_\up{max}\in\mathbb{N}$,
       ~$(\sigma_i)_{i\in [I_\up{max}]}$, 
       ~$\hat{\vect{q}}_\up{co}:\Omega\rightarrow\mathbb{C}^2$, 
        \\
       ~$U\leq N$, 
       ~$(\theta_{\up{U},1}, \varphi_{\up{U},1}),\ldots,(\theta_{\up{U},U}, \varphi_{\up{U},U})\in\Omega$, 
        \\
       ~$I\in\mathbb{Z}_{\geq 0}$, 
       ~$(\theta_{\up{I},1}, \varphi_{\up{I},1}),\ldots,(\theta_{\up{I},I}, \varphi_{\up{I},I})\in\Omega$ 
        \ENSURE~$\mat{Z}_\up{R}$,~$\mat{T}$
        \algrule
        \text{Initialization}:
        \algrule
        \STATE~$\mat{Z}_\up{R}^\up{best}\triangleq( Z_{\up{R},1}^\up{best}, \ldots, Z_{\up{R},R}^\up{best} )$ ~$\gets$~$(Z_\up{R}^\up{init}, \ldots, Z_\up{R}^\up{init})$
        \STATE~$\mathmakebox[\widthof{$\mathcal{Z}_\up{R}^\up{eval}$}][l]{\mat{T}^\up{best}} \gets \begin{bmatrix}\mat{I}_N & \mat{0}_{N,N-U}\end{bmatrix}^\T$
        \STATE~$\mathmakebox[\widthof{$\mathcal{Z}_\up{R}^\up{eval}$}][l]{f^\up{best}} \gets$~$0$ \label{al:init_f_min}
        \algrule
        \text{Iterative Process}:
        \algrule
        \FOR {$i = 1$ to~$I_\up{max}$}\label{al:main_loop_1}
            \STATE~$\mathmakebox[\widthof{$\mathcal{Z}_\up{R}$}][l]{\tilde{\mat{R}}} \gets$ random permutation of the elements of~$[R]$ \label{al:pseudo_random}
            \FOR {$r$ in~$\tilde{\mat{R}}$}\label{al:r_loop_1}
                \FOR {$z$ in~$\mathcal{Z}_\up{R}$}\label{al:z_loop_1}
                    \STATE~$\mathmakebox[\widthof{$\mathcal{Z}_\up{R}^\up{eval}$}][l]{\mat{Z}_\up{R}^\up{eval}} \gets$~$(\ldots;Z_{\up{R},r-1}^\up{best},z,Z_{\up{R},r+1}^\up{best},\ldots )$ \label{al:set_in_z}
                    \STATE~$\mathmakebox[\widthof{$\mathcal{Z}_\up{R}^\up{eval}$}][l]{\mat{H}_\up{U}}\gets$~$\mat{H}_\up{co} \big( \mat{Z}_\up{R}^\up{eval}\big)$ \label{al:channel_H}
                    \STATE~$\mathmakebox[\widthof{$\mathcal{Z}_\up{R}^\up{eval}$}][l]{\mat{T}^\up{ZF}} \gets$~$\mat{H}_\up{U}^\He \left(\mat{H}_\up{U} \mat{H}_\up{U}^\He\right)^{-1}$\label{al:zero_forcing}
                    \IF {$f \big( \mat{Z}_\up{R}^\up{eval},\sigma_i;\mat{T}^\up{ZF}\big)>f^\up{best}$}\label{al:inner_if}
                        \STATE~$\mathmakebox[\widthof{$f^\up{best}al$}][l]{\mat{Z}_\up{R}^\up{best}} \gets$~$\mat{Z}_\up{R}^\up{eval}$
                        \label{al:update_1}
                        \STATE~$\mathmakebox[\widthof{$f^\up{best}al$}][l]{\mat{T}^\up{best}} \gets$~$\mat{T}^\up{ZF}$ 
                        \label{al:update_2}
                        \STATE~$\mathmakebox[\widthof{$f^\up{best}al$}][l]{f^\up{best}} \gets$~$f \big( \mat{Z}_\up{R}^\up{best},\sigma_i;\mat{T}^\up{best}\big)$
                        \label{al:update_3}
                    \ENDIF
                \ENDFOR\label{al:z_loop_2}
            \ENDFOR\label{al:r_loop_2}
        \ENDFOR\label{al:main_loop_2}
        \algrule
        \RETURN~$\mat{Z}_\up{R}^\up{best}$,~$\mat{T}^\up{best}$
    \end{algorithmic} 
\end{algorithm}

Finally, a potential precoding matrix~$\mat{T}^\up{ZF}$, which is used for evaluation in the current coordinate ascent step, is calculated using zero forcing (see, e.g.,~\cite[Sec.~4]{bjoernson_massive_mimo_networks}) on line~\ref{al:zero_forcing}. 
Note that, for the calculation of the precoding matrix, only the channels to the primary users, but not to the secondary users, are considered.
The coordinate ascent objective function is given by 
\begin{align}
     f \!\left( \mat{Z}_\up{R},\sigma;\mat{T}^\up{ZF}\right)
     &\triangleq
     \frac{
        \tilde{P}_\up{signal}(\mat{Z}_\up{R},\mat{T}^\up{ZF})
        }{
            \tilde{P}_\up{interf.}(\mat{Z}_\up{R},\mat{T}^\up{ZF})
            +
            \tilde{P}_\up{second.}(\mat{Z}_\up{R},\mat{T}^\up{ZF})
            +\sigma
        },
\end{align}
where~$\sigma\geq0$ is a regularization parameter, which weights\footnote{A positive regularization parameter~$\sigma>0$ improves numerical stability.} the contribution of the quasi-power terms~$\tilde{P}_\up{signal}$,~$\tilde{P}_\up{interf.}$, \\and~$\tilde{P}_\up{second.}$.
On line~\ref{al:inner_if}, the algorithm checks whether the objective function, using~$\mat{Z}_\up{R}^\up{eval}$, is strictly greater than~$f_\up{best}$ and updates the respective parameters accordingly on lines~\ref{al:update_1}--\ref{al:update_3}.

\subsection{Numerical Simulations}\label{sec:casestudy}
\new{
We simulated our algorithm in three scenarios using the RRA depicted in~\fref{fig:case_study_RRA}. 
In all scenarios, we set the reference impedance to~$R_0=\SI{50}{\ohm}$ and assume that the reconfigurable impedances are implemented using varactors. 
}
Based on the specifications of the MCE Metelics \mbox{MGV-100-21} varactor~\cite{metelics_mgv_100}, we assume a constant resistance of $\SI{1.2}{\ohm}$ and an adjustable reactance within the range of~$\SI{-196}{\ohm}$ to~$\SI{-14}{\ohm}$ at the operating frequency of $\SI{5.4}{\giga\hertz}$.
The set of available impedance values,~$\mathcal{Z}_\up{R}$, \new{is} constructed by sampling~$32$ uniformly spaced values from this range. 
Furthermore, we set the number of main iterations of our algorithm to~$I_\up{max}=10$ and set the (iteration dependent) regularization parameter \mbox{to~$\sigma_i = 20(0.5)^i$ for~$i\in[I_\up{max}]$}. 
Motivated by the observation that both the two actively fed antennas and the passive \new{reflective} elements are aligned along the x-axis (in Cartesian coordinates), we provide the algorithm with the polarization estimation function~$\hat{\vect{q}}_\up{co}(\theta,\phi)=[\cos(\phi),-\sin(\phi)]^\T$, which corresponds to the co-polarization of a linearly polarized antenna along the x-axis.
To simplify visualization of the simulation results provided next, we introduce negative altitude angles~($\theta<0^\circ$). 
Specifically, for~$\theta\in[-180^\circ,0^\circ)$ and~$\varphi\in\mathbb{R}$, the direction~$(\theta,\varphi)$ refers to~$(-\theta,\varphi+180^\circ)$. 

\new{\subsubsection{Case Study Scenario 1}
In the first scenario, one primary user is placed \mbox{at~$\theta_{\rm{U},1}\!=\!30^\circ$} \mbox{and~$\varphi_{\rm{U},1}\!=\!0^\circ$} and one secondary user is placed \mbox{at~$\theta_{\rm{I},1}\!=\!15^\circ$} \mbox{and~$\varphi_{\rm{I},1}\!=\!0^\circ$}. 
The algorithm is then executed twice: once normally (incorporating both users) and once while ignoring the secondary user. 
\fref{fig:cs_fig_1} shows the resulting REMS gain. 
In both instances, our algorithm achieves a gain of more than $\SI{12}{\decibel}$ in the direction of the primary user. 
This demonstrates that the algorithm can suppress the transmitted radiation intensity in the direction of the secondary user while maintaining the REMS gain toward the primary user. 
}

\new{\subsubsection{Case Study Scenario 2}
In the second scenario, an additional primary user is placed \mbox{at~$\theta_{\rm{U},2}\!=\!0^\circ$} \mbox{and~$\varphi_{\rm{U},2}\!=\!0^\circ$}. 
\fref{fig:cs_fig_2} shows the resulting REMS gain achieved by our algorithm.  
Since there are now two primary users, we show the REMS gain for the case where~$\phv{v}_\up{Tx}=\mat{T}\vect{e}_1$ to visualize the beam when a signal is sent to the first primary user. 
Additionally, we show the REMS gain for the case where~$\phv{v}_\up{Tx}=\mat{T}\vect{e}_2$ to visualize the beam when a signal is sent to the second primary user.
The REMS gain in the direction of both primary users is more than $\SI{8.5}{\decibel}$. 
Moreover, the signals intended for the primary users interfere with each other with a REMS gain of less than $\SI{-25}{\decibel}$ and the REMS gain in the direction of the secondary user is mitigated to less than~$\SI{-25}{\decibel}$.
}
\begin{figure}[t]
    \centering
    {
    \small
    \begin{tikzpicture}
        \draw[white] (2cm,5.3cm) -- (3cm,5.3cm);
        \begin{axis}[%
            after end axis/.append code={
                  \pgfplotsset{
                      axis line style=opaque,
                      ticklabel style=opaque,
                      tick style=opaque,
                      grid=none
                  }
                  \csname pgfplots@draw@axis\endcsname},
            width=7cm,
            height=4.6cm,
            scale only axis,
            xmin=-90,
            xmax=90,
            xtick={-90,-60,-30,0,30,60,90},
            xlabel={$\theta$ (\si{\deg})},
            ymin=-30,
            ymax=25,
            ytick={-30,-20,-10,0,10,20,30},
            scaled ticks=false,
            ylabel ={$G_\up{REMS}(\phv{v}_\up{Tx}=\mat{T}\vect{e}_1;\theta,\varphi=0)$~$(\si{\decibel})$},
            ylabel near ticks,
            grid=both,
            minor tick num=1,
            minor grid style={color=grey_10},
            every minor tick/.style={minor tick length=0pt},
            axis background/.style={fill=white},
            every outer y axis line/.append style={black,very thick},
            legend style={at={(0.03,0.97)},anchor=north west,legend cell align=left, align=left, draw=black,thick, fill opacity=0.9}
            ]

            \draw[black_60, line width=.6, dash pattern=on 5pt off 1pt on 1pt off 1pt] (15,-29.8) -- (15,24.8);

            \draw[white, line width=.7] (30,-29.7) -- (30,23);
            \draw[black_60, line width=.6, dash pattern=on 5pt off 1pt on 1pt off 1pt] (30,-29.8) -- (30,24.8);
            
            \addplot [color=blue_120, line width=1pt] table [col sep=comma] {tikzfigs/data/cs_fig_1_proposed_with.dat};
            \addlegendentry{\footnotesize secondary user incorporated}

            
            \addplot [color=green_120, line width=1pt, dash pattern=on 3.5pt off 1pt] table [col sep=comma] {tikzfigs/data/cs_fig_1_proposed_without.dat};
            \addlegendentry{\footnotesize secondary user ignored}


            \begin{scope}
                \draw[white,fill=white] (-35,6) rectangle (5,9);
                \node[anchor=east] at (6,7.5) {\footnotesize \color{black_60} secondary user};
                \draw [color=black_60, line width=.5pt] plot[smooth, tension=1.2] coordinates {(5,7.5) (9,7) (13,6)};
            \end{scope}

            \begin{scope}
                \draw[white,fill=white] (35,-24) rectangle (80,-21);
                \node[anchor=west] at (38,-22.5) {\footnotesize \color{black_60} primary user};
                \draw [color=black_60, line width=.5pt] plot[smooth, tension=1.2] coordinates {(40,-22.5) (35,-23) (32,-24)};
            \end{scope}
        \end{axis}
    \end{tikzpicture}
    }
    \caption{
    \new{   
        Results for Case Study Scenario 1. 
        The REMS gain resulting from (i)~\fref{alg:1} running normally (incorporating the secondary user) and (ii)~\fref{alg:1} ignoring the secondary user is plotted in the~$\varphi=0^\circ$ slice.
        When the algorithm ignores the secondary user, it manages to achieve a REMS gain of more than $12$\,$\si{\decibel}$, which it can maintain even while the REMS gain toward the secondary user is suppressed to $-24$\,$\si{\decibel}$.
        Both traces were calculated using the proposed REMS modeling method with~\fref{eq:calculate_GA}.
    }
    }
    \label{fig:cs_fig_1}
    \vspace{0.59cm}
    \centering
    {
    \small
    \begin{tikzpicture}
        \begin{axis}[%
            after end axis/.append code={
                  \pgfplotsset{
                      axis line style=opaque,
                      ticklabel style=opaque,
                      tick style=opaque,
                      grid=none
                  }
                  \csname pgfplots@draw@axis\endcsname},
            width=7cm,
            height=4.6cm,
            scale only axis,
            xmin=-90,
            xmax=90,
            xtick={-90,-60,-30,0,30,60,90},
            xlabel={$\theta$ (\si{\deg})},
            ymin=-30,
            ymax=25,
            ytick={-30,-20,-10,0,10,20,30},
            scaled ticks=false,
            ylabel ={$G_\up{REMS}(\phv{v}_\up{Tx};\theta,\varphi=0)$~$(\si{ \decibel})$},
            ylabel near ticks,
            grid=both,
            minor tick num=1,
            minor grid style={color=grey_10},
            every minor tick/.style={minor tick length=0pt},
            axis background/.style={fill=white},
            every outer y axis line/.append style={black,very thick},
            legend style={at={(0.03,0.97)},anchor=north west,legend cell align=left, align=left, draw=black,thick, fill opacity=0.9}
            ]

            \draw[black_light_20, line width=.4] (-90,-20) -- (90,-20);
            \draw[black_light_20, line width=.4] (-90,-10) -- (90,-10);
            \draw[black_light_20, line width=.4] (-90,0) -- (90,0);
            \draw[black_light_20, line width=.4] (-90,10) -- (90,10);
            \draw[black_light_20, line width=.4] (-90,20) -- (90,20);

            \draw[black_light_20, line width=.4] (-60,-30) -- (-60,25);
            \draw[black_light_20, line width=.4] (-30,-30) -- (-30,25);
            \draw[black_light_20, line width=.4] (0,-30) -- (0,25);
            \draw[black_light_20, line width=.4] (30,-30) -- (30,25);
            \draw[black_light_20, line width=.4] (60,-30) -- (60,25);

             \draw[black_60, line width=.6, dash pattern=on 5pt off 1pt on 1pt off 1pt] (15,-29.8) -- (15,24.8);

            \draw[white, line width=.7] (30,-29.7) -- (30,23);
            \draw[black_60, line width=.6, dash pattern=on 5pt off 1pt on 1pt off 1pt] (30,-29.8) -- (30,24.8);

            \draw[white, line width=.7] (0,-29.7) -- (0,23);
            \draw[black_60, line width=.6, dash pattern=on 5pt off 1pt on 1pt off 1pt] (0,-29.8) -- (0,24.8);

            \addplot [color=blue_120, line width=1pt] table [col sep=comma] {tikzfigs/data/cs_fig_2_proposed_e2.dat};
            \addlegendentry{\footnotesize~$\phv{v}_\up{Tx}=\mat{T}\vect{e}_1$}
            
            \addplot [color=green_120, line width=1pt, dash pattern=on 3.5pt off 1pt] table [col sep=comma] {tikzfigs/data/cs_fig_2_proposed_e1.dat};
            \addlegendentry{\footnotesize~$\phv{v}_\up{Tx}=\mat{T}\vect{e}_2$}

            \begin{scope}
                \draw[white,fill=white] (-55,6) rectangle (-10,9);
                \node[anchor=east] at (6-14.5,7.5) {\footnotesize \color{black_60} primary user 1};
                \draw[white,fill=white] (-9,9) rectangle (-5,11);
                \draw [color=black_60, line width=.5pt] plot[smooth, tension=1.2] coordinates {(-12,8.5) (-7,10) (-2,11)};
            \end{scope}

            \begin{scope}
                \draw[white,fill=white] (35,-29+45) rectangle (80,-26+45);
                \draw[white,fill=white] (25,-29+45+1) rectangle (40,-26+45-1);
                \node[anchor=west] at (33,-27.5+45) {\footnotesize \color{black_60} secondary user};
                \draw [color=black_60, line width=.5pt] plot[smooth, tension=1.2] coordinates {(36,-27.5+45) (25,-28+45) (17,-29+45)};
            \end{scope}

            \begin{scope}
                \draw[white,fill=white] (35,-29) rectangle (80,-26);
                \node[anchor=west] at (34,-27.5) {\footnotesize \color{black_60} primary user 2};
                \draw [color=black_60, line width=.5pt] plot[smooth, tension=1.2] coordinates {(36,-27.5) (34,-27.7) (31.5,-28)};
            \end{scope}
        \end{axis}
    \end{tikzpicture}
    }
    \caption{
    \new{   
        Results for Case Study Scenario 2. 
        The REMS gain resulting from~\fref{alg:1} is plotted in the~$\varphi=0^\circ$ slice.
        The blue curve ($\phv{v}_\up{Tx}=\mat{T}\vect{e}_1$) at~$\theta=0^\circ$ and the green curve~($\phv{v}_\up{Tx}=\mat{T}\vect{e}_2$) at~$\theta=30^\circ$ show the REMS gain in the direction of both primary users is slightly more than~$8.5$\,$\si{\decibel}$. Moreover, we see from both curves at~$\theta\in\{0^\circ,15^\circ,30^\circ\}$ that (i) the signals intended for the primary users interfere with each other with a REMS gain of less than~$-25$\,$\si{\decibel}$, and (ii) the REMS gain in the direction of the secondary user is also mitigated to less than~$-25$\,$\si{\decibel}$.
        Both traces were calculated using the proposed REMS modeling method with~\fref{eq:calculate_GA}.
    }    
    }\label{fig:cs_fig_2}
\end{figure}
\begin{figure}[t]
    \centering
    {
    \small
    \begin{tikzpicture}      
        \draw[white] (2cm,5.3cm) -- (3cm,5.3cm);      
        \begin{axis}[%
            width=7cm,
            height=4.6cm,
            scale only axis,
            xmin=-90,
            xmax=90,
            xtick={-90,-60,-30,0,30,60,90},
            xlabel={$\theta$ (\si{\deg})},
            ymin=-30,
            ymax=25,
            ytick={-30,-20,-10,0,10,20,30},
            scaled ticks=false,
            ylabel ={$G_\up{REMS}(\phv{v}_\up{Tx}=\mat{T}\vect{e}_1;\theta,\varphi=0)$~$(\si{\decibel})$},
            ylabel near ticks,
            grid=both,
            minor tick num=1,
            minor grid style={color=grey_10},
            every minor tick/.style={minor tick length=0pt},
            axis background/.style={fill=white},
            every outer y axis line/.append style={black,very thick},
            legend style={at={(0.03,0.97)},anchor=north west,legend cell align=left, align=left, draw=black,thick, fill opacity=0.9}
            ]

            \draw[black_60, line width=.6, dash pattern=on 5pt off 1pt on 1pt off 1pt] (-40,-29.8) -- (-40,24.8);

            \draw[black_60, line width=.6, dash pattern=on 5pt off 1pt on 1pt off 1pt] (40,-29.8) -- (40,24.8);
            
            \addplot [color=blue_120, line width=1pt] table [col sep=comma] {tikzfigs/data/cs_fig_4_with_coupling.dat};
            \addlegendentry{\footnotesize coupling incorporated}

            \addplot [color=green_120, line width=1pt, dash pattern=on 3.5pt off 1pt] table [col sep=comma] {tikzfigs/data/cs_fig_4_no_coupling.dat};
            \addlegendentry{\footnotesize coupling ignored}

            \begin{scope}
                \draw[white,fill=white] (-30,6) rectangle (15,9);
                \node[anchor=east] at (15,7.5) {\footnotesize \color{black_60} secondary user};
                \draw [color=black_60, line width=.5pt] plot[smooth, tension=1.2] coordinates {(47-79,-23+30.5) (44-79,-22.7+30.5) (41-79,-22+30.5)};
            \end{scope}

            \begin{scope}
                \draw[white,fill=white] (45,-24) rectangle (88,-21);
                \node[anchor=west] at (48,-22.5) {\footnotesize \color{black_60} primary user};
                \draw [color=black_60, line width=.5pt] plot[smooth, tension=1.2] coordinates {(50,-22.5) (45,-23.2) (41,-24)};
            \end{scope}
        \end{axis}
    \end{tikzpicture}
    } 
    \caption{
    \new{   
        Results for Case Study Scenario 3. 
        The REMS gain resulting from (i)~\fref{alg:1} running normally and (ii)~\fref{alg:1} ignoring the coupling between the passive \new{reflective} elements is plotted in the~$\varphi=0^\circ$ slice.
        Ignoring the coupling results in a decrease in the REMS gain to the primary user by $1$\,$\si{\decibel}$, the beam directed toward the primary user deviating by $1.5$\,$\si{\degree}$, and the algorithm no longer being able to suppress the REMS gain toward the secondary user. 
        Both traces were calculated using the proposed REMS modeling method with~\fref{eq:calculate_GA}.
    }}
    \label{fig:cs_fig_3}
    \vspace{0.3cm}
    \centering
    {
    \small
    \begin{tikzpicture}      
        \draw[white] (2cm,5.3cm) -- (3cm,5.3cm);      
        \begin{axis}[%
            width=7cm,
            height=4.6cm,
            scale only axis,
            xmin=-90,
            xmax=90,
            xtick={-90,-60,-30,0,30,60,90},
            xlabel={$\theta$ (\si{\deg})},
            ymin=-30,
            ymax=25,
            ytick={-30,-20,-10,0,10,20,30},
            scaled ticks=false,
            ylabel ={$G_\up{REMS}(\phv{v}_\up{Tx}=\mat{T}\vect{e}_1;\theta,\varphi=0)$~$(\si{\decibel})$},
            ylabel near ticks,
            grid=both,
            minor tick num=1,
            minor grid style={color=grey_10},
            every minor tick/.style={minor tick length=0pt},
            axis background/.style={fill=white},
            every outer y axis line/.append style={black,very thick},
            legend style={at={(0.03,0.97)},anchor=north west,legend cell align=left, align=left, draw=black,thick, fill opacity=0.9}
            ]
            
            \addplot [color=blue_120, line width=1pt] table [col sep=comma] {tikzfigs/data/cs_fig_1_proposed_without.dat};
            \addlegendentry{\footnotesize proposed model}

            \addplot [only marks, color=blue_60, line width=1pt, mark=square*, mark size=1.5pt] table [col sep=comma] {tikzfigs/data/cs_fig_3_HFSS.dat};
            \addlegendentry{\footnotesize full-wave EM simulation}
        \end{axis}
    \end{tikzpicture}
    }
    \caption{
    \new{   
        Validation of the proposed REMS modeling method using the output of~\fref{alg:1} in Case Study Scenario 1. 
        The REMS gain obtained~(i) using the proposed REMS modeling method with~\fref{eq:calculate_GA} and (ii) from a full-wave EM simulation is plotted in the~$\varphi=0^\circ$ slice.
        The results of the two methods match each other.
    }
    }
    \label{fig:cs_fig_verify}
\end{figure}

\new{\subsubsection{Case Study Scenario 3}
In the third scenario, one primary user is placed \mbox{at~$\theta_{\rm{U},1}\!=\!40^\circ$} \mbox{and~$\varphi_{\rm{U},1}\!=\!0^\circ$} and one secondary user is placed \mbox{at~$\theta_{\rm{I},1}\!=\!-40^\circ$} \mbox{and~$\varphi_{\rm{I},1}\!=\!0^\circ$}. 
The algorithm is then executed twice: once normally and once while the algorithm ignores coupling between the $100$ ($\lambda/2$-separated) passive \new{reflective} elements. 
The algorithm ignores coupling between the \new{reflective} elements by treating the respective off-diagonal elements of the inter-element coupling matrix~$\mat{S}_{\up{R}_{\up{R}\up{R}}}$ as zero.}
\new{\fref{fig:cs_fig_3} shows the resulting REMS gain achieved by our algorithm. 
Ignoring coupling leads to a reduction in the REMS gain to the primary user by \SI{1}{\decibel}, a deviation of the beam direction toward the primary user by \SI{1.5}{\degree}, and prevents the algorithm from suppressing the REMS gain toward the secondary user. 
In other words, the algorithm that ignores coupling tries but is unable to perform joint beam- and null-forming because it is based on an incomplete model.
}
\begin{rem}
    \new{
        The simulation results of Case~Study~Scenario~3, depicted in~\fref{fig:cs_fig_3}, demonstrate that coupling between the $\lambda/2$-separated passive \new{reflective} elements has a significant impact on the behavior of the RRA. This finding disproves the widespread assumption that the effect of coupling can be neglected for an antenna-to-antenna separation of $\lambda/2$.
    } 
\end{rem}

\subsubsection{Further Validation}
\new{The REMS gain traces in~\fref{fig:cs_fig_1}-\ref{fig:cs_fig_3} were calculated using the proposed REMS modeling method with~\fref{eq:calculate_GA}. 
To validate that these REMS gain values are correct, we conduct a \emph{full-wave EM simulation} of the secondary-user-ignored case in Case~Study~Scenario~1. 
In the full-wave EM simulation, we input (i) the impedance values of the reconfigurable impedances~$\mat{Z}_\up{R}$ and (ii) the amplitudes and phases of~$\phv{v}_\up{Tx}$, as output by~\fref{alg:1}, directly into the simulation software. 
\fref{fig:cs_fig_verify} shows the REMS gain values obtained from both the proposed REMS modeling method and the full-wave EM simulation.
As can be seen, the results from both methods match well, validating the consistency of the proposed REMS modeling methods with physics.
}
%


\section{Limitations}
\label{sec:limitations}

While the proposed modeling method is both \emph{efficient} and \emph{physically consistent}, it has the following key limitations.

First,~\fref{asm:basic_assumption_radiating_part} limits our model to REMSs with no movable parts. 
This prevents our model from describing systems as the ones described in~\cite{lor_phon_lim_reconfigurable_transmissive_metasurface_with_a_combination_of_scissor_and_rotation_actuators_for_independently_controlling_beam_scanning_and_polarization_conversion,veljovic_skrivervik_ultralow_profile_circuilarly_polarized_reflectarray_antenna_for_cubesat_intersatellite_links_in_k_band,huang_pogorzelski_a_ka_band_microstrip_reflectarray_with_elements_having_variable_rotation_angles} or fluid antennas~\cite{wong_shojaeifard_tong_zhang_fluid_antenna_systems}. 
However, as movable parts tend to have a high propensity for mechanical degradation, it remains an open question what role such antenna systems will play in future wireless networks.
\new{Second}, as discussed in~\fref{rem:sampling}, we assumed that radiating structure kernels can be \new{efficiently} approximated arbitrarily well with sampled kernels.
However, we have neither discussed the number of required samples nor how an optimal sampling strategy would look like. A detailed investigation of both of these aspects, which relate to the mathematical complexity of a REMS's far field, is part of ongoing research.

\new{
Third, a complete description of the EM fields surrounding a REMS would encompass both the near- and far-field regions; however, our approach focuses solely on modeling the far-field interaction of REMSs.
Note that modeling the far-field interaction is sufficient for many applications in wireless communication and sensing systems, as the various entities (e.g., different REMSs) are typically located within each other's far-field region.
Nevertheless, there are also applications in which modeling the REMS-near-field interaction is of interest; see, e.g.,~\cite{jiang_shi_zhang_large_scale_ris_enabled_air_ground_channels}.
}

\new{Fourth, the model proposed in~\fref{sec:modeling} is designed for narrowband analysis of REMSs.
However, our model can readily be generalized for broadband analysis, e.g., by treating each frequency of interest separately.
It remains an open question whether the model parameters for all frequencies of interest must be stored individually or if more memory- and compute-efficient approaches exist.}

\section{Conclusions}
\label{sec:conclusions}
In this work, we have proposed an \emph{efficient} and \emph{physically consistent} modeling framework for reconfigurable electromagnetic structures (REMSs).
Our rigorous mathematical formalism combines circuit-theoretic models with our REMS-to-far-field-interaction approach, which enables us to model the interaction of arbitrary REMSs with (incoming and outgoing) electromagnetic waves in the far-field region using a \emph{single} full-wave electromagnetic (EM) simulation.
We have validated our framework in two scenarios by comparing it to Ansys HFSS simulations, which reveals the efficiency and physical accuracy of our model. In addition, we have demonstrated the effectiveness of our model via a case study of a reconfigurable reflectarray (RRA) that performs joint multiuser beam- and null-forming using a novel computationally efficient algorithm that simultaneously optimizes the digital beamforming matrix and the RRA's tunable parameters. All of these results confirm both the efficiency and physical consistency of our framework. 
There are many avenues for future work, some of which we outline next. 
First, systematically investigating how the kernels of the radiating structure should be sampled is an ongoing effort. 
Second, while our REMS model can be generalized to performing a wideband analysis, it is currently unclear whether the model parameters for all frequencies of interest must be stored or if more efficient representations exist.
Third, the algorithm proposed in~Section~\ref{sec:algorithm} was heuristically motivated. We believe that many refinements to our algorithm are possible, including alternative objective functions and more efficient parameter search strategies.

\section{Acknowledgments}
The authors would like to thank Dr.~Raphael Rolny, Gian Marti, and Carolina Nolasco Ferencikova for discussions and suggestions concerning this paper. 
The authors also thank Prof.~Hua Wang for pointing us \new{toward} reference~\cite{babakhani_etal_transmitter_architectures_based_on_near_field_direct_antenna_modulation} and Dr.~Ali Basem for his introduction to Ansys HFSS. 

The work of AST and CS was funded in part by armasuisse; the work of CS was also funded by an ETH Zurich Research Grant and by the Swiss State Secretariat for Education, Research, and Innovation (SERI) under the SwissChips initiative.

\vspace{-2mm}
\appendices
\newpage
\section{Proofs}\label{app:proofs}

\subsection{Proof of~\fref{prop:hilbert_space}}\label{app:proof:hilbert_space}
\begin{proofw}
    To proof that the outgoing far-field power wave pattern~$\phv{a}_\up{F}$ is an element of~$L^2$, we show that
    \begin{align}
        &\!\!\!\|\phv{a}_\up{F}\|_{L^2}^2 
        \nonumber
        \\
        \label{eq:proof_hilbert_1}
        &=
        \lim_{r\rightarrow\infty}
        \oiint\displaylimits_{\Omega} \frac{1}{Z_0}\Big\|\phv{E}^\nearrow(\theta,\varphi)\Big\|^2 \, \sin(\theta)\,\up{d}(\theta,\varphi)
        \\
        \label{eq:proof_hilbert_2}
        &=
        \lim_{r\rightarrow\infty}
        \oiint\displaylimits_{\Omega} \frac{1}{Z_0}\Big\|\phv{E}^\nearrow(\theta,\varphi)\frac{e^{-jkr}}{r}\Big\|^2 \,r^2 \sin(\theta)\, \up{d}(\theta,\varphi)
        \\
        \label{eq:proof_hilbert_3}
        &=
        P^\nearrow.
    \end{align}
    Here,~\eqref{eq:proof_hilbert_1} follows from~\fref{defi:far_field_power_wave_pattern} and~\eqref{eq:radial_component_is_zero};~\eqref{eq:proof_hilbert_2} follows from incorporating the factor~$\frac{e^{-jkr}}{r}$ to obtain the phasors of the electric fields, and~\eqref{eq:proof_hilbert_3} follows because the fields in the far-field region have plane-wave character. 
    Because, by assumption,~$P^\nearrow$ is finite, it follows that~$\phv{a}_\up{F}\in L^2$. 
    The proof for the incoming far-field power wave pattern~$\phv{b}_\up{F}$ proceeds analogously.
\end{proofw}

\subsection{Proof of~\fref{prop:calculate_channel}}\label{app:proof:proposition_channel}
\begin{proofw}
In order to apply our modeling framework to the two-REMSs scenario, we utilize the two spherical coordinate systems as depicted in~\fref{fig:proof_prop_sketch}. 
First, we derive the relationship between the outgoing far-field radiation pattern~$\phv{a}_\up{F}^{(1)}$ of the first REMS and the incoming far-field radiation pattern~$\phv{b}_\up{F}^{(2)}$ of the second REMS.
We use the superposition principle and assume that~$\phv{a}_\up{F}^{(2)}=\mat{0}$, in which case, in the far-field region of REMS~1, it holds that
\begin{align}
    \tilde{\phv{E}}^{(1)}
    \triangleq
    \begin{bmatrix}
        \left[\phv{E}\big(\vect{r}^{(1)}\big)\right]_{\hat{\vect{\theta}}^{(1)}}
        \\
        \left[\phv{E}\big(\vect{r}^{(1)}\big)\right]_{\hat{\vect{\varphi}}^{(1)}}
    \end{bmatrix}
&=
\sqrt{Z_0}
\frac{e^{-j k r^{(1)}}}{r^{(1)}}\phv{a}_\up{F}^{(1)}\!\big(\hat{\vect{r}}^{(1)}\big).
\label{eq:experiment_relationship_far_field_1}
\end{align}
%
We now use the relation~$\vect{r}^{(1)}=\vect{r}^{(2)}+\vect{d}$ to rewrite~\eqref{eq:experiment_relationship_far_field_1} as follows:
\begin{align}
    \tilde{\phv{E}}^{(1)}
    &=
    \sqrt{Z_0}
    \frac{e^{-j k \left\|\vect{r}^{(2)}+\vect{d}\right\|_2 }}{\|\vect{r}^{(2)}+\vect{d}\|_2}\phv{a}_\up{F}^{(1)}\!\!\left( \frac{\vect{r}^{(2)}+\vect{d}}{\|\vect{r}^{(2)}+\vect{d}\|_2} \right)\!.
    \label{eq:experiment_relationship_far_field_2}
\end{align}
Then, we utilize (i) the fact that we are interested in the case where the distance between the two REMSs approaches infinity, and (ii) the continuity assumption in the proposition, which implies that~$\phv{a}_\up{F}^{(1)}$ is a continuous function, to conclude that
\begin{align}
    \lim_{d\rightarrow\infty}
    \tilde{\phv{E}}^{(1)}
    &=
    \sqrt{Z_0}
    \frac{e^{-j k d}}{d}
    \phv{a}_\up{F}^{(1)}\!\big( \hat{\vect{d}} \big)
    e^{-j k  r^{(2)}(\hat{\vect{r}}^{(2)})^\T\hat{\vect{d}}}.
    \label{eq:experiment_relationship_far_field_2_approx}
\end{align}
By comparing the right-hand side of~\eqref{eq:experiment_relationship_far_field_2_approx} with~\eqref{eq:plane_wave}, one can conclude that the electric field in the vicinity to the second REMS has the form of a plane wave traveling in the direction given by the unit vector~$\hat{\vect{d}}$.
It follows from~\eqref{eq:plane_wave_scattering_as_spherical_waves} that, from the perspective of the second REMS, the electric field can be treated as the superposition of an incoming and an outgoing 
\newpage
\noindent
spherical wave of the form
\begin{align}
    \lim_{d\rightarrow\infty}
    \tilde{\phv{E}}^{(1)}
    &=
    \sqrt{Z_0}
    \frac{2\pi}{j k}
    \frac{e^{-j k d}}{d}
    \phv{a}_\up{F}^{(1)}\!\big( \hat{\vect{d}} \big)
    \delta\big(\hat{\vect{r}}^{(2)}+\hat{\vect{d}}\big)
    \frac{e^{+jkr^{(2)}}}{r^{(2)}}
    \nonumber
    \\
    &\;\;\;-
    \sqrt{Z_0}
    \frac{2\pi}{j k}
    \frac{e^{-j k d}}{d}
    \phv{a}_\up{F}^{(1)}\!\big( \hat{\vect{d}} \big)
    \delta\big(\hat{\vect{r}}^{(2)}-\hat{\vect{d}}\big)
    \frac{e^{-jkr^{(2)}}}{r^{(2)}}.
    \label{eq:experiment_plane_wave_as_two_spherical_waves}
\end{align}
From the definition of spherical coordinates, it follows that for positions on the line between the two coordinate origins, it holds that 
\begin{align}
    \begin{bmatrix}
        \left[\phv{E}\big(\vect{r}^{(2)}\big)\right]_{\hat{\vect{\theta}}^{(2)}}
        \\
        \left[\phv{E}\big(\vect{r}^{(2)}\big)\right]_{\hat{\vect{\varphi}}^{(2)}}
    \end{bmatrix}
    =
    \begin{bmatrix}
        1 & 0  \\
        0 & -1 
    \end{bmatrix}
    \begin{bmatrix}
        \left[\phv{E}\big(\vect{r}^{(1)}\big)\right]_{\hat{\vect{\theta}}^{(1)}}
        \\
        \left[\phv{E}\big(\vect{r}^{(1)}\big)\right]_{\hat{\vect{\varphi}}^{(1)}}
    \end{bmatrix}.
    \label{eq:experiment:basis_transformation_matrix}
\end{align}
We can now combine~\eqref{eq:experiment_plane_wave_as_two_spherical_waves}  and~\eqref{eq:experiment:basis_transformation_matrix} with our definition of the far-field power wave pattern in~\eqref{eq:definition_far_field_power_waves_a} and~\eqref{eq:definition_far_field_power_waves_b} to arrive at the desired relationship 
\begin{align} 
    \lim_{d\rightarrow\infty}
    \phv{b}_\up{F}^{(2)}
    \big(\hat{\vect{r}}^{(2)}\big)
    &=
    \delta\big(\hat{\vect{r}}^{(2)}+\hat{\vect{d}}\big)
    \,\mat{C}\,
    \phv{a}_\up{F}^{(1)}\!\big( \hat{\vect{d}} \big),
    \label{eq:experiment_final_relationship_in_out}
\intertext{
where we used the matrix~$\mat{C}$ from~\eqref{eq:definition_matrix_c}.
Similarly, one can show that
}
    \lim_{d\rightarrow\infty}
    \phv{b}_\up{F}^{(1)}
    \big(\hat{\vect{r}}^{(1)}\big)
    &=
    \delta\big(\hat{\vect{r}}^{(1)}-\hat{\vect{d}}\big)
    \,\mat{C}\,
    \phv{a}_\up{F}^{(2)}\!\big( -\hat{\vect{d}} \big).
    \label{eq:experiment_final_relationship_in_out_2}
\end{align}

\begin{figure}[tp]
    \centering
    \begin{tikzpicture}

        \draw [shift={(-2.5,1.5)}, line width=1pt,pattern=south east lines my] plot[smooth cycle] coordinates {(-.5,-.3) (-.7,-.01) (-.6,.3) (0,.7) (.5,.6) (.6,-.2) (.3,-.5) (0,-.5)};
        \node [shift={(-2.5,2.2)},anchor=south west, align=left, fill=white] at (20:0.3) {\small REMS 1};
        \draw [shift={(-2.5,1.5)}, line width=.5pt] plot[smooth, tension=1.3] coordinates {(.9,.9) (.8,.7) (.6,.6)};

        \node [shift={(.1,0.2)},anchor=south west, align=left, fill=white] at (20:0.3) {\small REMS 2};
        \draw [shift={(1.61,-.5)}, line width=1pt,pattern=south east lines my] plot[smooth cycle] coordinates {(-.4,-.3) (-.5,-.01) (-.4,.2) (0,.5) (.4,.9) (.9,.8) (.6,-.2) (.3,-.5) (0,-.5)};
        \draw [shift={(1.61,.5)}, line width=.5pt] plot[smooth, tension=1.3] coordinates {(-.6,-.1) (-.4,-.3) (0,-.4)};
        
        \centerOfC at (-2.5,1.5)
        \centerOfC at (1.61,-.5)

        \draw [line width=0.3pt, color=black_light, arrows = {-Stealth[inset=0, length=6pt, angle'=25]}] (-2.5,1.5) -- (-1,-.9);
        \draw [line width=0.3pt, color=black_light, arrows = {-Stealth[inset=0, length=6pt, angle'=25]}] (1.61,-.5) -- (-1,-.9);
        \draw [line width=0.3pt, color=black_light, arrows = {-Stealth[inset=0, length=6pt, angle'=25]}] (-2.5,1.5) -- (1.61,-.5);

        \node [color=black_light] at (-.5,.95) {\small$\vect{d} = d\hat{\vect{d}}$};
        \node [color=black_light] at (-2.6,0) {\small$\vect{r}^{(1)} = r^{(1)}\hat{\vect{r}}^{(1)}$};
        \node [color=black_light] at (.2,-1.1) {\small$\vect{r}^{(2)} = r^{(2)}\hat{\vect{r}}^{(2)}$};

    \end{tikzpicture}
    \caption{Two REMSs in empty space, each with the origin of a spherical coordinate system located at its center. The spatial displacement of these origins, and therefore of the REMS, is characterized by the vector~$\vect{d}$, which we further factorize as the product between the distance~$d$ and the unit vector~$\hat{\vect{d}}$. As illustrated, any point in space can be characterized by both coordinate systems using the vectors~$\vect{r}^{(1)}$ and~$\vect{r}^{(2)}$. Each of these position vectors can also be factorized into the scalar~$r^{(\cdot)}$ and the unit vector~$\hat{\vect{r}}^{(\cdot)}$.
}
\label{fig:proof_prop_sketch}
\end{figure}
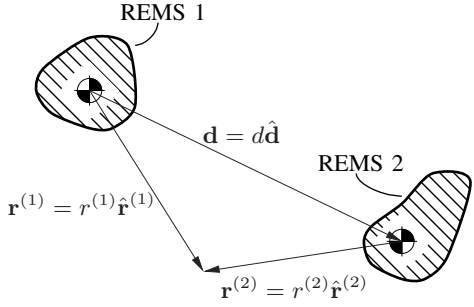

To complete the proof, we now assume that all circuit-theoretic ports of the radiating structures of both REMSs are perfectly matched and that only the~$\ell^{(1)}$th port of the first REMS is stimulated with a non-zero input signal. 
Next, we combine equations~\eqref{eq:matrix_SF},~\eqref{eq:basic_form_SFFT},~\eqref{eq:basic_form_SFTF},~\eqref{eq:basic_form_SFFF},~\eqref{eq:experiment_final_relationship_in_out}, and~\eqref{eq:experiment_final_relationship_in_out_2}, and utilize the geometric series\footnote{This geometric series converges for sufficiently large distances~$d$, as both singular values of~$\mat{M}$ approach~$0$ for~$d\rightarrow\infty$. Consequently, the inverse of~$\mat{I}-\mat{M}$ is guaranteed to exist.} generated by the matrix~$\mat{M}$ from~\eqref{eq:definition_matrix_m} to arrive at 
\begin{align}\label{eq:last_step_proof_prop} 
    \lim_{d\rightarrow\infty}[\phv{b}_{\tilde{R}}^{(2)}]_{\ell^{(2)}}
    =\,&
    \Big(
        \vect{s}_{\up{R}_{\up{R}\up{F}}}^{(2)}
        (\ell^{(2)};-\hat{\vect{d}})
    \Big)^\T
    \mat{C}
    \big(
        \mat{I}_2-\mat{M}
    \big)^{-1}
    \nonumber
    \\
    &\cdot
    \vect{s}_{\up{R}_{\up{F}\up{R}}}^{(1)}
    (\ell^{(1)};\hat{\vect{d}})
    [\phv{a}_{\tilde{R}}^{(1)}]_{\ell^{(1)}},
\end{align}
which concludes the proof. 
\end{proofw}

\subsection{Proof of~\fref{prop:reciprocity}}\label{app:proof_reciprocity}
\begin{proofw}
    By construction, the volume~$V_R$ depicted in~\fref{fig:scetch_reciprocity} contains no sources.
    Because we furthermore assume Lorentz reciprocity holds for the volume~$V_R$ as~$R\rightarrow\infty$, it follows by definition that~\eqref{eq:lorenz_reciprocity_definition} holds for~$V=V_R$ when~$R$ is sufficiently large.
    We divide the surface~$\partial V_R$ into~$S_R$ and~$S_{\up{port},m}$,~$\tilde{S}_{\up{port},m}$ \mbox{for~$m\in[M]$} as depicted in \fref{fig:scetch_reciprocity}.
    Consequently, one can write 
    \begin{align}
        \SI[per-mode = symbol]{0}{\watt}
        &=
        \lim_{R\rightarrow\infty}
        \oiint_{\partial V_R}
        \hat{\vect{n}}^\T
        \big(
        \mat{E}_1 \times \mat{H}_2 
        -
        \mat{E}_2 \times \mat{H}_1 
        \big)
        \up{d} S
        \\
        &=
        \lim_{R\rightarrow\infty}
        \oiint_{S_R}
        \hat{\vect{r}}^\T
        \big(
        \mat{E}_1 \times \mat{H}_2 
        -
        \mat{E}_2 \times \mat{H}_1 
        \big)
        \up{d} S
        \nonumber
        \\
        &\;\;\;+
        \sum_{m\in[M]}
        \iint_{S_{\up{port},m}}
        \hat{\vect{n}}_m^\T
        \big(
        \mat{E}_1 \times \mat{H}_2 
        -
        \mat{E}_2 \times \mat{H}_1 
        \big)
        \up{d} S
        \nonumber
        \\
        &\;\;\;+
        \sum_{m\in[M]}
        \iint_{\tilde{S}_{\up{port},m}}
        \hat{\vect{n}}^\T
        \underbrace{
        \big(
        \mat{E}_1 \times \mat{H}_2 
        -
        \mat{E}_2 \times \mat{H}_1 
        \big)
        }
        _{=\,\SI{0}{\W\per\meter^2}}
        \up{d} S,
        \label{eq:reci_proof_divide_surface}
    \end{align}
    where the last term can be deduced because, inside a perfectly conducting shield, the electric field strength is equal to zero.
    We now analyze the two remaining integral and sum terms in~\eqref{eq:reci_proof_divide_surface} separately.
    It holds that 
    \begin{align}
        &
        \lim_{R\rightarrow\infty}
        \oiint_{S_R}
        \hat{\vect{r}}^\T
        \big(
        \mat{E}_1 \times \mat{H}_2 
        -
        \mat{E}_2 \times \mat{H}_1 
        \big)\,
        \up{d} S
        \nonumber
        \\
        &\quad
        =
        -\frac{2}{Z_0}
        \iint_{\Omega}
        \left(\mat{E}_1^\swarrow\right) ^\T \mat{E}^\nearrow_2 
        -
        \left(\mat{E}_2^\swarrow\right)^\T \mat{E}^\nearrow_1 
        \up{d} \Omega
        \label{eq:reci_proof_far_field_algebra}
        \\
        &\quad
        =
        -2
        \iint_{\Omega}
        \left(\phv{b}_\up{F}^{(1)}\right)^\T
        \phv{a}_\up{F}^{(2)} 
        -
        \left(\phv{b}_\up{F}^{(2)} \right)^\T
        \phv{a}_\up{F}^{(1)} 
        \up{d} \Omega.
        \label{eq:reci_proof_insert_def_far_field_waves}
    \end{align}
    Here,~\eqref{eq:reci_proof_far_field_algebra} follows from the Maxwell-Faraday equa-\\tion \mbox{$\nabla\times\mat{E}= -j\omega\mu\mat{H}$} and the fact that for large~$R$, the electric field on~$S_R$ can be written as a decomposition into an outgoing and an incoming wave, as in~\eqref{eq:far_field_superposition}.
    Equation~\fref{eq:reci_proof_insert_def_far_field_waves} follows directly from~\eqref{eq:definition_far_field_power_waves_a} and~\eqref{eq:definition_far_field_power_waves_b}. 
    We mainly follow the derivation in~\cite[Ch.~1,~Sec.~4]{kerns_plane_wave_scattering_matrix_theory_of_antennas_and_antenna_antenna_interactions}, using our definition of power waves in~\eqref{eq:definition_power_waves_a} and~\eqref{eq:definition_power_waves_b}, along with the assumption that each port supports exactly one propagation mode, to derive that
    \begin{align}
        &\!\!\!\sum_{m\in[M]}
        \iint_{S_{\up{port},m}}
        \hat{\vect{n}}_m\cdot
        \big(
        \mat{E}_1 \times \mat{H}_2 
        -
        \mat{E}_2 \times \mat{H}_1 
        \big)\,
        \up{d} S
        \nonumber
        \\
        &=
        -2
        \hspace{-1.3mm}
        \sum_{m\in[M]}
        \phs{a}_{\tilde{\up{R}},m}^{(1)} 
        \phs{b}_{\tilde{\up{R}},m}^{(2)} 
        -
        \phs{a}_{\tilde{\up{R}},m}^{(2)} 
        \phs{b}_{\tilde{\up{R}},m}^{(1)} 
        \\
        &=
        -2\left(
        \left(\phv{a}_{\tilde{\up{R}}}^{(1)}\right)^\T
        \phv{b}_{\tilde{\up{R}}}^{(2)}  
        -
        \left(\phv{a}_{\tilde{\up{R}}}^{(2)}\right)^\T
        \phv{b}_{\tilde{\up{R}}}^{(1)}\right)
        .
        \label{eq:reci_proof_sum_T_ports}
    \end{align}
By inserting~\eqref{eq:reci_proof_insert_def_far_field_waves} and~\eqref{eq:reci_proof_sum_T_ports} back into~\eqref{eq:reci_proof_divide_surface}, we arrive at
\begin{align}
    \oiint_{\Omega}
    \left(\phv{b}_\up{F}^{(1)}\right)^\T 
    \hspace{-1.5mm}
    \hspace{-0.5mm}
    \phv{a}_\up{F}^{(2)} 
    \!-\!
    \left(\phv{b}_\up{F}^{(2)}\right)^\T
    \hspace{-1.5mm}
    \hspace{-0.5mm}
    \phv{a}_\up{F}^{(1)} 
    \up{d} \Omega
    &=
    \left(\phv{b}_{\tilde{\up{R}}}^{(1)}\right)^\T 
    \hspace{-1.5mm}
    \hspace{-0.5mm}
    \phv{a}_{\tilde{\up{R}}}^{(2)}  
    \!-\!
    \left(\phv{b}_{\tilde{\up{R}}}^{(2)}\right)^\T 
    \hspace{-1.5mm}
    \hspace{-0.5mm}
    \phv{a}_{\tilde{\up{R}}}^{(1)}
    .
\end{align}
Since the incoming waves~$\phv{b}_{\tilde{\up{R}}}$ and~$\phv{b}_\up{F}$ can be chosen arbitrarily and the corresponding outgoing waves~$\phv{a}_{\tilde{\up{R}}}$ and~$\phv{a}_\up{F}$ are fully determined by~\eqref{eq:matrix_SF},~\eqref{eq:basic_form_SFFT},~\eqref{eq:basic_form_SFTF}, and~\eqref{eq:basic_form_SFFF}, it holds that 
\begin{align}
    \SI[per-mode = symbol]{0}{\watt}&=
    \left(\phv{a}_{\tilde{\up{R}}}^{(1)}\right)^\T 
    \Big(
        \mat{S}_{\up{R}_{\up{R}\up{R}}}
        -
        \mat{S}^\T_{\up{R}_{\up{R}\up{R}}}
    \Big)
    \phv{a}_{\tilde{\up{R}}}^{(2)}
\label{eq:proof_reci_last_SFTT}
\\
    \SI[per-mode = symbol]{0}{\watt}&=
    \hspace{-1.3mm}
    \sum_{m\in[M]}
    \phs{a}_{\tilde{\up{R}},m}^{(2)} 
    \big\langle\phv{b}_\up{F}^{(1)},
    \overline{\vect{s}}_{\up{R}_{\up{F}\up{R}}}(m;\cdot)
    -
    \overline{\vect{s}}_{\up{R}_{\up{R}\up{F}}}(m;\cdot)
    \big\rangle_{L^2}
\label{eq:proof_reci_last_SFFT}
\\
    \SI[per-mode = symbol]{0}{\watt}&=\oiint_{\Omega}
    \oiint_{\Omega'} \hspace{-1mm}
    \left(\phv{b}_\up{F}^{(2)}(\hat{\vect{r}})\right)^\T 
    \Big(
    \mat{S}_{\up{R}_{\up{F}\up{F}}}(\hat{\vect{r}};\hat{\vect{r}}')
    \nonumber
    \\
    &\quad -
    \mat{S}^\T_{\up{R}_{\up{F}\up{F}}}(\hat{\vect{r}}';\hat{\vect{r}})
    \Big)
    \phv{b}_\up{R}^{(1)}(\hat{\vect{r}}')
    \,\up{d}\Omega'
    \up{d}\Omega.
\label{eq:proof_reci_last_SFFF}
\end{align}
Here,~\eqref{eq:proof_reci_last_SFTT} follows from setting~$\phv{b}_\up{F}^{(1)}$ and~$\phv{b}_\up{F}^{(2)}$ to zero,~\eqref{eq:proof_reci_last_SFFT} follows from setting~$\phv{a}_{\tilde{\up{R}}}^{(1)}$ and~$\phv{b}_\up{F}^{(2)}$ to zero, and~\eqref{eq:proof_reci_last_SFFF} follows from setting~$\phv{a}_{\tilde{\up{R}}}^{(1)}$ and~$\phv{a}_{\tilde{\up{R}}}^{(2)}$ to zero. 
Finally, we use the fact that~$\phv{a}_{\tilde{\up{R}}}$ and~$\phv{b}_\up{F}$ can be chosen arbitrarily to conclude from~\eqref{eq:proof_reci_last_SFTT},~\eqref{eq:proof_reci_last_SFFT}, and~\eqref{eq:proof_reci_last_SFFF} that the Proposition holds.
\end{proofw}

\subsection{Proof of Equation~\fref{eq:basic_form_SFFT}}\label{app:proof:eq:basic_form_SFFT}
\begin{proofw}[Proof of~\eqref{eq:basic_form_SFFT}]
Since~$\oper{S}_{\up{R}_{\up{F}\up{R}}}$ is a finite-rank operator with rank not larger than~$M$, it can be represented using the following canonical form of finite-rank operators (cf.~\cite[Thm.~6.1]{weidmann_linear_operators_in_hilbert_spaces}):
\begin{align}
    \oper{S}_{\up{R}_{\up{F}\up{R}}}\vect{v}
    &=
    \sum_{n\in[M]} \alpha_n \langle \vect{v} , \vect{g}_n \rangle \tilde{\vect{g}}_n.
\end{align}
Here,~$\{\vect{g}_n\}_{n\in[M]}$,~$\vect{g}_n\!\!\in\!\mathbb{C}^M$ and~$\{\tilde{\vect{g}}_n\}_{n\in[M]}$,~$\tilde{\vect{g}}_n\!\!\in\! L^2$ are orthonormal bases, and~$\{\alpha_n\}_{n\in[M]}$ are coefficients in~$\mathbb{C}$.
We now expand each element~$\vect{g}_n$ into the standard basis of~$\mathbb{R}^M$, denoted by~$\{\vect{e}_{m}\}_{m\in[M]}$, and write
\begin{align}
    &\!\!\!\big(\oper{S}_{\up{R}_{\up{F}\up{R}}}\vect{v}\big)(\theta,\varphi)
    =
    \sum_{n\in[M]} \alpha_n \langle \vect{v} , \vect{g}_n \rangle 
    \tilde{\vect{g}}_n(\theta,\varphi)
    \nonumber
    \\
    &=
    \sum_{n\in[M]} \alpha_n \Big\langle \vect{v} , \sum_{m\in[M]} \langle \vect{g}_n , \vect{e}_{m}\rangle \vect{e}_{m} \Big\rangle 
    \tilde{\vect{g}}_n(\theta,\varphi)
    \label{eq:proof_basic_form_SFFT_expansion_step}
    \\
    &=
    \sum_{m\in[M]} 
    \sum_{n\in[M]}\alpha_n \langle \vect{v} , e_{m} \rangle  \overline{\langle \vect{g}_n,\vect{e}_m\rangle} 
    \tilde{\vect{g}}_n(\theta,\varphi)
    \label{eq:proof_basic_form_SFFT_linearity_step}
    \\
    &=
    \sum_{m\in[M]} 
    \underbrace{
    \sum_{n\in[M]}\alpha_n  \overline{[\vect{g}_n]}_m 
    \tilde{\vect{g}}_n(\theta,\varphi) 
    }_{\triangleq\,\vect{s}_{\up{R}_{\up{F}\up{R}}}(m;\theta,\varphi)}
    [\vect{v}]_m,
    \label{eq:proof_basic_form_SFFT_reordering_step}
\end{align}
where we perform the above-mentioned extension to obtain~\eqref{eq:proof_basic_form_SFFT_expansion_step}. Here,~\eqref{eq:proof_basic_form_SFFT_linearity_step} is a consequence of the linearity characteristics of inner products and where~\eqref{eq:proof_basic_form_SFFT_reordering_step} simply involves reordering terms.
\end{proofw}


\subsection{Proof of Equation~\fref{eq:basic_form_SFTF}}\label{app:proof:eq:basic_form_SFTF}
\begin{proofw}
We exploit that~$\oper{S}_{\up{R}_{\up{R}\up{F}}}$ is a finite-rank operator with rank not larger than~$M$. 
Therefore we can represent the operator using the following canonical form of finite-rank operators (cf.~\cite[Thm.~6.1]{weidmann_linear_operators_in_hilbert_spaces}):
\begin{align}
    \big[\oper{S}_{\up{R}_{\up{R}\up{F}}}\vect{f}\big]_m
    &=
    \Big[\sum_{n\in[M]} \alpha_n \langle \vect{f} , \tilde{\vect{g}}_n \rangle \vect{g}_n\Big]_m
\end{align}
with~$m\in[M]$. 
Here,~$\{\vect{g}_n\}_{n\in[M]}$,~$\vect{g}_n\!\!\in\! \mathbb{C}^M$ and $\{\tilde{\vect{g}}_n\}_{n\in[M]}$, $\tilde{\vect{g}}_n\!\!\in\! L^2$ are orthonormal bases and~$\{\alpha_n\}_{n\in[M]}$ are coefficients in~$\mathbb{C}$.
We now expand each element~$\vect{g}_n$ into the standard basis of~$\mathbb{R}^M$, denoted by~$\{\vect{e}_{m'}\}_{m'\in[M]}$, and write
\begin{align}
    \big[\oper{S}_{\up{R}_{\up{R}\up{F}}}\vect{f}\big]_m
    &=
    \Big[\sum_{n\in[M]} \alpha_n \langle \vect{f} , \tilde{\vect{g}}_n \rangle_{L^2} \vect{g}_n\Big]_m
    \nonumber
    \\
    &=
    \Big[\sum_{n\in[M]} \alpha_n \langle \vect{f} , \tilde{\vect{g}}_n \rangle_{L^2} \sum_{m'\in[M]} \langle \vect{g}_n , \vect{e}_{m'}\rangle_{\mathbb{C}^M} \vect{e}_{m'} \Big]_m
    \label{eq:proof_basic_form_SFTF_expansion_step}
    \\
    &=
    \sum_{n\in[M]} \alpha_n \langle \vect{f} , \tilde{\vect{g}}_n \rangle_{L^2}  [\vect{g}_n]_m  
    \label{eq:proof_basic_form_SFTF_filter_step}
    \\
    &=
    \big\langle \vect{f} ,
    \underbrace{
    \sum_{n\in[M]} \overline{(\alpha_n [\vect{g}_n]_m)}
    \tilde{\vect{g}}_n 
    }_{\triangleq\,\overline{\vect{s}}_{\up{R}_{\up{R}\up{F}}}(m;\theta,\varphi)}
    \big\rangle_{L^2},
    \label{eq:proof_basic_form_SFTF_linearity_step}
\end{align}
where we perform the above-mentioned extension to obtain~\eqref{eq:proof_basic_form_SFTF_expansion_step}; 
where~\eqref{eq:proof_basic_form_SFTF_filter_step} follows from the filter characteristics of the standard basis;
and where~\eqref{eq:proof_basic_form_SFTF_linearity_step} is a consequence of the linearity characteristics of inner products.
\end{proofw}


\subsection{Proof of Right-Hand Side Equalities in~\fref{defi:basic_power_metrics}}\label{app:proof_power_metrics_basic}
\begin{proofw}
    It follows from~\fref{defi:basic_power_metrics} and the definition of power waves in~\eqref{eq:definition_power_waves_a} and~\eqref{eq:definition_power_waves_b} that 
    \begin{align}
        P_\up{T}
        &\triangleq
        \Re\{\phv{v}_\up{T}^H\phv{i}_\up{T}\}
        \\
        &=
        \Re\{\sqrt{R_0}(\phv{a}_\up{T}+\phv{b}_\up{T})^H(\sqrt{R_0})^{-1}(\phv{a}_\up{T}-\phv{b}_\up{T})\}
        \\
        &=
        \Re\big\{(\phv{a}_\up{T}+\phv{b}_\up{T})^H(\phv{a}_\up{T}-\phv{b}_\up{T})\big\}
        \\
        &=
        \|\phv{a}_\up{T}\|^2_2
        -\|\phv{b}_\up{T}\|^2_2
        +
        \Re\Big\{
            2
            j
            \Im\big\{
            \phv{b}_\up{T}^H
            \phv{a}_\up{T}
            \big\}
        \Big\}
        \\
        &=
        \|\phv{a}_\up{T}\|^2_2
        -\|\phv{b}_\up{T}\|^2_2.
    \end{align}
    The right-hand side of~\eqref{eq:defi_PR} can be proven analogously, and the right-hand side of~\eqref{eq:defi_PF} follows directly from the proof of~\fref{prop:hilbert_space}.
\end{proofw}


\subsection{Proof of Right-Hand Side Equality in~\fref{defi:power_metrics_PA}}\label{app:proof_power_metrics_PA}
\begin{proofw}
    Let~$\phv{v}_\Gamma=\,\mat{0}$,~$\phv{i}_\Gamma=\,\mat{0}$,~$\phv{v}_\Upsilon=\,\mat{0}$, and~$\phv{b}_\up{F}=\,\mat{0}$. 
    We will now demonstrate that the term~$P_\up{T}$ is upper bounded by 
    \begin{align}
        \label{eq:proof_PA:bound_1}
        P_\up{T}
        &=
        \sum_{n\in [N]}
        \Re
        \big\{
            \phv{v}_{\up{T},n}\phv{i}_{\up{T},n}
        \big\}
        =
        \sum_{n\in [N_\up{Tx}]}
        \Re
        \big\{
            \phv{v}_{\up{T},n}\phv{i}_{\up{T},n}
        \big\}
        \\
        &=
        \sum_{n\in [N_\up{Tx}]}
        \frac{1}{4 \Re\{\mat{Z}_{\up{Tx},n}\}}
        \Big(
            \left|
                \phv{v}_{\up{T},n}
                +
                \mat{Z}_{\up{Tx},n}
                \phv{i}_{\up{T},n}
            \right|^2
            \nonumber
            \\
            &\;\;\;
            -
            \left|
                \phv{v}_{\up{T},n}
                -
                \mat{Z}_{\up{Tx},n}
                \phv{i}_{\up{T},n}
            \right|^2
        \Big)
        \\
        \label{eq:proof_PA_leq}
        &\leq
        \sum_{n\in [N_\up{Tx}]}
        \frac{1}{4 \Re\{\mat{Z}_{\up{Tx},n}\}}
        \left|
            \phv{v}_{\up{T},n}
            +
            \mat{Z}_{\up{Tx},n}
            \phv{i}_{\up{T},n}
        \right|^2
        \\
        &=
        \sum_{n\in [N_\up{Tx}]}
        \!
        \frac{1}{4}
        \overline{
            \left(
                \phv{v}_{\up{T},n}
                \! 
                +
                \!
                \mat{Z}_{\up{Tx},n}
                \phv{i}_{\up{T},n}
            \right)
        }
        \Re\{\mat{Z}_{\up{Tx},n}\}^{-1}\!
        \left(
            \phv{v}_{\up{T},n}
            \!
            +
            \! 
            \mat{Z}_{\up{Tx},n}
            \phv{i}_{\up{T},n}
        \right)
        \\
        &=
        \sum_{n\in [N_\up{Tx}]}
        \frac{1}{4}
        \overline{
            \phv{v}
        }_{\up{Tx},n}
        \Re\{\mat{Z}_{\up{Tx},n}\}^{-1}
        \phv{v}_{\up{Tx},n}
        \\
        &=
        \label{eq:proof_PA:bound}
        \frac{1}{4}
        \phv{v}_\up{Tx}^H
        \Re\{\mat{Z}_\up{Tx}\}^{-1}
        \phv{v}_\up{Tx}.
    \end{align}
Here, the second equality in~\eqref{eq:proof_PA:bound_1} follows from the fact that the LNA blocks are passive systems, as we assumed that~\mbox{$\phv{v}_\Gamma=\,\mat{0}$} and~\mbox{$\phv{i}_\Gamma=\,\mat{0}$,} and~\eqref{eq:proof_PA_leq} follows from~\fref{asm:basic_assumption_PA_LNA}.
To show that the bound in~\eqref{eq:proof_PA:bound} is tight, we now demonstrate that for~$\mat{S}_\up{T}=
\begin{bmatrix}
    \mat{S}_{\up{T}_\up{TT}}^\up{Tx} & \mat{0} \\ \mat{0} & \mat{0}
\end{bmatrix}$ \\and~$\mat{S}_{\up{T}_\up{TT}}^\up{Tx}=(\overline{\mat{Z}}_\up{Tx}+R_0\mat{I}_{N})^{-1}(\overline{\mat{Z}}_\up{Tx}-R_0\mat{I}_{N})$,
it holds that
\begin{align}
        P_\up{T}
    &=
        \label{eq:proof_PA_existence_1}
        \Re
        \big\{
            \phv{v}_\up{T}^H\phv{i}_\up{T}
        \big\}
    \\
    &=
        \Re
        \big\{
            (\overline{\mat{Z}}_\up{Tx}\phv{i}_\up{T})^H\phv{i}_\up{T}
        \big\}
    \\
    &=
        \frac{1}{4}
        \Re
        \big\{
            (\overline{\mat{Z}}_\up{Tx}\Re\{\mat{Z}_\up{Tx}\}^{-1}
            \phv{v}_\up{Tx})^H
            \Re\{\mat{Z}_\up{Tx}\}^{-1}
            \phv{v}_\up{Tx}
        \big\}
    \\
    &=
        \label{eq:proof_PA_last}
        \frac{1}{4}
        \phv{v}_\up{Tx}^H
        \Re\{\mat{Z}_\up{Tx}\}^{-1}
        \phv{v}_\up{Tx},
\end{align}
where~\eqref{eq:proof_PA_existence_1} follows from recognizing that, for the chosen scattering matrix~$\mat{S}_\up{T}$, the input impedance of the tuning network seen by the PAs is equal to~$\overline{\mat{Z}}_\up{Tx}$. 
This concludes the proof.
\end{proofw}

\balance
\bstctlcite{IEEEexample:BSTcontrol} 
\bibliographystyle{IEEEtran}
\bibliography{bib/publishers,bib/IEEE_abbr,bib/journals_proceedings_ect,bib/library}
\balance

\end{document}